\documentclass[11pt]{article}
\usepackage{latexsym}
\usepackage{amsmath}
\usepackage{multibox}
\usepackage{verbatim}
\usepackage{mathrsfs}

\usepackage{amssymb,amsmath}
\usepackage{amsthm}
\usepackage{amscd}
\usepackage{amssymb}
\usepackage{amsfonts}
\usepackage{url}
\usepackage{slashed}
\usepackage[latin1]{inputenc}
\usepackage[english]{babel}

 \csname
@addtoreset\endcsname{equation}{section}

\def\gz0{\gamma^{0}}

 \def\det{{\rm det\,}}

\def\ket#1{|#1\rangle}
\def\scs#1{\section{\sc #1}}
\def\scss#1{\subsection{\sc #1}}
\def\scsss#1{\subsubsection{\sc #1}}

\def\a{\alpha}
\def\b{\beta}
\def\g{\gamma}
\def\G{\Gamma}
\def\d{\delta}

\def\e{\epsilon}

\def\z{\zeta}

\def\l{\lambda}
\def\L{\Lambda}
\def\m{\mu}
\def\n{\nu}
\def\x{\xi}

\def\p{\pi}

\def\r{\rho}

\def\s{\sigma}

\def\f{\phi}

\def\cA{{\cal A}}

\def\cD{{\cal D}}

\def\cF{{\cal F}}
\def\cG{{\cal G}}

\def\cI{{\cal I}}
\def\cJ{{\cal J}}

\def\cL{{\cal L}}

\def\cP{{\cal P}}

\def\cS{{\cal S}}
\def\cT{{\cal T}}

\def\cV{{\cal V}}

\def\be{\begin{equation}}
\def\ee{\end{equation}}
\def\bs{\begin{split}}
\def\es{\end{split}}
\def\bea{\begin{eqnarray}}
\def\eea{\end{eqnarray}}
\def\ba{\begin{array}}
\def\ea{\end{array}}
\def\bec{\begin{center}}
\def\ec{\end{center}}
\def\ba{\begin{align}}
\def\ena{\end{align}}

\def\12{\frac{1}{2}}

\def\bra{\langle \,}
\def\ket{\, \rangle}

\def\ra{\rightarrow}

\def\lra{\leftrightarrow}

\def\psisl{\not {\!\! \psi}}

\begin{document}

\begin{flushright}
{\today}
\end{flushright}

\vspace{25pt}

\begin{center}
{\Large\sc String Lessons for Higher-Spin Interactions}\\
\vspace{25pt}
{\sc A.~Sagnotti \ and \ M.~Taronna}\\[15pt]
{\sl\small
Scuola Normale Superiore and INFN\\
Piazza dei Cavalieri, 7\\I-56126 Pisa \ ITALY \\
e-mail: {\small \it
sagnotti@sns.it, m.taronna@sns.it}}\vspace{10pt}
\vspace{35pt}

{\sc\large Abstract}

\end{center}
{String Theory includes a plethora of higher--spin excitations, which clearly
lie behind its most spectacular properties, but whose detailed behavior is
largely unknown. Conversely, string interactions contain much useful
information on higher--spin couplings, which can be very valuable in current
attempts to characterize their systematics. We present a simplified form for
the three--point (and four--point) amplitudes of the symmetric tensors belonging
to the first Regge trajectory of the open bosonic string and relate them to local
couplings and currents. These include the cases first discussed, from a field
theory perspective, by Berends, Burgers and van Dam, and generalize their
results in a suggestive fashion along lines recently explored by Boulanger, Metsaev and others. We also comment on the recovery of gauge symmetry in the low--tension limit, on the current--exchange amplitudes that can be built from these couplings and on the extension to mixed--symmetry states.}

\setcounter{page}{1}
\pagebreak
{\linespread{0.75}\tableofcontents}
\newpage


\scs{Introduction}\label{sec:intro} String Theory \cite{stringtheory} is
regarded as a promising scheme for the Fundamental Interactions, with the
peculiar and enticing feature that gravity is somehow forced into the
picture. Yet, our understanding is still rather primitive, so much so that even some serious
difficulties that are to be confronted when trying to link the whole framework to Particle Physics might find better qualifications once the roots of the subject
will rest on firmer grounds. Surely enough, string dualities have led to the
emergence of the remarkable M--theory picture, and non--perturbative methods are
being developed starting from the first--quantized formulation, but the
mechanical model of strings, while strikingly convenient to set up a
perturbative expansion, shows clear limitations when one tries to arrive at
background independent statements.

The typical choice of Planck--scale tensions has long limited the analysis of
string excitations to the massless sector, which is held directly responsible
for long--range interactions. Still, the very consistency of String Theory rests
on the presence of infinite towers of massive modes and thus, a fortiori, on
the presence of a plethora of \emph{massive higher--spin (HS) excitations}
\cite{solvay}. To lowest order, these dispose themselves along linear Regge
trajectories in the mass--squared--spin plane, so that for instance, for the
first Regge trajectory of the open bosonic string,
\be \a^{\,\prime}\,M_s^{\,2}\,=\,s\,-\,1\ , \ee
where $\a^{\,\prime}$, the so--called Regge slope, is inversely proportional to the string tension.
It is hard to resist the feeling that a closer look at the HS
excitations be the key for a deeper understanding of String Theory. Conversely,
String Theory adds important motivations for the study of HS theories, and can
provide a fruitful laboratory to probe their unusual dynamics. It has long been
appreciated by a number of authors \cite{Gross,AmatiCiafaloniVeneziano} that the massive HS excitations play an important role in the still largely unexplored high--energy regime of String Theory (for recent developments along these lines, see \cite{Taiwan}), but the question that we have in mind here has perhaps a wider scope. It concerns the actual place of String Theory within the possibly far wider realm of HS Gauge Theory.

A key lesson of lower--spin theories has been that massive gauge fields
are naturally associated with broken symmetries. This correspondence underlines the Standard Model of Particle Physics and has long
served as a key source of inspiration, with two main consequences in this
context. The first is a cogent motivation to concentrate on \emph{massless} HS
theories, the very case that in flat space is to confront long--recognized
difficulties. In analogy with the more familiar low--spin cases, one would
like to recover from massless HS theories their massive counterparts,
particularly in connection with string spectra, focussing on symmetry breaking
mechanisms and on the nature of the available spontaneously broken phases. The
second is that these ideas foster a long--held feeling that our current
understanding of String Theory is seriously incomplete. For instance,
the string tension determines the masses of the HS excitations, and yet it
lacks a dynamical origin, a familiar fact that has nonetheless a peculiar and puzzling manifestation in some relatively simple orientifold models \cite{desc} where supersymmetry is broken, in an apparently rigid fashion, at the string scale \cite{bsb}. The ensuing non--linear realization in the low--energy field theory \cite{nonlin} goes in hand with our lack of control of the phenomenon, a superHiggs--like mechanism whose order parameter appears inaccessible. At the same time, a number of qualitative arguments, suggested by the AdS/CFT correspondence \cite{adscft} and stimulated by the original observations in \cite{Sezgin:2002rt}, including the interesting analysis of \cite{Bianchi:2003wx}, lend further support to the symmetry--breaking scenario, and yet lack a clear dynamical setting within String Theory.

We have at present a satisfactory understanding of free HS
fields. Two distinct approaches have come to terms with a problem that, in some respects, dates back to the early days of Quantum Field Theory (see, \emph{e.g.}, \cite{Majorana:1932rj}).
The first is the ``metric--like'' approach: it builds upon the work of Singh and Hagen \cite{SH} and Fronsdal \cite{Fronsdal}, and more recently was connected to the geometric ideas of \cite{de Wit:1979pe}. All in all, the non--Lagrangian equations \cite{FranciaSagnotti}
\be
\frac{1}{\square^{\,n}\!\!}\ \,\partial\cdot{\mathcal{R}^{[n]}}_{;\,\m_1\ldots\,\m_{2n+1}}\,=\,0\
 \label{oddnl} \ee
for odd spins $s\,=\,2n+1$, and
\be \frac{1}{\square^{\,n-1}\!\!}\ \,{\mathcal{R}^{[n]}}_{;\, \m_1\ldots\,\m_{2n}}\,=\,0\ \label{evenl} \ee
for even spins $s\,=\,2n$, link the linearized HS curvatures to the dynamics of an infinite set of free symmetric tensors $\phi_{\m_1 \ldots \m_s}$. Their fermionic counterparts for multi--symmetric spinor tensors $\psi_{\m_1 \ldots \m_s}$ were further developed in \cite{franciamass}, and make use of fermionic analogs of the HS curvatures.
Notice that eqs.~\eqref{oddnl} and \eqref{evenl} are \emph{non local} for spin $s \geq 3$ but include three familiar \emph{local} low--spin cases, the Maxwell equation for $s=1$, the linearized Einstein equation for $s=2$, and formally even the Klein--Gordon equation for $s=0$. One has thus a glimpse of how the geometric framework of Einstein gravity might eventually encompass HS fields. More recently, these results have found their place within a ``minimal'' local formulation of symmetric HS fields that rests, for any given $s$, on at most two additional fields \cite{minimal}\footnote{These local Lagrangians may be regarded
as simplified forms of the previous BRST constructions of
\cite{Buchbinder:2001bs,fotopoulos}.}. The constructions for symmetric (spinor~-~)tensors that we have just described afford
interesting generalizations to mixed--symmetry fields \cite{mixed}. To begin
with, non--local equations for multi--symmetric mixed--symmetry fields of the type
$\phi_{\m_1 \ldots \m_{s_1} \, ;\, \n_1 \ldots \n_{s_2}\,;\, \ldots}$ were
first discussed in \cite{Bekaert:2002dt}. Moreover, the minimal local
Lagrangians also generalize to these systems \cite{mixed}, in a way that
extends and completes the pioneering work done by Labastida and others in the
1980's \cite{labastida}. On the other hand the second approach, usually
referred to as ``frame--like'' \cite{Vasiliev:1988sa}, aims at extending to HS
fields the Cartan--Weyl framework and has led already to the Vasiliev systems,
consistent non--linear equations coupling together an infinite set of symmetric
tensors of arbitrary ranks. The Vasiliev setting, originally defined in four
dimensions and more recently extended, up to some technical subtleties, to
space times with arbitrary numbers of dimensions, provides a remarkable
counterexample to long--held difficulties and no--go theorems and takes a
deceptively simple form in an extended non--commutative space time. However, the
equations are of first order, which has long hampered their association to a
satisfactory action principle and, perhaps more importantly  for our present
purposes, in its bosonic incarnation String Theory is naturally closer to the
metric--like form.

Despite the success of the Vasiliev approach, our current understanding of the
systematics of HS interactions remains very limited, although explicit
investigations of HS vertices have by now a long history, that dates back to
classic works of the 1980's, by Bengtsson, Bengtsson and Brink \cite{b3} (in
the light--cone formulation) and by Berends, Burgers and van Dam \cite{bbvd} (in
the covariant formulation, that is closer in spirit to the present work). The
covariant approach has been extended over the years by a number of authors, and
most notably by Boulanger and collaborators \cite{cubic}. A crucial new
addition to this picture is the identification, in \cite{Boulanger:2008tg}, of
higher--derivative ``seeds'' for the cubic couplings, whose structure conforms
to the important light--cone analysis of Metsaev \cite{Metsaev:2007rn}. This is
an important step, since following the original observation of
\cite{arag_deser} that the minimal coupling to gravity is problematic for HS
fields, Fradkin and Vasiliev \cite{Fradkin:1986qy} showed long ago that the
problem can be cured, in the presence of a cosmological term $\Lambda$, via a
chain of higher--derivative interactions sized by negative powers of $\Lambda$.
Technically, what \cite{Boulanger:2008tg} added to this picture is the
identification of a key scaling limit for $\Lambda$ and the gravitational
coupling that wipes out the lower members of the tail. The surviving
higher--derivative ``seeds'' have in general a multi--polar character,
precisely as needed to circumvent long--recognized difficulties with long--range
HS forces \cite{NoGo}, and can give rise to the lower terms by Kaluza--Klein
reductions along the lines of \cite{SS} or by radial reductions along the lines
of \cite{radial}. As we shall see, String Theory has important lessons on the
actual nature of these basic couplings. While some
obstructions seem to arise for \emph{odd--spin} HS fields, recently
complete cubic vertices for \emph{even--spin} HS fields were obtained, in
covariant form, by Manvelyan, Mkrtchyan and R\"uhl \cite{Manvelyan}. Moreover,
some peculiar scattering amplitudes involving exotic \emph{massless} HS states,
that can somehow arise in String Theory within a decoupled sector, were
computed in \cite{DimaPolyakov}. There is thus a real chance, at this time, to
gain some control of quartic and higher vertices, coming also to terms with the
non--local nature that one expects to emerge in the massless limit. It is
perhaps worth stressing, in this respect, that the results of
\cite{FranciaSagnotti,franciamass}, and the subsequent analysis of current
exchanges of \cite{minimal}, can be regarded as simple and yet concrete
indications to the effect that HS symmetries can make special non--local
couplings fully compatible with the locality of physical observables.

This paper is aimed at improving our understanding of the deep and far
reaching connections between String Theory and HS Gauge Theory. Important facts
have emerged over the years \cite{Buchbinder:1999ar}, but so far they have mainly concerned free fields.
Thus, for instance, the ``triplets'' \cite{triplets,FranciaSagnotti,francia10} emerging in the low--tension limit of String Theory have long served as a source of inspiration for geometric forms of HS dynamics, but far more can be gained from a
closer look at tree--level string interactions. The potential lessons that we
are after in this investigation concern both the actual form of the cubic (and
higher--order) interactions between massive HS modes and the currents that
determine them. For simplicity, we shall deal mostly with the symmetric tensors
of the first Regge trajectory of the bosonic string, the class of fields for
which more information is available from direct vertex constructions
\cite{cubic,Zinoviev,Boulanger:2008tg,Metsaev:2007rn,Manvelyan}, and for which, as we shall see, the string amplitudes also result in particularly handy expressions. Our emphasis will be on the important, if hidden, role of corresponding Noether symmetries, with an eye to
the long--held feeling that String Theory should ultimately draw its origin from
a generalized Higgs effect responsible for its massive excitations. Hopefully, this work can provide some additional evidence for it.

It is instructive to introduce our arguments starting from a simple
circumstance where a Stueckelberg symmetry can show up
explicitly in String Theory. It concerns the second massive level of the open
bosonic string, where the most general vertex involving only $X^\m$ for a physical state
takes the form
\be
\cV^{\,(2)}\,=\,\oint\, dz\, \left[\,h_{\m\n}(p)\,\partial X^\m\,\partial X^\n+b_\m(p)\,\partial^{\,2}
X^{\m}\right]\, e^{ip\cdot X}\ .
\ee
This extended vertex operator is invariant under the Stueckelberg symmetry
transformations
\be
\d\, h_{\,\m\n}(p)\,=\,-\,i\,p_{\,\m}\,\L_{\,\m}(p)\,-\,i\,p_{\,\n}\,\L_{\,\m}(p)\ ,\qquad \d\, b_\m(p)\,=\,-\,2\,\L_{\,\m}(p)\ ,
\ee
in contrast with the usual representative for the massive spin--2 state,
\be
\cV^{\,(2)\,\prime}\,=\,\oint\, dz\, h_{\m\n}(p)\,\partial X^\m\,\partial X^\n\, e^{\,ip\,\cdot X}\ ,
\ee
and similar relations can be presented for all massive vertex operators. Above and beyond this simple game with free fields, we shall see a more important fact that resonates with this viewpoint, a conventional gauge invariance with a neat space--time rationale that lies well hidden in the couplings that we shall describe in the next sections.

A useful perspective on HS interactions dates back to the classic works of
Berends, Burgers and Van Dam \cite{bbvd} that we have already mentioned and was
further elaborated upon in many subsequent works, including
\cite{Bekaert:2007mi}. It is directly inspired by the Noether procedure that
played a key role in the construction of supergravity \cite{supergravity}, so that its starting
point are Poincar\'e--invariant local deformations of the form
\be S\,[\f]\,=\,\sum_s\, S^{\,(0)}\,[\f_{\m_1\ldots\,\m_s}]\,+\,\e\, S^{\,(1)}\,[\{\f_{\m_1\ldots\,\m_s}\}]\,+\,O\left(\e^2\right)\ , \ee
including at least one field of spin $s>2$, that are accompanied by deformations of the local gauge symmetries of the type
\be \d_{\x} \ \f_{\m_1\ldots\,\m_s}\,=\  \d^{\,(0)}_{\x} \f_{\m_1\ldots\,\m_s}
\,+\,\e\,\d^{\,(1)}_{\x} \f_{\m_1\ldots\,\m_s}\,+\,O\left(\e^2\right) \ . \ee
Out of these, the relevant ones should not correspond to local field redefinitions
\be\begin{split}  \f_{\m_1\ldots\,\m_s}\ &\ra\ \ \f_{\m_1\ldots\,\m_s}\,+\,\e\,
f(\f)_{\m_1\ldots\,\m_s} \,+\,O\left(\e^2\right)\ ,\\
 \xi_{\m_1\ldots\,\m_{s-1}}\ &\ra\ \ \xi_{\m_1\ldots\,\m_{s-1}}\,+\,\e\,
\z(\f,\xi)_{\m_1\ldots\,\m_{s-1}}\,+\,O\left(\e^2\right)\ ,
\end{split} \ee
of gauge fields and parameters, and generally acquire a non--abelian character
already at the first order in $\e$. In particular,
the first Lagrangian deformations $S^{\,(1)}$, cubic in the fields, always
result from Noether couplings of the type
\be \sum_s\  \f_{\m_1\ldots\,\m_s} \, J^{\,\m_1 \ldots\, \m_s} \, , \ee
where the nature of the currents $J^{\,\m_1 \ldots\, \m_s}$ is instrumental for
determining whether or not the gauge algebra is actually deformed. It is
\emph{not deformed} when the current is \emph{identically conserved}, a
circumstance whose simplest manifestation is perhaps Pauli's dipole coupling of
a Dirac spinor $\psi$ to electromagnetism, that can be presented in the form
\be A_\m \, \partial_\n \, \left(\bar{\psi} \, \gamma^{\,\m\n}\, \psi\right) \, ,\ee
to be contrasted with the usual minimal coupling
\be A_\m \, \left( \bar{\psi} \, \gamma^{\,\m}\, \psi \right) \, ,\ee
whose consistency with the gauge symmetry rests on the Dirac equation. This
perturbative approach can be used to deal systematically with HS interactions,
order by order in the couplings. There are different ways of presenting the
procedure, including the elegant BV--based formalism, originally developed in \cite{bh} and
widely used in a number of subsequent works, and in particular in
\cite{Boulanger:2008tg}.

With this perspective in mind, the present work is devoted to the tree--level
scattering amplitudes involving three or four external states belonging to the
first Regge trajectory of the bosonic string. We follow and extend the results
contained in the Master's Thesis of the junior author \cite{taronna09}. Our
derivations proceed along lines that are closely related to the string vertices
\cite{oldcubic69,oldcubic70,oldcubic,west} that were developed over the years
in various forms, but as we shall see remarkable simplifications are possible.
These amplitudes are analyzed in detail and are turned, via some more or less
standard $\star$--product techniques, into cubic couplings for massive and
massless higher--spin excitations. Most of the analysis rests on the
construction of generating functions for these couplings. These make it
possible to compute, in a simple and manifestly $SL(2,R)$ invariant fashion,
relatively handy expressions embodying all three and four--point amplitudes for
this class of states. Introducing the ``symbols'' $\xi^{\,\m}$ of the ordinary
$\a^{\,\m}$ oscillators of String Theory, one can group totally symmetric
momentum--space fields into generating functions of the type
\be
{\phi}_{\,i}(p_{\,i}\,,\,\xi_{\,i})\,=\,\sum_{n=0}^{\infty}\,\frac{1}{n!} \ {\phi}_{\,i\,\m_1\ldots\,\m_n}(p_{\,i})\ \xi_{\,i}^{\,\m_1}\ldots\, \xi_{\,i}^{\,\m_n}\ ,
\ee
where the $p_{\,i}$ are the corresponding momenta.
Referring to these quantities, one can compute
and simplify three-- and four--point scattering amplitudes at tree level for all
open--string external states of the first Regge trajectory.

Our main results are simple expressions for the three--point amplitudes, that can be cast in the rather compact form
\be
\cA\,=\,i\,\frac{g_o\!}{\a^{\,\prime}\!\!}\ (2\pi)^{\,d}\,\delta^{\,(d)}(p_{\,1}+p_{\,2}+p_{\,3})\,
\Big\{\cA^{\,+}\ Tr[\L^{a_1}\L^{a_2}\L^{a_3}]\,+\,\cA^{\,-}\ Tr[\L^{a_2}\L^{a_1}\L^{a_3}]\Big\}\ ,
\ee
where $g_o$ denotes the open string coupling and where
\begin{multline}
\cA^{\,\pm}\,=\vphantom{\sqrt{\frac{\a^{\,\prime}\!\!}{2}}}\exp{\Big[\,\partial_{\xi_{\,1}} \cdot\partial_{\xi_{\,2}}\,+\,\partial_{\xi_{\,2}}\cdot\partial_{\xi_{\,3}}\,+\,\partial_{ \xi_{\,3}}\cdot\partial_{ \xi_{\,1}}\Big]} \\ \times\,{\phi}_{\,1}\left(p_{\,1},\,\xi_{\,1}\,\pm\,\sqrt{\frac{\a^{\,\prime}\!\!}{2}}\,p_{\,23}\right)\, {\phi}_{\,2}\left(p_{\,2},\,\xi_{\,2}\,\pm\,\sqrt{\frac{\a^{\,\prime}\!\!}{2}}\,p_{\,31}\right)\, {\phi}_{\,3}\left(p_{\,3},\,\xi_{\,3}\,\pm\,\sqrt{\frac{\a^{\,\prime}\!\!}{2}}\,p_{\,12}\right) \Bigg|_{\xi_{\,i}\,=\,0} \!\!\!\!\! ,\label{116}
\end{multline}
or equivalently
\begin{multline}
\cA^{\,\pm}\,=\,{\phi}_{\,1}\left(p_{\,1},\ \partial_{\xi}\,\pm\,\sqrt{\frac{\a^{\,\prime}\!\!}{2}}\,p_{\,31}\right)\\
\times\,{\phi}_{\,2}\left(p_{\,2},\ \xi\,+\,\partial_{\xi}\,\pm\,\sqrt{\frac{\a^{\,\prime}\!\!}{2}}\,p_{\,23}\right)\, {\phi}_{\,3}\left(p_{\,3},\ \xi\,\pm\,\sqrt{\frac{\a^{\,\prime}\!\!}{2}}\,p_{\,12}\right)\,\Bigg|_{\xi=0},\label{117}
\end{multline}
where
\be
p_{\,ij}\,=\,p_{\,i}\,-\,p_{\,j}\ .
\ee
In presenting these results, we have also taken into account the Chan--Paton factors \cite{Paton:1969je} of the external states, and the derivative $\partial_{\,\xi}$ always acts to the right, with no ordering ambiguity for traceless $\phi_{\,i}$. From eqs.~\eqref{116} or \eqref{117}, one can then compute explicitly the current generating function,
\be
\cJ(x,\xi)\,=\,i\,\frac{g_o\!}{\a^{\,\prime}\!}\ \Big\{J^{\,+}(x,\xi)\ Tr\left[\ \cdot\ \L^{a_2}\L^{a_3}\right]\,+\,J^{\,-}(x,\xi)\
Tr\left[\ \cdot\ \L^{a_3}\L^{a_2}\right]\Big\}\ ,
\ee
with
\begin{multline}
J^{\,\pm}(x,\xi)\,=\,{\Phi}_{\,2}\left(x\ \mp\ i\sqrt{\frac{\a^{\,\prime}\!\!}{2}}\,\xi,\ \partial_{\chi}+\,\xi\,\mp\, i\sqrt{2\a^{\,\prime}}\,\partial_{\,3}\right)\\\times\,{\Phi}_{\,3}\left(x\ \pm\ i\sqrt{\frac{\a^{\,\prime}\!\!}{2}}\,\xi,\ \chi\,+\,\xi\,\pm\, i\sqrt{2\a^{\,\prime}}\,\partial_{\,2}\right)\Bigg|_{\chi=0}\ .
\end{multline}
In this paper we refer to the generating functions of fields and currents in coordinate space with the upper--case letters $\Phi$ and $J^{\,\pm}$, and to their momentum--space counterparts with the corresponding lower--case letters $\phi$ and $j^{\,\pm}$, so that
\begin{multline}
{j}^{\,\pm}=\,\exp\left\{\pm\, \sqrt{\frac{\a^{\,\prime}\!\!}{2}}\ \xi\cdot p_{\,23}\right\}\\\times\,{\phi}_{\,2}\left(p_{\,2},\ \partial_{\,\chi}+\,\xi\,\pm\,\sqrt{\frac{\a^{\,\prime}\!\!}{2}}\,p_{\,31}\right)\,{\phi}_{\,3}\left(p_{\,3},\ \chi\,+\,\xi\,\pm\,\sqrt{\frac{\a^{\,\prime}\!\!}{2}}\,p_{\,12}\right)\,\Bigg|_{\chi=0}\ .
\end{multline}
This notation is meant to distinguish spatial Fourier transforms from others
with respect to the symbols, that will be widely used in the following sections
and will be indicated by a  ``$\ \ \widetilde{}\ \ $".

The couplings thus extracted from String Theory are analyzed in a number of
cases, in order to identify the off--shell currents that lie behind string
interactions. The main obstacle is, as expected, that these couplings emerge in
an on--shell form that rests on the field equations for the external states. One
can see nonetheless signs of an emerging gauge symmetry that is exactly
recovered in the massless limit. Interestingly, it \emph{is not} confined to
the higher--derivative terms identified in \cite{Gross}, but extends to
subleading contributions that deform the free abelian HS gauge symmetry. In
this respect, we were able to extract from the massive result \eqref{116} a
simple generating function for all gauge invariant three--point abelian and non--abelian couplings involving HS states,
\begin{multline} \label{mass0cubic}
\cA^{\, [0] \,\pm}\,=\,\vphantom{\int}\exp{\left\{\pm\sqrt{\frac{\a^{\,\prime}\!\!}{2}}\ \Big[(\partial_{\xi_{\,1}}\cdot\partial_{\xi_{\,2}})
(\partial_{\xi_{\,3}}\cdot p_{\,12})\,+\,(\partial_{\xi_{\,2}}\cdot\partial_{\xi_{\,3}})(\partial_{\xi_{\,1}}\cdot p_{\,23})\,+\,(\partial_{\xi_{\,3}}\cdot\partial_{\xi_{\,1}})(\partial_{\xi_{\,2}}\cdot
p_{\,31})\Big]\right\}}\\ \vphantom{\int} \times\, \phi_{\,1}\left(p_{\,1}\,,\,\xi_{\,1}\,\pm\,\sqrt{\frac{\a^{\,\prime}\!\!}{2}}\,p_{\,23}\right)\,
\phi_{\,2}\left(p_{\,2}\,,\,\xi_{\,2}\,\pm\,\sqrt{\frac{\a^{\,\prime}\!\!}{2}}\,p_{\,31}\right)\, \phi_{\,3}\left(p_{\,3}\,,\,\xi_{\,3}\,\pm\,\sqrt{\frac{\a^{\,\prime}\!\!}{2}}\,p_{\,12}\right)\,\Bigg|_{\xi_{\,i}\,=\,0} \ .
\end{multline}
Interestingly, this generating function rests on the single differential
operator
\be
\cG\,=\,\sqrt{\frac{\a^{\,\prime}\!\!}{2}}\ \Big[(\partial_{\xi_{\,1}}\cdot\partial_{\xi_{\,2}})(\partial_{\xi_{\,3}}\cdot p_{\,12})\,+\,(\partial_{\xi_{\,2}}\cdot\partial_{\xi_{\,3}})(\partial_{\xi_{\,1}}\cdot p_{\,23})\,+\,(\partial_{\xi_{\,3}}\cdot\partial_{\xi_{\,1}})(\partial_{\xi_{\,2}}\cdot
p_{\,31})\Big]\ ,
\ee
which commutes with the linearized gauge transformations up to the linearized
field equations, as pertains to couplings that, in the spirit of the previous
discussion, generally deform the gauge algebra. The corresponding
\emph{conserved} currents, that for brevity we display here up to de Donder terms and multiple traces and more completely in Appendix A, read
\begin{multline} \label{cons_current_int}
J^{\,[0]\, \pm}(x\,;\,\xi)\,=\, \exp\left(\mp\  i\,\sqrt{\frac{\a^{\,\prime}\!\!}{2}}\ \xi_{\,\a}
\left[\partial_{\zeta_1}\cdot\partial_{\zeta_2}\,\partial^{\,\a}_{\,12}-2\,\partial^{\,\a}_{\zeta_1}\,
\partial_{\zeta_2}\cdot \partial_{\,1}
+\,2\,\partial^{\,\a}_{\zeta_2}\,\partial_{\zeta_1}\cdot \partial_{\,2}\right]\right)\\\times\,\Phi_{\,1}\left(x\,\mp\,i\sqrt{\frac{\a^{\,\prime}\!\!}{2}}\ \xi\, ,\,\zeta_1\,\mp\,i\,\sqrt{2\a^{\,\prime}}\,\partial_{\,2}\right)\,\Phi_{\,2}\left(x\,\pm\,i\sqrt{\frac{\a^{\,\prime}\!\!}{2}}\ \xi \,,\,\zeta_2\,\pm\,i\,\sqrt{2\a^{\,\prime}}\,\partial_{\,1}\right)\Bigg|_{\zeta_i\,=\,0}\ ,
\end{multline}
where for instance $\partial_{\,12}=\partial_{\,1} - \partial_{\,2}$. As a
result, all consistent abelian and non--abelian cubic interactions are somehow
on the same footing, since all of them, including the ``seeds'' of
\cite{Boulanger:2008tg}, descend via $\cG$ from the remarkably simple highest--derivative couplings
\be
\cA \ \sim \ \phi_{\,1}\left( p_{\,1}\right)\cdot \left(p_{\,23}\right)^{s_1} \ \phi_{\,2}\left( p_{\,2}\right)\cdot \left(p_{\,31}\right)^{s_2}\ \phi_{\,3}\left( p_{\,3}\right)\cdot \left(p_{\,12}\right)^{s_3} \ ,
\ee
where the indices carried by any of the fields are contracted with combinations of the other momenta. While String Theory makes use of the exponential of
$\cG$, we have no way at present to exclude that other functions of $\cG$ would
result in equally consistent choices, leaving way to a multitude of possible
cubic HS interactions. These findings thus provide a rationale, for
instance, for the important recent results of
\cite{Boulanger:2008tg,Manvelyan,Bekaert:2007mi} and, as we have stressed,
resonate with the long--held feeling that string spectra result from the
spontaneous breaking of HS gauge symmetries.

In agreement with the large number of no--go results on minimal couplings in
flat space, the interactions that we identify are generally of multi--polar
type, as in \cite{Boulanger:2008tg}. From them one can also build generating
functions of four--point exchanges for totally symmetric fields, which open the
way to the construction of interesting four--point amplitudes in Field Theory
describing the exchange of infinitely many massive and massless higher--spin
particles in various dimensions. One can thus analyze their high--energy
behavior, along the lines of the beautiful recent work of Bekaert, Joung and
Mourad \cite{Bekaert:2009ud}, making also some further steps toward an eventual
deconstruction of the Veneziano amplitude.

The plan of the paper is as follows. Section
\ref{sec:tools} is devoted to a brief review of basic
ingredients that are needed to construct the string
$S$--matrix and the corresponding physical states. In
Section \ref{sec:ampl} we clarify the links between
amplitudes for the first Regge trajectory of the open
bosonic string and free field theory generating functions.
We then introduce the symbols of the ordinary $\a$
oscillators of String Theory and construct a generating
function that subsumes, for the first Regge trajectory, all
tree--level three--point correlation functions. For
three--point and four--point amplitudes we also describe
how to recast the result in a manifestly
$SL(2,R)$--invariant fashion for all states belonging to
the first Regge trajectory of the open bosonic string. In
Section \ref{sec:couplings} we describe in some detail the
explicit form of a number of cubic couplings, with special
attention to the gauge symmetry that emerges in the
massless limit. In Section \ref{sec:exchanges} we study
current exchanges along the lines of
\cite{minimal,Bekaert:2009ud} and then in Section
\ref{sec:QFTamplitudes}, following \cite{Bekaert:2009ud},
we present some applications of the couplings extracted
from String Theory, computing on the Field Theory side
scattering amplitudes that involve infinitely many massive
or massless HS exchanges and studying their behavior in
some interesting limits. Our Conclusions are summarized in
Section \ref{sec:conclusions}. Finally, following a request
of the referee, in the Appendix we display how to complete
the cubic couplings off--shell. Moreover, we generalize the
conserved higher--spin bosonic currents to the fermionic
case, and discuss how all these cubic couplings can provide
interesting information on quartic HS interactions.

\vskip 36pt


\scs{Open--String Scattering Amplitudes}\label{sec:tools}


In this section we review a few basic properties of the tree--level amplitudes
of the open bosonic string and the standard construction of its physical
states. This warm--up exercise is meant to recall some tools that are
instrumental for obtaining compact forms of the scattering amplitudes for
the first Regge trajectory.

\vskip 24pt


\scss{Open String $S$--matrix and physical states}\label{sec:physicalstates}


In String Theory, the $S$--matrix can be conveniently constructed starting from two basic ingredients. The first is the Polyakov path integral, based on
\begin{equation}
S_P[X,\g]\,=\,-\, \frac{1}{4\pi\a^{\,\prime}}\int_M d^{\,2} \xi\ \sqrt{\g}\, \g^{\,ab}\,\partial_{\,a} X^\m \partial_{\,b} X_\m+\l\, \chi\ ,\label{Polyakov}
\end{equation}
where the $X^{\m}$ are the string coordinates, $\g_{ab}$ is the world--sheet
metric, to be regarded as an independent field, and $\chi$ denotes the Euler
character of the Riemann surface spanned by the $\xi^a$. The
second ingredient is a collection of vertex operators reflecting the asymptotic
states of the string spectrum \cite{stringtheory}. As is well known, in the
critical dimension, after fixing the local gauge symmetries the functional
integral collapses to a finite--dimensional one over the moduli space of the
(punctured) Riemann surfaces of interest.

The reduction of the path integral can be simply attained for the tree--level
amplitudes of interest in this work, that for open strings can be defined with
reference to the upper half--plane. The end results are Koba--Nielsen--like
integrals of correlation functions of local operators of the type
\begin{multline}
S_{j_{\,1}\ldots j_{\,n}}(p_{\,1},\ldots p_{\,n})\,=\,\sum\int d y_{\,4}\ldots d y_{\,n}\ |\,\hat{y}_{12}\,\hat{y}_{13}\,\hat{y}_{23}|\\\times\bra \cV_{j_{\,1}}(\hat{y}_{\,1},p_{\,1})\,\cV_{j_{\,2}}(\hat{y}_{\,2},p_{\,2})\,\cV_{j_{\,3}}(\hat{y}_{\,3},p_{\,3}) \,\ldots\,\cV_{j_{\,n}}(y_{\,n},p_{\,n})\ket\ Tr(\L^{a_1}\ldots\, \L^{a_n})\ ,\label{Smatrix}
\end{multline}
where $\cV_{j}(y,p)$ are vertex operators corresponding to the given asymptotic states, the $y_{\,i}$'s are real variables,
\be
y_{\,ij}\,=\,y_{\,i}-y_{\,j} \ ,
\ee
and the three $\hat{y}_{\,i}$'s have been fixed taking into account the $SL(2,R)$
or M\"obius invariance, thus bringing about the measure factor
$|\,\hat{y}_{12}\,\hat{y}_{13}\,\hat{y}_{23}|$. Finally, the sum is over all
cyclically inequivalent orderings of the insertion points and the $\L^{a_i}$
denote the familiar Chan--Paton matrices \cite{Paton:1969je}.

There is a systematic procedure for associating vertex
operators to open--string states. The construction rests,
say, on the mode expansions of the string coordinates
$X^\m$ with Neumann boundary conditions,
\be
X^\m(z)\,=\,x^{\,\m}-i\a^{\,\prime}p^{\,\m}\ln \left(z \, \bar{z} \right)\,+\, i\sqrt{\frac{\a^{\,\prime}\!\!}{2}}\,\sum_{n\neq 0}\,\frac{\a_{\,n}^{\,\m}}{n}\, \left(z^{-n}+\bar{z}^{-n}\right)\ ,
\ee
where $z$ is the standard complex coordinate of the two--dimensional
world--sheet, $x^\m$ is the center--of--mass coordinate and the $\a_{\,n}^{\,\m}$ are the bosonic string oscillators obeying the conditions
\begin{equation}
(\a_{\,m}^{\,\m})^{\dagger}\, \,=\,\, \a_{-m}^{\,\m}\ , \quad
\left(x^{\,\m}\right)^\dagger\, \,=\,\, x^{\,\m}
\end{equation}
and the commutation relations
\begin{equation}
[\a_{\,m}^{\,\m},\a_{\,n}^{\,\n}]\, \,=\,\, m\,\eta^{\,\m\n}\,\delta_{\,m+n,0}\ ,
\end{equation}
since in this paper we work with a Minkowski metric of mostly positive signature. Moreover
\be
[x^{\,\m}, p^{\,\n}]\, \,=\,\, i\, \eta^{\,\m\n}\ ,
\ee
with $\sqrt{\,2\a^{\,\prime}}\,p^{\,\m}\,\equiv \,\a_{\,0}^{\,\m}$. The $\a_{\,m}^{\,\m}$ build the
Fock space of string excitations as
\begin{equation}
|\l_{\,i};p\ket \,=\,\prod_{n>0}\,(\a_{-n}^{\,\m_{i_n}})^{\,\l_n}\,|\,0\,;\,p\ket\ ,\label{stato}
\end{equation}
where $p$ denotes the string momentum, so that
\begin{equation}
|\,0\,; p\ket\,=\,e^{\,i\,p\,\cdot x}\,|\,0\,;\,0\ket\ .
\end{equation}

In what follows it will suffice to refer to the standard covariant
quantization, avoiding to work explicitly with ghosts. The key ingredients are
then the Virasoro operators
\begin{equation}
L_{\,m}\,=\,\frac{1}{2}\sum_{n\,=\,-\infty}^{\infty}:\a_{\,m-n}\cdot \a_{\,n}:\ ,
\end{equation}
that satisfy the Virasoro algebra
\begin{equation}
[L_{\,m},L_{\,n}]\,=\,(m-n)\,L_{\,m+n}+\frac{D}{12}\,m\,(m^2-1)\,\delta_{\,m+n,\,0}\ ,
\end{equation}
while physical states are to satisfy the conditions
\begin{equation}
\begin{split}
L_{\,m}\,|\,\phi\ket\,&=\,0\ ,\ m>0\ ,\\
(L_0-1)\,|\,\phi\ket\,&=\,0\ .\label{Physical1}
\end{split}
\end{equation}
The last condition identifies the mass shell, while
$|\,\phi\ket$ is obtained, in general, combining string
oscillators with momentum--space fields of the type
\be
\phi_{\,\m_1\ldots\,\m_{s_1},\,\n_1\ldots\,\n_{s_2},\,\ldots}(p)\ . \ee
These are usually referred to as multi--symmetric mixed--symmetry fields: indices
belonging to any given set accompany identical string oscillators and are thus
totally symmetric, while no symmetry is present under interchanges between
different sets. Actually, the $L_{\,n}$ with $n>0$ can be generated starting
from $L_{\,1}$ and $L_{\,2}$, so that it suffices to impose the conditions
\begin{equation}
L_{\,1}\,|\,\phi\ket\,=\,0\ ,\ \ L_{\,2}\,|\,\phi\ket\,=\,0\ ,\ \ (L_0-1)\,|\,\phi\ket\,=\,0\ ,
\end{equation}
that can be put in direct correspondence with the Fierz--Pauli conditions for
massive HS fields. This simple observation will prove very useful in the
following. Finally, vertex operators are in one--to--one correspondence with physical states, which are linked in their turn to the oscillators via the
correspondence
\begin{equation}
\a_{-m}^{\,\m}|\,0\ket \ \lra \ \frac{1}{\sqrt{2\a^{\,\prime}}}\,\frac{i}{(m-1)!}\
\partial^{\,m} X^\m(0)\ ,\ \ m\geq1\ ,
\end{equation}
that becomes manifest if the oscillators are recovered as Laurent coefficients
of the holomorphic currents ${\partial X}^\m$,
\begin{equation}
\a_{-m}^{\,\m}\,=\,\frac{1}{\sqrt{2\a^{\,\prime}}}\oint \frac{dz}{2\pi} \ z^{-m}\, \partial X^\m(z)\ .
\end{equation}

\vskip 24pt


\scss{The first Regge Trajectory}\label{sec:FirstRegge}


The states
\begin{equation}
|\,\phi\ket\,=\,\phi_{\m_1\ldots\,\m_s}(p) \ \a_{-1}^{\,\m_1}\ldots\,\a_{-1}^{\,\m_s}\ |\,0\,;\,p\ket\label{first}
\end{equation}
of the first Regge trajectory comprise the highest--rank tensors available at each mass level.
They describe symmetric tensors in space time that are built out of $\a_{-1}^{\,\m}$ only, whose vertex operators are thus of the form
\begin{equation}
\cV_{\,\phi}(p,y)\,=\,\left(\frac{i}{\sqrt{2\a^{\,\prime}}}\right)^s\phi_{\m_1\ldots\,\m_s}(p)\
:\partial X^{\m_1}(y)\ldots\partial X^{\m_s}(y)\, e^{\,ip\,\cdot X}:\ ,
\label{o.s.states}
\end{equation}
where $y$ can be taken to lie on the real axis.

The expansion in the $\a^{\,\m}_{\,n}$ operators can be converted into a more convenient one in terms of their symbols, a collection of vectors $\xi^{(n)\, \m}$, so that for instance
\begin{equation}
\phi(p)\,=\,\phi_{\m_1\ldots\,\m_s}(p)\ \xi^{\,\m_1}\ldots\, \xi^{\,\m_s}\ ,\label{genfunc}
\end{equation}
where for brevity we have let $\xi^{(1)}=\xi$, a convention that we shall try
to abide to insofar as possible. On the states \eqref{o.s.states}, the Virasoro
constraints translate directly into the mass--shell conditions
\begin{equation}
M^{\,2}\,=\, -\, p^{\,2}\,=\,\frac{s-1}{\a^{\,\prime}}
\end{equation}
and the Fierz--Pauli conditions
\be
p\,\cdot\,\partial_{\xi}\ \phi(p\,,\xi)\,=\,0\ ,\qquad  \ \partial_{\xi}\,\cdot\,
\partial_{\xi}\ \phi(p\,,\xi) \,=\,0\ ,\label{constr}
\ee
so that the corresponding momentum--space fields $\phi_{\m_1\ldots\,\m_s}(p)$ are
traceless and transverse to the momentum $p$.

While most of the literature has focussed over the years on the low--lying modes, some
previous works, including \cite{Buchbinder:1999ar}, have already dealt in some detail
with some aspects of the massive string excitations. The main purpose of this paper is to
proceed further along the lines of \cite{west}: as we shall see, some simplifications are
possible, which lead to relatively handy expressions for all three--point and four--point
amplitudes for the symmetric tensors of the first Regge trajectory. The explicit results
contain interesting lessons on the properties of these HS fields.

\vskip 24pt


\scss{A Generating Function for string amplitudes}\label{sec:GenFunc}


In this section we discuss the cubic couplings between massive open--string states. In
principle, these results are closely related to the vertex that was discussed in various
forms, starting from the early years of String Theory \cite{oldcubic70}, and was then
widely elaborated upon by a number of authors in the eighties \cite{oldcubic,west}. In
practice, however, one can simplify matters to a large extent, taking as a starting point
for the derivation the simple generating function
\be
\begin{split}
\mathbf{Z}\left[J(\s)\right]\,&=\,\int\cD X^\m\ \exp\left(-\frac{1}{4\pi \a^{\,\prime}}\int_M
d^{\,2}\s\ \partial X^\m \, \bar{\partial}X_\m+\,i\int_M d^{\,2}\s\ J(\s)\cdot
X(\s)\right)\\\,&=\,i\, (2\pi)^{\,d}\, \delta^{\,(d)}(J_{\,0})\,
\left[\det'\left(-\frac{\,
\partial\,\bar{\partial}}{4\pi^2\a^{\,\prime}}\right)\right]^{\,-\,
\frac{d}{2}} \, \exp\left(-\frac{1}{2}\int d^{\,2}\s\, d^{\,2}\s'\ J(\s)\cdot
J(\s')\,G(\s,\s')\right)\ ,\label{Gen}
\end{split}
\ee
and working with general external currents of the form
\begin{equation}
J(\s)\,=\,\sum_{i\,=\,1}^N\left(p_{\,i}\,\delta^{\,2}(\s-\s_i)\,
+\, \sum_{m=1}^\infty (-1)^{\,m}\, \frac{\xi^{\,(m)}_{\,i}}{\sqrt{2\a^{\,\prime}}\, (m-1)!}
\ \partial^{\,(m)}\delta^{\,2}(\s-\s_i)\right)\ ,\label{curr}
\end{equation}
that associate sources for the local operators $\partial^{\,m} X^\m(y_i)$ to the
individual $\xi_{\,i}^{\,(m)\m}$ symbols.

In the following we shall mostly deal with correlation functions of states
belonging to the first Regge trajectory of the open bosonic string. As we have
seen, these correspond to the simplest class of HS fields, symmetric tensors of
arbitrary rank, for which more results are currently available
\cite{cubic,Zinoviev,Boulanger:2008tg,Manvelyan} both from direct constructions and from
the Vasiliev setting \cite{Vasiliev:1988sa}. Retaining only terms proportional
to $\xi_{\,i}^{\,(1)}$, that as we remarked we shall simply call $\xi_{\,i}$
for brevity, one can begin by obtaining the reduced form of $\mathbf{Z}$,
\begin{equation}
\mathbf{Z}\,=\,i\, g_o\, \frac{(2\pi)^{\,d}}{\a^{\,\prime}}\
\delta^{\,(d)}(J_{\,0})\,\exp\left[\sum_{i\neq j}^N\left(\a^{\,\prime}p_{\,i}\cdot p_{\,j}
\,\ln|y_{ij}|+\sqrt{2\a^{\,\prime}}\ \frac{\xi_{\,i}\cdot p_{\,j}}{y_{ij}}+\frac{1}{2}\ \frac{\xi_{\,i}
\cdot \xi_j}{y_{ij}^2}\right)\right]\ ,\label{Gen3}
\end{equation}
which encodes the correlation functions for the states of the first Regge
trajectory in the coefficients of the $\xi$--monomials.

In order to recover the correlation functions of vertex operators, one is to
combine the generating function $\mathbf{Z}$ with corresponding
expressions in momentum space like \eqref{genfunc}, and to this end it is
convenient to introduce the $\star$--product
\begin{equation}
\star\,:\ \left(\phi(p\,,\xi)\,,\,\psi(q\,,\chi)\right) \ \ra \ \phi\,\star\,\psi\,=\,\exp\Big(\partial_{ \xi}\cdot\partial_{\chi}\Big)\,\phi(p\,,\xi)\ \psi(q\,,\chi)\Big|_{\xi\,=\,\chi\,=\,0}\ ,\label{contra}
\end{equation}
where $\xi$ and $\chi$ are commuting symbols. Once the generating functions are chosen to be of the form
\begin{equation}
\phi(\xi)\,=\,\sum_{n\,=\,0}^{\infty}\frac{1}{n!}\ \phi_{\m_1\ldots\,\m_n}\,\xi^{\,\m_1}\ldots\, \xi^{\,\m_n}\ ,
\end{equation}
the $\star$--product of a pair of them is in fact
\begin{equation}
\phi\star\psi\,=\,\sum_{n\,=\,0}^{\infty}\frac{1}{n!}\ \phi_{\m_1\ldots\,\m_n} \psi^{\,\m_1\ldots\,\m_n}\ .
\end{equation}
A similar, albeit more complicated, product can be defined when more
index families are present, and this relation then generalizes to a sum of
contractions between the index families associated to the various types of
symbols. For a given set of vertex operators
\be \cV_{\,\phi}(p_{\,i},y_{\,i})\,=\,\sum_{s=0}^\infty \frac{1}{s!}\ \phi_{\,i\,\m_1\ldots\,\m_s}(p_i):\partial
X^{\m_1}\ldots\, \partial X^{\m_s}\, e^{\,i\,p_{\,i}\cdot X}: \ , \ee
the corresponding symbols are
\be
\phi_{\,i}(p_{\,i},\xi_{\,i})\,=\,\sum_{s=0}^\infty \frac{1}{s!}\  \phi_{\,i\,\m_1\ldots\,\m_s}(p_{\,i})\ \xi_{\,i}^{\,\m_1}\ldots\, \xi_{\,i}^{\,\m_s}\ ,
\ee
while the sought correlation function takes the form
\begin{equation}
\left\langle\cV_{\,\phi{\,1}}(p_{\,1},y_{\,1})\,\ldots\,\cV_{\,\phi_{\,n}}(p_{\,n},y_{\,n})\right\rangle\,
=\,\left[\,{\phi_{\,1}(p_{\,1},\xi_{\,1})}\ldots\,{\phi_{\,n}(p_{\,n},\xi_{\,n}})\right]
\star\, \mathbf{Z}(y_{\,1},\ldots,y_{\,n};\xi_{\,1},\ldots,\xi_{\,n})\ ,
\end{equation}
where the multiple $\star$--product combines with $\mathbf{Z}$ the various
${\phi_{\,i}}$ symbols.

This $\star$--product admits an integral representation that allows one to
simplify considerably the amplitudes. Let us recall this important, if
well--known fact, referring again for brevity to the first Regge trajectory.
One can begin by considering the Fourier transform of the generating function
$\phi(p\,,\xi)$
\begin{equation}
\tilde{\phi}(p\,,\pi)\,=\,\int \frac{d^d \xi}{(2\pi)^{\,d/2}}\ e^{- i \pi\cdot \xi}\, \phi(p\,,\xi)\ ,\label{F}
\end{equation}
and its inverse
\begin{equation}
\phi(p\,,\xi)\,=\,\int \frac{d^d\pi}{(2\pi)^{\,d/2}}\ e^{i\pi\cdot \xi}\, \tilde{\phi}(p\,,\pi)\ .\label{inversF}
\end{equation}
The end result follows once \eqref{inversF} is substituted in \eqref{contra}, which yields
\begin{equation}
\begin{split}
\phi(p)\star\psi(q)&\,=\,\int \frac{d^d\xi}{(2\pi)^{\,d/2}}\ \exp\Big(\partial_{\zeta}\cdot\partial_{\chi}\Big)\, e^{i\zeta\cdot \xi}\,\tilde{\phi}(p\, ,\xi)\, \psi(q,\chi)\Big|_{\chi\,=\,\zeta\,=\,0}\\
&\,=\,\int \frac{d^d\xi}{(2\pi)^{\,d/2}}\ \exp\Big(i\xi\cdot \partial_{\chi}\Big)\,\tilde{\phi}(p\, ,\xi)\,\psi(q,\chi)\Big|_{\chi\,=\,\zeta\,=\,0}\\
&\,=\,\int \frac{d^d\xi}{(2\pi)^{\,d/2}}\ \tilde{\phi}(p\, ,\xi)\,\psi(q,i\xi)\ .\label{ContrFormula}
\end{split}
\end{equation}
In other words, the $\star$--product results in an integral over the symbols
$\xi$ of two factors. The first is the Fourier transform of the symbol of the
first factor, $\phi$, while the second is the symbol of the second factor,
$\psi$, evaluated at purely imaginary arguments.

\vskip 36pt


\scs{Disk Amplitudes}\label{sec:ampl}


In this section we describe how to obtain simple and explicit forms for all
disk amplitudes involving three or four external states belonging to the first
Regge trajectory of the open bosonic string.

\vskip 24pt


\scss{Three--Point functions}\label{sec:threepoint}


In this case one can begin by parametrizing the masses of the three external
states as
\begin{align}
-p_{\,1}^{\,2}\,=\,&\frac{n_1-1}{\a^{\,\prime}}\ ,& -p_{\,2}^{\,2}\,=\,&\frac{n_2-1}{\a^{\,\prime}}\ ,& -p_{\,3}^{\,2}\,=\,&\frac{n_3-1}{\a^{\,\prime}}\ ,
\end{align}
and momentum conservation then translates into the three conditions
\begin{eqnarray}
&&2\a^{\,\prime} p_{\,1}\cdot p_{\,2}=n_1+n_2-n_3-1\ ,\qquad\qquad 2\a^{\,\prime} p_{\,1}\cdot p_{\,3}=n_1+n_3-n_2-1\ ,\nonumber\\ \vphantom{\int}&&\qquad\qquad\qquad\qquad\quad 2\a^{\,\prime} p_{\,2}\cdot p_{\,3}=\,n_2+n_3-n_1-1\ .
\end{eqnarray}
As a result, the portion of the generating function $\mathbf{Z}$ that only
depends on the momenta takes the form
\begin{equation}
|\,y_{12}\,y_{13}\,y_{23}|\,\exp\left[\a^{\,\prime}\sum_{i\neq j}p_{\,i}\cdot p_{\,j}\ln|y_{ij}|\right]\,=\,\left|\frac{y_{12}y_{13}}{y_{23}}\right|^{n_1} \left|\frac{y_{12}y_{23}}{y_{13}}\right|^{n_2}\left|\frac{y_{13}y_{23}}{y_{12}}\right|^{n_3}\ ,
\end{equation}
and therefore
\begin{multline}
\mathbf{Z} \,=\,i\,g_o\frac{(2\pi)^{\,d}}{\a^{\,\prime}}\ \delta^{\,(d)}(p_{\,1}+p_{\,2}+p_{\,3})\,\left|\frac{y_{12}y_{13}}{y_{23}}\right|^{n_1}
\left|\frac{y_{12}y_{23}}{y_{13}} \right|^{n_2}\left|\frac{y_{13}y_{23}}{y_{12}}\right|^{n_3} \\\times\exp\left[\sum_{i\neq j}^3\left(\frac{1}{2}\frac{\xi_{\,i}\cdot \xi_{\,j}}{y_{ij}^{\,2}}
+\sqrt{2\a^{\,\prime}}\ \frac{\xi_{\,i}\cdot p_{\,j}}{y_{ij}}\right)\right]\ .
\end{multline}
Using again momentum conservation this result can be further simplified and turned into the form
\begin{equation}
\begin{split}
\mathbf{Z}\,=\,&i\,g_o\frac{(2\pi)^{\,d}}{\a^{\,\prime}}\ \delta^{\,(d)}(p_{\,1}+p_{\,2}+p_{\,3})\,\left|\frac{y_{12}y_{13}}{y_{23}} \right|^{n_1}\left|\frac{y_{12}y_{23}}{y_{13}}\right|^{n_2}\left|\frac{y_{13}y_{23}}{y_{12}}\right|^{n_3} \\&\times\exp\left[\left(\frac{\xi_{\,1}\cdot \xi_{\,2}}{y_{12}^{\,2}}+\frac{\xi_{\,1}\cdot \xi_{\,3}}{y_{13}^{\,2}}+\frac{\xi_{\,2}\cdot \xi_{\,3}}{y_{23}^{\,2}}\right)\right.\\&\qquad\quad-\sqrt{\frac{\a^{\,\prime}\!\!}{2}}\left(\xi_{\,1}\cdot p_{\,23}\frac{y_{32}}{y_{21}y_{31}}+\xi_{\,2}\cdot p_{\,31}\frac{y_{13}}{y_{12}y_{32}}+\xi_{\,3}\cdot p_{\,12}\frac{y_{21}}{y_{13}y_{23}}\right.\\&\left.\left.\qquad\quad -\,\xi_{\,1}\cdot p_{\,1}\frac{y_{31}+y_{21}}{y_{31}y_{21}}-\xi_{\,2}\cdot p_{\,2}\frac{y_{12}+y_{32}}{y_{32}y_{12}}-\xi_{\,3}\cdot p_{\,3}\frac{y_{23}+y_{13}}{y_{23}y_{13}}\right)\right]\ ,\label{Gen2}
\end{split}
\end{equation}
where
\be
p_{\,ij}=p_{\,i}-p_{\,j} \ . \label{pij}
\ee

While this formal expression embodies all \emph{on--shell} three--point
amplitudes, these are mingled with a lot spurious information. The
actual physical amplitudes can be extracted from it combining, for any given
choice of the integers $n_1$, $n_2$ and $n_3$ that determine the masses of the
external states, precisely $n_1$, $n_2$ and $n_3$ copies of the corresponding symbols
arising from the exponential, that determine their tensor structures. The result is
\begin{multline}
\mathbf{Z}\,=\,i\,g_o\frac{(2\pi)^{\,d}}{\a^{\,\prime}} \
\delta^{\,(d)}(p_{\,1}+p_{\,2}+p_{\,3})\exp\left\{\sqrt{\frac{\a^{\,\prime}\!\!}{2}}\left(\xi_{\,1}\cdot
p_{\,23}\,\left\langle\frac{y_{23}}{y_{12}y_{13}}\right\rangle+\,\xi_{\,2}\cdot
p_{\,31}\,\left\langle\frac{y_{13}}{y_{12}y_{23}}\right\rangle\right.\right.\\\left.\left.+\,\xi_{\,3}\cdot
p_{\,12}\,\left\langle\frac{y_{12}}{y_{13}y_{23}}\right\rangle\right)+(\xi_{\,1}\cdot
\xi_{\,2}+\xi_{\,1}\cdot \xi_{\,3}+\xi_{\,2}\cdot \xi_{\,3})\vphantom{\sqrt{\frac{\a^{\,\prime}\!\!}{2}}}\right\}\ , \label{Zphys}
\end{multline}
where
\be \left\langle x\right\rangle\,=\,\text{sign}(x)\ ,\ee
and where we have also used the Fierz--Pauli conditions implied by the
Virasoro constraints to remove some terms proportional to $\xi_{\,i}\cdot
p_{\,i}$.

All dependence on the world--sheet $y_{ij}$ variables has thus disappeared, up
to some signs that reflect the flip properties of the external states and thus
their cyclic ordering, precisely as expected in view of the M\"obius invariance
of physical amplitudes. This is of course to happen in general, even for
amplitudes involving states of lower Regge trajectories, but in order to attain
comparable simplifications one should identify a convenient basis for these
states, that in general do not correspond to simple monomials in the
$\xi^{\,(i)}$. Lo and behold, subsidiary conditions on the momentum--space
tensors, or on the generating function, appear unavoidable to arrive at
M\"obius--invariant results in the general case. While one can see this
mechanism at work in special classes of examples, so far we have not been able
to obtain handy generalizations of eq.~\eqref{Zphys} that apply to arbitrary
trajectories. Still, in some cases one can simply go further, and for instance
\begin{equation}
\phi\left(p\,,\, \xi^{(n)}\right)\,=\,\phi_{\m_1^1\ldots\,\m_{s_1}^1,\,\ldots\,,\,\m_1^n\ldots\, \m_{s_n}^n}(p)\ \xi^{\,(1)}_{\m_1^1}\ldots\, \xi^{\,(1)}_{\m_{s_1}^1}\,\ldots\,\, \xi^{\,(n)}_{\m_1^n}\ldots\, \xi^{\,(n)}_{\m_{s_n}^n}\ ,
\end{equation}
describes a simple class of states belonging to subleading Regge trajectories
that solves the Virasoro constraints, provided the multi--symmetric tensor
$\phi_{\m_1^1\ldots\,\m_{s_1}^1;\,\ldots\,;\,\m_1^n\ldots\, \m_{s_n}^n}$ is
projected onto the irreducible representation corresponding to the $n$--row
Young tableau of type $(s_1,\,\ldots,s_n)$, with $s_1\geq \,\ldots\, \geq s_n$.
While states of this type do not exhaust the string spectrum, they do
correspond nonetheless to arbitrary Young tableaux. Therefore, their behavior
can in principle shed some light on generic features of the corresponding
couplings. We leave a detailed analysis of these more general amplitudes for
future work, but in eq.~\eqref{guess} we shall be able to propose a motivated
guess for the cubic couplings of generic mixed--symmetry states.

\vskip 24pt


\scss{Three--point scattering amplitudes for the first Regge trajectory}\label{sec:threepointampl}


Combining the tools of the previous sections one can compute
explicitly three--point scattering amplitudes for arbitrary external states
belonging to the first Regge trajectory. In this case no integrations are left once all insertions are fixed taking M\"obius invariance into account.
Therefore the generating function, when contracted with physical states, can
depend at most on the choice of the cyclically inequivalent
orderings of the insertions. The general formula for the three--point scattering
amplitudes is
\begin{equation}
\cA_3(p_{\,1},p_{\,2},p_{\,3})\,=\,\left[\phi_{\,1}(p_{\,1},\xi_{\,1})\,
\phi_{\,2}(p_{\,2},\xi_{\,2})\, \phi_{\,3}(p_{\,3},\xi_{\,3})\right] \, \star \,
\mathbf{Z}\,(p_{\,1},p_{\,2},p_{\,3};\,\xi_{\,1},\xi_{\,2},\xi_{\,3})\ ,\label{star}
\end{equation}
with
\be
\begin{split}
\mathbf{Z}\,&=\,i\,g_o\,\frac{(2\pi)^{\,d}}{\a^{\,\prime}}\
\delta^{\,(d)}(p_{\,1}+p_{\,2}+p_{\,3})
\exp\left\{\sqrt{\frac{\a^{\,\prime}\!\!}{2}}\left(\xi_{\,1}\cdot p_{\,23}\,
\left\langle\frac{y_{23}}{y_{12}y_{13}}\right\rangle\right.\right.\\&+\left.\left.
\xi_{\,2}\cdot
p_{\,31}\,\left\langle\frac{y_{13}}{y_{12}y_{23}}\right\rangle+\xi_{\,3}\cdot
p_{\,12}\,\left\langle\frac{y_{12}}{y_{13}y_{23}}\right\rangle\right)+(\xi_{\,1}\cdot
\xi_{\,2}+\xi_{\,1}\cdot \xi_{\,3}+\xi_{\,2}\cdot
\xi_{\,3})\vphantom{\sqrt{\frac{\a^{\,\prime}\!\!}{2}}}\right\}\Bigg|_{y_1<y_2<y_3}\\&+\ (1\lra 2)\vphantom{\sqrt{\frac{\a^{\,\prime}\!\!}{2}}}\ ,
\end{split}
\ee
and dressing eq.~\eqref{star} with Chan--Paton factors \cite{Paton:1969je} one finally obtains
\be
\begin{split}
\cA\,=\,\vphantom{\frac{d^dp_{\,i}}{(2\pi)^{\,d/2}}}i\,\frac{g_o}{\a^{\,\prime}}\,(2\pi)^{\,d}\, \delta^{\,(d)}(p_{\,1}+p_{\,2}+p_{\,3}) &\left\{\cA^{\,+}(p_{\,1},p_{\,2},p_{\,3})\  Tr[\L^{a_1}\L^{a_2}\L^{a_3}]\vphantom{\frac{d^dp_{\,i}}{(2\pi)^{\,d/2}}}\right.\\ &\ +\left. \,\cA^{\,-}(p_{\,1},p_{\,2},p_{\,3})\ Tr[\L^{a_2}\L^{a_1}\L^{a_3}]\vphantom{\frac{d^dp_{\,i}}{(2\pi)^{\,d/2}}}\right\}\ ,\label{ampl}
\end{split}
\ee
with
\begin{multline}
\cA^{\,\pm}(p_{\,1},p_{\,2},p_{\,3}) \,=\, \int \prod_{i\,=\,1}^3\frac{d^d\p_{\,i}}{(2\pi)^{\,d/2}}\ \mathbf{\tilde{Z}}_\pm(p_{\,1},p_{\,2},p_{\,3};\p_{\,1},\p_{\,2},\p_{\,3})\\ \times\,{\phi}_{\,1}(p_{\,1},i\p_{\,1})\, {\phi}_{\,2}(p_{\,2},i\p_{\,2})\, {\phi}_{\,3}(p_{\,3},i\p_{\,3})\ ,
\end{multline}
where the $\mathbf{\tilde{Z}}^{\,\pm}$ are Fourier transforms of
\begin{equation}
\mathbf{Z}^{\,\pm}\,=\,\exp\left\{\pm\sqrt{\frac{\a^{\,\prime}\!\!}{2}}\left( \xi_{\,1}\cdot p_{\,23}+ \xi_{\,2}\cdot p_{\,31}+\xi_{\,3}\cdot p_{\,12}\right)+\left(\xi_{\,1}\cdot \xi_{\,2}+\xi_{\,1}\cdot \xi_{\,3}+\xi_{\,2}\cdot \xi_{\,3}\right)\right\}
\end{equation}
with respect to the $\xi_{\,i}$.

The crucial remark at this point is that $\textbf{Z}^{\,\pm}$ are exponential functions, so that
their Fourier transforms can be simply related to distributions, with the end
result
\begin{multline}
\mathbf{\tilde{Z}}^{\,\pm}(\p_{\,1},\p_{\,2},\p_{\,3})\,=\,(2\pi)^{3d/2}\exp\left\{\pm \, i\,\sqrt{\frac{\a^{\,\prime}\!\!}{2}}(\partial_{\p_{\,1}}\cdot p_{\,23}+ \partial_{\p_{\,2}}\cdot p_{\,31}+\partial_{\p_{\,3}}\cdot p_{\,12})\right.\\\left.\vphantom{\pm i\sqrt{\frac{\a^{\,\prime}\!\!}{2}}}-(\partial_{\p_{\,1}}\cdot \partial_{\p_{\,2}}+\partial_{\p_{\,1}}\cdot \partial_{\p_{\,3}}+\partial_{\p_{\,2}}\cdot \partial_{\p_{\,3}})\right\}\,\delta(\p_{\,1})\,\delta(\p_{\,2})\,\delta(\p_{\,3})\ ,\label{Ztransform}
\end{multline}
where the $\delta$ denote $d$--dimensional $\d$--functions. Finally, the two
exponentials $e^{\,a\,\cdot\,\partial_\p}$ and $e^{\,\partial_\p\,\cdot\,
\partial_\p}$ act on generating functions respectively as a translation
operator and as a contraction operator, so that $\cA^{\,\pm}$ can be finally
presented in the form
\begin{multline}
\cA^{\,\pm}\,=\,\exp{\Big[\,\partial_{\xi_{\,1}}\cdot\partial_{\xi_{\,2}}\,+\,\partial_{ \xi_{\,2}}\cdot\partial_{\xi_{\,3}}\,+\,\partial_{\xi_{\,3}} \cdot\partial_{\xi_{\,1}}\Big]} \vphantom{{\phi}_{\,1}\left(p_{\,1},\,\xi_{\,1}\,\pm\,\sqrt{\frac{\a^{\,\prime}\!\!}{2}}\,p_{\,23}\right)} \\\times\,{\phi}_{\,1}\left(p_{\,1},\,\xi_{\,1}\,\pm\,\sqrt{\frac{\a^{\,\prime}\!\!}{2}}\,p_{\,23}\right)\,{\phi}_{\,2}\left( p_{\,2},\,\xi_{\,2}\,\pm\,\sqrt{\frac{\a^{\,\prime}\!\!}{2}}\,p_{\,31}\right)\, {\phi}_{\,3}\left(p_{\,3},\,\xi_{\,3}\,\pm\,\sqrt{\frac{\a^{\,\prime}\!\!}{2}}\,p_{\,12}\right)\Bigg|_{\xi_{\,i}\,=\,0}\ ,\label{Apiumeno}
\end{multline}
or equivalently in the form
\begin{equation}
\cA^{\,\pm}\,=\,{\phi}_{\,1}\left(p_{\,1}\,,\,\partial_{ \xi}\pm\sqrt{\frac{\a^{\,\prime}\!\!}{2}}\,p_{\,23}\right)\,{\phi}_{\,2}\left(p_{\,2}\,,\,\xi+\partial_{ \xi}\pm\sqrt{\frac{\a^{\,\prime}\!\!}{2}}\,p_{\,31}\right) \,{\phi}_{\,3}\left(p_{\,3}\,,\,\xi\pm\sqrt{\frac{\a^{\,\prime}\!\!}{2}}\,p_{\,12}\right)\Bigg|_{\xi\,=\,0}\ ,\label{Apiumeno2}
\end{equation}
where derivatives are always acting toward the right. No ordering ambiguities are present in the second factor, for traceless $\phi_{\,i}$.

\vskip 24pt


\scss{Four--point amplitudes for the First Regge
trajectory}\label{sec:fourpoint}


In this section we turn to four--point functions. Following steps similar to those that we have just illustrated for the three--point case, their general expressions can be considerably simplified when the external states belong to the first Regge
trajectory. The first step entails the explicit solution of the
$L_0$ constraint, that can be attained parametrizing the masses of the four
external states in terms of four integers $n_1$, $n_2$, $n_3$ and $n_4$, as
\begin{align}
-p_{\,1}^{\,2}&\,=\,\frac{n_1-1}{\a^{\,\prime}}\ ,& -p_{\,2}^{\,2}&\,=\,\frac{n_2-1}{\a^{\,\prime}}\ ,& -p_{\,3}^{\,2}&\,=\,\frac{n_3-1}{\a^{\,\prime}}\ ,& -p_{\,4}^{\,2}&\,=\,\frac{n_4-1}{\a^{\,\prime}}\ .
\end{align}
A closer look at the four--particle kinematics allows again a drastic simplification of the part of the generating function that only depends on the external momenta, so that
\begin{multline}
|\,y_{12}\,y_{13}\,y_{23}|\exp\left[\a^{\,\prime}\sum_{i\neq j}p_{\,i}\cdot p_j\ln|y_{ij}|\right]\,=\,\left|\frac{y_{13}y_{24}}{y_{12}y_{34}}\frac{y_{23}}{y_{24}y_{34}}\right| \left|\frac{y_{13}y_{14}}{y_{34}}\right|^{n_1} \\\times\left|\frac{y_{23}y_{24}}{y_{34}}\right|^{n_2}\left|\frac{y_{13}y_{23}}{y_{12}}\right|^{n_3} \left|\frac{y_{14}y_{24}}{y_{12}}\right|^{n_4} \left|\frac{y_{13}y_{24}}{y_{12}y_{34}}\right|^{-\a^{\,\prime}t-2}\left|\frac{y_{14}y_{23}}{y_{12}y_{34}}\right|^{-\a^{\,\prime}u-2}\ ,\label{fac}
\end{multline}
We take all momenta to be ingoing, so that the three Mandelstam variables are defined as
\be
s=-(p_{\,1}+p_{\,2})^2 \ , \quad t=-(p_{\,1}+p_{\,3})^2 \ , \quad u=-(p_{\,1}+p_{\,4})^2 \ , \label{mandelstam}
\ee
and are subject to the constraint
\begin{equation}
\a^{\,\prime}(s+t+u)\,=\,n_1+n_2+n_3+n_4-4\ .
\end{equation}

Making use of transversality and momentum conservation the generating function of the four--point correlations thus reduces to
\begin{equation}
\begin{split}
\mathbf{Z}\,&=\
i\,\frac{g_o^2}{\a^{\,\prime}}\ (2\pi)^{\,d}\,\delta^{\,(d)}(p_{\,1}+p_{\,2}+p_{\,3}+p_{\,4}) \left|\frac{y_{13}y_{24}}{y_{12}y_{34}}\right|\left|\frac{y_{23}}{y_{24}y_{34}}\right|\\
\times&\exp\left\{\vphantom{\sqrt{\frac{\a^{\,\prime}\!\!}{2}}}\sqrt{2\a^{\,\prime}}\,
\left[\left\langle\frac{y_{34}}{y_{13}y_{14}}\right\rangle \left(p_{\,3}\frac{y_{14}y_{23}}{y_{12}y_{34}}+p_{\,4}\frac{y_{13}y_{24}}{y_{12}y_{34}}\right)\cdot \xi_{\,1}\!-\left\langle\frac{y_{34}}{y_{23}y_{24}}\right\rangle\left(p_{\,3}\frac{y_{13}y_{24}}{y_{12}y_{34}}+ p_{\,4}\frac{y_{14}y_{23}}{y_{12}y_{34}}\right)\!\cdot \xi_{\,2}\right.\right.\\
&\qquad\qquad\ \ \left.+\left\langle\frac{y_{12}}{y_{13}y_{23}}\right \rangle\left(p_{\,1}\frac{y_{14}y_{23}}{y_{12}y_{34}}+ p_{\,2}\frac{y_{13}y_{24}}{y_{12}y_{34}}\right)\cdot \xi_{\,3}\!-\left\langle\frac{y_{12}}{y_{14}y_{24}}\right\rangle\left(p_{\,1}\frac{y_{13}y_{24}}{y_{12}y_{34}} +p_{\,2}\frac{y_{14}y_{23}}{y_{12}y_{34}}\right)\!\cdot \xi_{\,4}\right]\\
&\qquad\qquad\qquad\qquad\ \ +\left[\,\left|\frac{y_{13}y_{24}}{y_{12}y_{34}}\frac{y_{14}y_{23}}{y_{12}y_{34}}\right|(\xi_{\,1}\cdot \xi_{\,2}+\xi_{\,3}\cdot \xi_{\,4})+\left|\frac{y_{14}y_{23}}{y_{12}y_{34}}\right|(\xi_{\,1}\cdot \xi_{\,3}+\xi_{\,2}\cdot \xi_{\,4})\right.\\
&\qquad\qquad\qquad\qquad\ \ +\left.\left.\left|\frac{y_{13}y_{24}}{y_{12}y_{34}}\right|(\xi_{\,1}\cdot \xi_{\,4}+\xi_{\,2}\cdot \xi_{\,3})\,\right]\vphantom{\sqrt{\frac{\a^{\,\prime}\!\!}{2}}}\right\}\left|\frac{y_{13}y_{24}}{y_{12}y_{34}}\right|^{-\a^{\,\prime}t-2} \left|\frac{y_{14}y_{23}}{y_{12}y_{34}}\right|^{-\a^{\,\prime}u-2}\ ,\label{fourpoint}
\end{split}
\end{equation}
where, as before,
\be \langle x\rangle\,=\,\text{sign}(x)\ . \ee
Let us stress that eq.~\eqref{fourpoint} is a manifestly M\"obius invariant
measure, so that a M\"obius invariant result obtains after integrating over a
single $y_{\,i}$ variable, like for the four--tachyon Veneziano amplitude. As
for the more familiar low--lying states, one can thus fix the positions of three
insertions, say letting $y_{1}\,=\,0$, $y_{\,2}\,=\,1$ and
$y_{\,3}\,=\,\infty$, to integrate $y_{\,4}$, that we shall call $y$ for
brevity in what follows, over the three intervals of the real axis thus identified, associating
to the resulting amplitudes Chan--Paton factors that reflect the ordering of the
external states along the boundary. This, however, does now exhaust the
cyclically inequivalent configurations, so that one is also to repeat the
construction after reversing the order of a pair, say $y_{\,1}$ and
$y_{\,2}$. All in all, one is thus led to the following expression for the
amplitude,
{\allowdisplaybreaks
\begin{eqnarray}
\cA &=&\,i\,\frac{g_o}{\a^{\,\prime}}\,(2\pi)^{\,d}\,\delta^{\,(d)}(p_{\,1}+p_{\,2}+p_{\,3}+p_{\,4})\int_0^1\,dy\\
&\times &\left[\vphantom{\cA^{(1)}_+(y)} \right.(1-y)^{-\a^{\,\prime}t-2}y^{-\a^{\,\prime}u-2}\left(\cA^{(1)\, +}(y)\,Tr[\L^{a_1}\L^{a_4}\L^{a_2}\L^{a_3}]\,+\cA^{(1)\,-}(y) \,Tr[\L^{a_2}\L^{a_4}\L^{a_1}\L^{a_3}]\right)\nonumber\\
&+&(1-y)^{-\a^{\,\prime}s-2}y^{-\a^{\,\prime}t-2}\left(\cA^{(2)\,+}(y)\,Tr[\L^{a_1}\L^{a_2}\L^{a_4}\L^{a_3}] \,+\cA^{(2)\,-}(y)\,Tr[\L^{a_2}\L^{a_1}\L^{a_4}\L^{a_3}]\right)\nonumber\\
&+&\!\!\left.(1-y)^{-\a^{\,\prime}u-2}y^{-\a^{\,\prime}s-2}\left(\cA^{(3)\,+}(y)\,Tr[\L^{a_4}\L^{a_1}\L^{a_2}\L^{a_3}] \,+\cA^{(3)\,-}(y)\,Tr[\L^{a_4}\L^{a_2}\L^{a_1}\L^{a_3}]\right)\right]\ ,\nonumber\label{fouramplitude}
\end{eqnarray}}
\!\!where
\begin{equation}
\cA^{(i)\,\pm}(\l)\,=\,\left[\phi_{\,1}(p_{\,1},\xi_{\,1})\,\phi_{\,2}(p_{\,2},\xi_{\,2})\, \phi_{\,3}(p_{\,3},\xi_{\,3})\,\phi_{\,4}(p_{\,4},\xi_{\,4})\right]
\, \star \,
{\mathbf{Z}}^{(i)\,\pm}(\xi_{\,1},\xi_{\,2},\xi_{\,3},\xi_{\,4};\,p_{\,1},p_{\,2},p_{\,3},p_{\,4})\
,\label{Ai}
\end{equation}
and
\begin{equation}
\begin{split}\label{Z1}
\mathbf{Z}^{(1)\,\pm}\,=&\,\vphantom{\int_0^1}\exp\left\{\vphantom{\int_0^1}\sqrt{2\a^{\,\prime}} \,\pm\,\Big[(-p_{\,3}\,y+p_{\,4}\,(1-y))\cdot \xi_{\,1}+(p_{\,3}\,(1-y)-p_{\,4}\,y)\cdot \xi_{\,2}\right.\\+\vphantom{\int_0^1}&\,(p_{\,1}\,y-p_{\,2}\,(1-y))\cdot \xi_{\,3}+\,(-\,p_{\,1}\,(1-y)+p_{\,2}\,y)\cdot \xi_{\,4}\Big]\\
+&\left.\vphantom{\int_0^1}\Big[\,y(1-y)\,(\xi_{\,1}\cdot \xi_{\,2}+\xi_{\,3}\cdot \xi_{\,4})+y\,(\xi_{\,1}\cdot \xi_{\,3}+\xi_{\,2}\cdot \xi_{\,4})+(1-y)\,(\xi_{\,1}\cdot \xi_{\,4}+\xi_{\,2}\cdot \xi_{\,3})\,\Big]\right\}\ ,
\end{split}
\end{equation}
\begin{equation}
\begin{split}\label{Z2}
\mathbf{Z}^{(2)\,\pm}\,=&\,\exp\left\{\vphantom{\int_0^1}\sqrt{2\a^{\,\prime}}\,\pm\,\Big[-(p_{\,3}+p_{\,4}\,y)\cdot \xi_{\,1}+(p_{\,3}\,y+p_{\,4})\cdot \xi_{\,2}\right.\\ +&\, \vphantom{\int_0^1}(p_{\,1}+p_{\,2}\,y)\cdot \xi_{\,3}-(p_{\,1}\,y+p_{\,2})\cdot \xi_{\,4}\Big]\\
+&\left.\vphantom{\int_0^1}\Big[\,y\,(\xi_{\,1}\cdot \xi_{\,2}+\xi_{\,3}\cdot \xi_{\,4})+(1-y)\,(\xi_{\,1}\cdot \xi_{\,3}+\xi_{\,2}\cdot \xi_{\,4})+y(1-y)\,(\xi_{\,1}\cdot \xi_{\,4}+\xi_{\,2}\cdot \xi_{\,3})\,\Big]\right\}\ ,
\end{split}
\end{equation}
\begin{equation}
\begin{split}\label{Z3}
\mathbf{Z}^{(3)\,\pm}\,=&\,\exp\left\{\vphantom{\int_0^1}\sqrt{2\a^{\,\prime}}\,\pm\,\Big[-(p_{\,3}\,(1-y)+p_{\,4})\cdot \xi_{\,1}+(p_{\,3}+p_{\,4}\,(1-y))\cdot \xi_{\,2}\right.\\ -&\,\vphantom{\int_0^1}(p_{\,1}\,(1-y)-p_{\,2})\cdot \xi_{\,3}+(p_{\,1}+p_{\,2}\,(1-y))\cdot \xi_{\,4}\Big]\\
\vphantom{\int_0^1}+&\left.\Big[\,(1-y)\,(\xi_{\,1}\cdot \xi_{\,2}+\xi_{\,3}\cdot \xi_{\,4})+y(1-y)\,(\xi_{\,1}\cdot \xi_{\,3}+\xi_{\,2}\cdot \xi_{\,4})+y\,(\xi_{\,1}\cdot \xi_{\,4}+\xi_{\,2}\cdot \xi_{\,3})\,\Big]\vphantom{\int_0^1}\right\}\ .
\end{split}
\end{equation}
Finally, using eq.~\eqref{ContrFormula} one can simplify further eq.~\eqref{Ai}
with techniques similar to those illustrated for the three-point case. In this way $\cA^{(i)\,\pm}$ can be finally recast in the form
{\allowdisplaybreaks
\begin{eqnarray}
\cA^{(1)\,\pm}(y) \!\!&=&\vphantom{\int_0^1}\!\!\exp\Big\{y\,(1-y)\,(\partial_{\xi_{\,1}}\cdot
\partial_{\xi_{\,2}}+\partial_{\xi_{\,3}}\cdot \partial_{\xi_{\,4}})+y\,
(\partial_{\xi_{\,1}}\cdot \partial_{\xi_{\,3}}+
\partial_{\xi_{\,2}}\cdot \partial_{\xi_{\,4}}) \label{A} \nonumber\\ &&\ \ \ \ +\,\vphantom{\int_0^1}(1-y)\,(\partial_{\xi_{\,1}}\cdot \partial_{\xi_{\,4}}+
\partial_{\xi_{\,2}}\cdot \partial_{\xi_{\,3}})\Big\} \\
\!\!\!\!\!\! \times&& \vphantom{\int_0^1}\!\!\!\!\!\!\!\!\!\!\!\!\!\!{\phi}_{\,1}\left(p_{\,1},\ \xi_{\,1}\,\pm\,\sqrt{2\a^{\,\prime}}\,
(-p_{\,3}\,y\,+\,p_{\,4}\,(1-y))\right)\,{\phi}_{\,2}\left(p_{\,2},\
\xi_{\,2}\,\pm\, \sqrt{2\a^{\,\prime}}\,(p_{\,3}\,(1-y)\,-\,p_{\,4}\,y)\right)\nonumber \\
\!\!\!\!\!\! \times&& \vphantom{\int_0^1}\!\!\!\!\!\!\!\!\!\!\!\!\!\!{\phi}_{\,3}\left(p_{\,3},\
\xi_{\,3}\,\pm\,\sqrt{2\a^{\,\prime}}\,(p_{\,1}\,y-p_{\,2}\,
(1-y)\right)\,{\phi}_{\,4}\left(p_{\,4},\
\xi_{\,4}\,\pm\,\sqrt{2\a^{\,\prime}}\,(-p_{\,1}\,(1-y)\,+
\,p_{\,2}\,y)\right)\Big|_{\xi_{\,i}\,=\,0}\ .\nonumber\\
\nonumber\\
\cA^{(2)\,\pm}(y) \!\!&=&\vphantom{\int_0^1}\!\!\exp\Big\{y\,(\partial_{\xi_{\,1}}\cdot
\partial_{\xi_{\,2}}+\partial_{\xi_{\,3}}\cdot \partial_{\xi_{\,4}})+(1-y)\,
(\partial_{\xi_{\,1}}\cdot \partial_{\xi_{\,3}}+
\partial_{\xi_{\,2}}\cdot \partial_{\xi_{\,4}})  \nonumber\\ &&\ \ \ \ +\,\vphantom{\int_0^1}y\,(1-y)\,(\partial_{\xi_{\,1}}\cdot \partial_{\xi_{\,4}}+\nonumber
\partial_{\xi_{\,2}}\cdot \partial_{\xi_{\,3}})\Big\} \\
\!\!\!\!\!\! \times&& \vphantom{\int_0^1}\!\!\!\!\!\!\!\!\!\!\!\!\!\!{\phi}_{\,1}\left(p_{\,1},\ \xi_{\,1}\,\pm\,\sqrt{2\a^{\,\prime}}\,
(-p_{\,3}\,-\,p_{\,4}\,y)\right)\,{\phi}_{\,2}\left(p_{\,2},\
\xi_{\,2}\,\pm\, \sqrt{2\a^{\,\prime}}\,(p_{\,3}\,y\,+\,p_{\,4})\right)\nonumber \\
\!\!\!\!\!\! \times&& \vphantom{\int_0^1}\!\!\!\!\!\!\!\!\!\!\!\!\!\!{\phi}_{\,3}\left(p_{\,3},\
\xi_{\,3}\,\pm\,\sqrt{2\a^{\,\prime}}\,(p_{\,1}\,+\,p_{\,2}\,y\right)\,{\phi}_{\,4}\left(p_{\,4},\
\xi_{\,4}\,\pm\,\sqrt{2\a^{\,\prime}}\,(-p_{\,1}\,y\,-
\,p_{\,2})\right)\Big|_{\xi_{\,i}\,=\,0}\ ,\nonumber\\
\nonumber\\
\cA^{(3)\,\pm}(y) \!\!&=&\vphantom{\int_0^1}\!\!\exp\Big\{(1-y)\,(\partial_{\xi_{\,1}}\cdot
\partial_{\xi_{\,2}}+\partial_{\xi_{\,3}}\cdot \partial_{\xi_{\,4}})+y\,(1-y)\,
(\partial_{\xi_{\,1}}\cdot \partial_{\xi_{\,3}}+
\partial_{\xi_{\,2}}\cdot \partial_{\xi_{\,4}}) \nonumber\\ &&\ \ \ \ +\,\vphantom{\int_0^1}y\,(\partial_{\xi_{\,1}}\cdot \partial_{\xi_{\,4}}+\nonumber
\partial_{\xi_{\,2}}\cdot \partial_{\xi_{\,3}})\Big\} \\
&&\!\!\!\!\!\!\!\!\!\!\!\!\!\!\!\!\!\!\!\!\!\!\!\!\!\!\!\!\!\!\!\!\!\! \times \vphantom{\int_0^1}{\phi}_{\,1}\left(p_{\,1},\ \xi_{\,1}\,\pm\,\sqrt{2\a^{\,\prime}}\,
(-p_{\,3}\,(1-y)\,-\,p_{\,4})\right)\,{\phi}_{\,2}\left(p_{\,2},\
\xi_{\,2}\,\pm\, \sqrt{2\a^{\,\prime}}\,(p_{\,3}\,+\,p_{\,4}\,(1-y))\right)\nonumber \\
&&\!\!\!\!\!\!\!\!\!\!\!\!\!\!\!\!\!\!\!\!\!\!\!\!\!\!\!\!\!\!\!\!\!\! \times \vphantom{\int_0^1}{\phi}_{\,3}\left(p_{\,3},\
\xi_{\,3}\,\pm\,\sqrt{2\a^{\,\prime}}\,(-\,p_{\,1}\,(1-y)\,+\,p_{\,2}\right)\,{\phi}_{\,4}\left(p_{\,4},\
\xi_{\,4}\,\pm\,\sqrt{2\a^{\,\prime}}\,(p_{\,1}\,+
\,p_{\,2}\,(1-y))\right)\Big|_{\xi_{\,i}\,=\,0}\ .\nonumber
\end{eqnarray}}
\!\!where we have replaced $\xi_{\,i}$ with $\partial_{\xi_{\,i}}$ in
eqs.~\eqref{Z1}, \eqref{Z2} and \eqref{Z3}, letting them act on the four
generating functions for the external states.

\vskip 24pt


\scss{String Currents}\label{sec:stringcurr}\label{sec:Currents}


One can also turn the three--point amplitudes of Section
\ref{sec:threepointampl} into current generating functions whose Fourier
transforms encode the open--string couplings in the form $\Phi\star \cJ$, in
terms of the generating function $\Phi$ of coordinate--space fields. To this
end, one can define the current generating function in momentum space,
\begin{equation}
{\mathcal{J}}(\xi)\,=\,i\,\frac{g_o}{\a^{\,\prime}}\,(2\pi)^{\,d}\, \delta^{\,(d)}(p_{\,1}+p_{\,2}+p_{\,3})\,\Big\{{j}^{\,+}(\xi)\ Tr[\ \cdot\ \L^{a_2}\L^{a_3}]+
{j}^{\,-}(\xi)\ Tr[\ \cdot\ \L^{a_3}\L^{a_2}]\Big\}\ ,\label{Curr}
\end{equation}
that takes a rather compact form in terms of the auxiliary currents
\be
{j}^{\,\pm}(\xi)=\int
\frac{d^d\pi_{\,2}}{(2\pi)^{\,d/2}}\frac{d^d\pi_{\,3}}{(2\pi)^{\,d/2}}\
\mathbf{\tilde{Z}}^{\,\pm}(-p_{\,2}-p_{\,3},p_{\,2},p_{\,3};\xi,\pi_{\,2},\pi_{\,3})\,{\phi}_{\,2}
(p_{\,2},i\pi_{\,2})\,{\phi}_{\,3}(p_{\,3},i\pi_{\,3})\ .
\ee
The Fourier transform of $\mathbf{Z}^{\,\pm}$ with respect to $\xi_{\,2}$ and $\xi_{\,3}$ is then
\begin{multline}
\mathbf{\tilde{Z}}^{\,\pm}(\xi,\p_{\,2},\p_{\,3})\,=\,(2\pi)^{\,d}\exp\left\{\pm\, i\,\sqrt{\frac{\a^{\,\prime}\!\!}{2}}\Big(-i\xi\cdot p_{\,23}+ \partial_{\p_{\,2}}\cdot p_{\,31}+\partial_{\p_{\,3}}\cdot p_{\,12}\Big)\right.\\\left.\vphantom{i\sqrt{\frac{\a^{\,\prime}\!\!}{2}}}-\Big(-i\xi\cdot \partial_{\pi_{\,2}}-i\xi\cdot \partial_{\p_{\,3}}+\partial_{\p_{\,2}}\cdot \partial_{\p_{\,3}}\Big)\right\}\,\delta^{\,(d)}(\p_{\,2})\,\delta^{\,(d)}(\p_{\,3})\ ,
\end{multline}
and the other integrals in \eqref{Curr} yield finally
\begin{multline} \label{current+-}
{j}^{\,\pm}\,=\,\exp\left(\pm \sqrt{\frac{\a^{\,\prime}\!\!}{2}}\
\xi\cdot p_{\,23}\right)\exp\left(\partial_{\pi_{\,2}}\cdot
\partial_{\pi_{\,3}}\right)\\\times {\phi}_{\,2}\left(p_{\,2},\ \pi_2+\xi
\pm\sqrt{\frac{\a^{\,\prime}\!\!}{2}}\,p_{\,31}\right)\,{\phi}_{\,3}\,
\left(p_{\,3},\ \pi_3+\xi\pm\sqrt{\frac{\a^{\,\prime}\!\!}{2}}\,
p_{\,12}\right)\Bigg|_{\pi_{\,2}\,=\,\pi_{\,3}\,=\,0}\ ,
\end{multline}
where in order to simplify the last expression we have performed the further redefinitions
\begin{equation}
\p_{\,1}\ra -\, i\, \p_{\,1}\ ,\qquad
\p_{\,2}\ra -\, i\, \p_{\,2}\ .
\end{equation}
The second factor in eq.~\eqref{current+-} can now be regarded as a translation
operator, so that the generating function ${j}^{\,\pm}$ can be turned into the
equivalent expression
\begin{multline}
{j}^{\,\pm}\,=\,\exp\left\{\pm \sqrt{\frac{\a^{\,\prime}\!\!}{2}}\ \xi\cdot
p_{\,23}\right\}\ {\phi}_{\,2}\left(p_{\,2},\,\partial_\p+\xi\pm\sqrt{\frac{\a^{\,\prime}\!\!}{2}}\,p_{\,31}\right)
{\phi}_{\,3}
\left(p_{\,3},\,\p+\xi\pm\sqrt{\frac{\a^{\,\prime}\!\!}{2}}\,p_{\,12}
\right)\Bigg|_{\p\,=\,0}\ .
\end{multline}
The coordinate--space current can then be recovered by a Fourier transform,
\begin{multline}
J^{\,\pm}(x,\xi)\,=\,\int \frac{d^dp_{\,2}}{(2\pi)^{\,d}}\
\frac{d^dp_{\,3}}{(2\pi)^{\,d}}\ e^{ix\cdot
(p_{\,2}+p_{\,3})}{j}^{\,\pm}(-\,p_{\,2}\,-\,p_{\,3},\,p_{\,2},\,p_{\,3};\,\xi)\\\,=\,\int
\frac{d^d p_{\,2}}{(2\pi)^{\,d}} \ \frac{d^d p_{\,3}}{(2\pi)^{\,d}}\
\exp\left[i\left(x\,\mp\, i\sqrt{\frac{\a^{\,\prime}\!\!}{2}}\ \xi\right)\cdot
p_{\,2}\right]\exp\left[i\left(x\, \pm\,
i\sqrt{\frac{\a^{\,\prime}\!\!}{2}}\ \xi\right)\cdot p_{\,3}\right]\\\times
{\phi}_{\,2}\left(p_{\,2}\,,\,\partial_{\,\chi}+\xi\pm\sqrt{2\a^{\,\prime}}p_{\,3}\right) {\phi}_{\,3}
\left(p_{\,3}\,,\,\chi+\xi\mp\sqrt{2\a^{\,\prime}}\,p_{\,2}\right)\Bigg|_{\chi\,=\,0}\
,
\end{multline}
and making use of standard identities one finally obtains
\begin{multline}
J^{\,\pm}(x,\xi)\,=\,{\Phi}_{\,2}\left(x\,\mp\,i\sqrt{\frac{\a^{\,\prime}\!\!}{2}}\ \xi,\,
\partial_{\chi}+\xi\mp
i\sqrt{2\a^{\,\prime}}\,\partial_{\,3}\right)\\\times{\Phi}_{\,3}\left(x\,
\pm\,i\sqrt{\frac{\a^{\,\prime}\!\!}{2}}\ \xi,\
\chi+\xi\pm i\sqrt{2\a^{\,\prime}}\,\partial_{\,2}\right)\Bigg|_{\chi\,=\,0}\ ,
\label{offshell}
\end{multline}
where the ${\Phi}_{\,i}$ are coordinate--space generating functions for the
fields. As usual, the complete result for the current follows after combining,
as in eq.~\eqref{Curr}, $J^{\,+}$ and $J^{\,-}$ with the Chan--Paton factors that
reflect the internal symmetry group. The couplings then obtain combining the
generating function \eqref{offshell} with $\Phi_{\,1}$, the corresponding one for the remaining set of fields.

In general, the current \eqref{Curr} is \emph{not} conserved. For the
generating function, current conservation would translate into the statement
that, on shell,
\begin{equation}
\left(\eta^{\,\m\n}\frac{\partial}{\partial x^{\,\m}}\ \frac{\partial}{\partial \xi^{\n}}\right)\cJ(x,\xi)\ = \ 0\ ,\label{Cons}
\end{equation}
and in order to see explicitly what happens in our case it is convenient to define $a^{\pm}$, $b^{\pm}$ and $c^{\pm}$ in such a way that
\begin{equation}
J^{\,\pm}(x,\xi)\,=\,{\Phi}_{\,2}(a^{\mp},b^\mp)\,{\Phi}_{\,3}(a^{\pm},c^{\pm})\ .
\end{equation}
As a result
{\allowdisplaybreaks
\begin{eqnarray}
\partial_{\xi}\cdot\partial_x\,J^{\,\pm}(x,\xi)&\!\!\!=&\!\!\!\mp\ \, i\,\sqrt{\frac{\a^{\,\prime}\!\!}{2}}\left[\partial_{a^\mp}^{\,2}{\Phi}_{\,2}(a^{\mp},b^\mp)\right] {\Phi}_{\,3}(a^{\pm},c^{\pm})+ \left[\partial_{a^\mp}\cdot \partial_{b^\mp}{\Phi}_{\,2}(a^{\mp},b^\mp)\right]{\Phi}_{\,3}(a^{\pm},c^{\pm})\nonumber\\
\pm&&\!\!\!\!\!\!\!\!\!\!\! i\sqrt{\frac{\a^{\,\prime}\!\!}{2}}\left[\partial_{a^\mp}{\Phi}_{\,2}(a^{\mp},b^\mp)\right] \left[\partial_{a^\pm}{\Phi}_{\,3}(a^{\pm},c^{\pm})\right]+\left[\partial_{a^\mp} {\Phi}_{\,2}(a^{\mp},b^\mp)\right]\left[\partial_{c^\pm}{\Phi}_{\,3}(a^{\pm},c^{\pm})\right]\nonumber\\
\mp&&\!\!\!\!\!\!\!\!\!\!\!  i\sqrt{\frac{\a^{\,\prime}\!\!}{2}}\left[\partial_{a^\mp}{\Phi}_{\,2}(a^{\mp},b^\mp)\right] \left[\partial_{a^\pm}{\Phi}_{\,3}(a^{\pm},c^{\pm})\right]+ \left[\partial_{b^\mp}{\Phi}_{\,2}(a^{\mp},b^\mp)\right] \left[\partial_{a^\pm}{\Phi}_{\,3}(a^{\pm},c^{\pm})\right]\nonumber\\
\pm&&\!\!\!\!\!\!\!\!\!\!\!  i\sqrt{\frac{\a^{\,\prime}\!\!}{2}}\ {\Phi}_{\,2}(a^\mp,b^\mp)
\left[\partial_{a^\pm}^{\,2}{\Phi}_{\,3}(a^\pm,c_\pm)\right]+ {\Phi}_{\,2}(a^\mp,b^\mp) \left[\partial_{a^\pm}\cdot\partial_{c^\pm}{\Phi}_{\,3}(a^\pm,c^\pm)\right]\ ,\label{var}
\end{eqnarray}}
\!\!where pairs of derivatives are always contracted together, even when this is not explicitly indicated.
One can now use the on--shell Klein--Gordon and transversality relations, that in this notation read
\begin{equation}
\begin{split}
(\partial_a^{\,2}+m^2)\,\Phi(a,b)& \,= \, 0\ ,\\
\partial_a\cdot \partial_b\ \Phi(a,b)&\, =\, 0\ ,
\end{split}
\end{equation}
to finally turn eq.~\eqref{var} into
\be
\partial_{\xi}\cdot\partial_x\ J^{\,\pm}(x,\,\xi)\, =\, \left[\partial_{a^\mp}{\Phi}_{\,2}(a^{\mp},b^\mp)\right] \left[\partial_{c^\pm}{\Phi}_{\,3}(a^{\pm},c^{\pm})\right] +\left[\partial_{b^\mp}{\Phi}_{\,2}(a^{\mp},b^\mp)\right] \left[\partial_{a^\pm}{\Phi}_{\,3}(a^{\pm},c^{\pm})\right]\ .
\ee

Therefore, in general the currents are not
conserved. In the next section we shall see how such residual contributions can be ascribed to the external masses in such a way that, once they are ignored, conserved currents emerge. At any rate, one can readily extract from eq.~\eqref{Curr} a portion of the current that is \textit{manifestly conserved},
\begin{multline}
J^{\,\pm}(x,\xi)\,=\,{\Phi}_{\,2}\left(x\,\mp\,i\sqrt{\frac{\a^{\,\prime}\!\!}{2}}\,\xi,\,\partial_{\chi}\mp\,
i\sqrt{2\a^{\,\prime}}\,\partial_{\,3}\right)\\\times\,{\Phi}_{\,3}
\left(x\,\pm\,i\sqrt{\frac{\a^{\,\prime}\!\!}{2}}\,\xi,\,\chi\,\pm\,
i\sqrt{2\a^{\,\prime}}\,\partial_{\,2}\right)\Bigg|_{\chi\,=\,0}\
,\label{off-shell}
\end{multline}
while non--conserved terms show up whenever $\xi$ appears in the second argument of the ${\Phi}_{\,i}$. The high energy limit corresponds to $\a^{\,\prime}\ra\infty$, and performing the redefinition $\xi_{\,i}\,\rightarrow \sqrt{\frac{1}{2\a^{\,\prime}\!\!\!}}\ \,\xi_{\,i}$ turns \eqref{off-shell} into
\begin{multline}
J^{\,\pm}(x,\xi)\,=\,{\Phi}_{\,2}\left(x\,\mp\,i\,\frac{\xi}{2}\,,\,\partial_{ \chi}+\frac{\xi}{\sqrt{2\a^{\,\prime}}}\,\mp\, i\sqrt{2\a^{\,\prime}}\,\partial_{\,3}\right)\\\times\,{\Phi}_{\,3}\left(x\,\pm\,i\,\frac{\xi}{2}\,,\ \chi\,+\frac{\xi}{\sqrt{2\a^{\,\prime}}}\,\pm\, i\sqrt{2\a^{\,\prime}}\,\partial_{\,2}\right)\Bigg|_{\chi\,=\,0}\ ,\label{born}
\end{multline}
so that the relative weights of the various contributions become manifest.

Let us stress that the conventional high--energy limit of String Theory
identified in \cite{Gross} corresponds to retaining only the highest--derivative terms present in the amplitudes, and results from the manifestly conserved
portion of the current \eqref{born}. The other terms in the current can be
related to interesting off--shell couplings after carefully taking into account
the massive Klein--Gordon equation that the external states are to
satisfy. The highest derivative terms thus provide only a partial view of the
high--energy behavior of the amplitudes, and actually leave out a most
interesting datum, their non--abelian structure. In the following sections we
shall analyze in detail some of these couplings in order to turn them, insofar
as possible, into an off--shell Lagrangian form that can conveniently provide
useful indications on HS couplings.

In principle, one can extend this type of analysis to all states of the open
string spectrum. However, complications arise when considering subleading Regge
trajectories: the cancelation of all $y_{\,i}$--dependent terms is not
manifest anymore for these more general states, since it rests on the complete
Virasoro constraints. This problem reflects the technical difficulties that are
met when trying to construct a complete generating function containing, in a
covariant form, representatives for all physical states of the string spectrum.
We have only partial results in this respect, and therefore we leave this
problem for future work, confining our attention to special cases. One of
these, as we shall see in Section \ref{sec:00s}, is the scattering of
arbitrary physical states off a pair of tachyons, but in eq.~\eqref{guess} we
shall also present a motivated guess for the limiting behavior of generic
three--point functions of mixed--symmetry states.

Summarizing, in this section we have shown that for the first Regge trajectory
of the open bosonic string all cubic couplings can be subsumed in the
expression $\Phi \star \cJ$, where the \emph{partially conserved} currents
$\cJ$ of eq.~\eqref{Curr} have a relatively simple form. The emphasis on the
$SL(2,R)$ symmetry, that we kept manifest at all stages, is perhaps the key
difference with respect to previous approaches and was instrumental in attaining this simplification.

\vskip 36pt


\scs{Explicit String Couplings}\label{sec:couplings}


In this section we study the amplitudes \eqref{Apiumeno} and the currents
\eqref{offshell} in order to extract from them off--shell couplings. Our aim is to
identify, on the field theory side, explicit string--inspired cubic interactions of the
form $\Phi\star \cJ$, with special emphasis on their massless limit, where the gauge
symmetry of the HS modes is restored and where, at the same time, long--recognized
difficulties with massless HS fields are eventually expected to show up. The limiting
forms of the amplitudes and the corresponding conserved currents will be denoted by a
$[0]$ superscript.

The starting point is the expansion of eq.~\eqref{Apiumeno}, computed using the standard multinomial formula,
\begin{multline}
\cA^{\,\pm}\,=\,{\phi}_{\,1}\,{\phi}_{\,2}\,{\phi}_{\,3}\!\!\!\sum_{i\,,\,j\,,\,k\,\in\,\cI}\,
\left(\!\pm\sqrt{\frac{\a^{\,\prime}\!\!}{2}}\,\right)^{s_1+s_2+s_3-2i-2j-2k} \\
\times\,\left[
\frac{s_1!\, s_2!\,s_3!\ p_{\,23}^{\,s_1-j-k}\ p_{\,31}^{\,s_2-k-i}\ p_{\,12}^{\,s_3-i-j}}{i!\,j!\,k!\,(s_1\,-\,j\,-\,k)!\,(s_2\,-\,k\,-\,i)!\,(s_3\,-\,i\,-\,j)!}\
\delta_{\,23}^{\,i}\
\delta_{\,31}^{\,j}\ \delta_{\,12}^{\,k}\right]\ ,\label{expl}
\end{multline}
where ${\phi}_{\,1}$, ${\phi}_{\,2}$ and ${\phi}_{\,3}$ are totally symmetric momentum--space fields of spins $s_1$, $s_2$ and $s_3$ respectively, the $p_{\,ij}$ are
defined in eq.~\eqref{pij} and
\be
\cI\,=\,\left\{i\,,\,j\,,\,k\,\in\,\mathbb{N}\,\Big|\ s_1-j-k\,\geq\,0\,;\,s_2-k-i\,\geq\,0\,;\,s_3-i-j\,\geq\,0\,\right\}\ .
\ee

For brevity we are leaving some space--time indices implicit, and as a result we are not
indicating that each of the $p_{\,ij}$ is contracted with a corresponding ${\phi}_{\,k}$,
with $k$ different from $i$ and $j$. Finally, the power, say, of $\d_{\,ij}$, indicates
the contractions between ${\phi}_{\,i}$ and ${\phi}_{\,j}$.

\vskip 24pt


\scss{0--0--$s$ Couplings}\label{sec:00s}


In this case the string current \eqref{Curr} reduces to
\begin{equation}
\cJ(x,\xi)\,=\,i\,\frac{g_o}{\a^{\,\prime}}\,\Big\{J^{\,+}(x,\xi )\ Tr\,[\ \cdot\ \L^{a_1}\L^{a_2}]+J^{\,-}(x,\xi)\ Tr\,[\ \cdot\ \L^{a_2}\L^{a_1}]\Big\}\ ,
\end{equation}
with
\begin{equation}
J^{\, \pm}(x,\xi)\,=\,\Phi\left(x\,\pm\, i\,\sqrt{\frac{\a^{\,\prime}\!\!}{2}}\ \xi\right)\,
\Phi\left(x\,\mp\, i\,\sqrt{\frac{\a^{\,\prime}\!\!}{2}}\ \xi\right)\ ,\label{scalar}
\end{equation}
while the corresponding amplitude is
\be
\cA^{\,\pm}_{\,0-0-s}\,=\,\left(\!\pm\sqrt{\frac{\a^{\,\prime}\!\!}{2}}\,\right)^{s}\phi_{\,1}\,\phi_{\,2}\ \phi_{\,3}\cdot p_{\,12}^{\,s}\ ,
\ee
where $\phi_{\,1}$ and $\phi_{\,2}$ are scalar fields and $\phi_{\,3}$ is a spin--$s$ field.
The scalar currents \eqref{scalar} were identified long ago by Berends, Burgers and van Dam
\cite{bbvd} and were recently reconsidered by Bekaert, Joung and Mourad in
\cite{Bekaert:2009ud}. Here we see them emerging as effective tree--level
0--0--$s$ couplings of the open bosonic string. They are inevitably
\emph{conserved}, since in this case they are bound to coincide with the
highest derivative portion, simply because the coupling between a pair of
scalar fields and a spin--$s$ tensor is to involve at least $s$ derivatives.
One can also obtain, in a similar fashion, general couplings between pairs of
tachyons and arbitrary mixed--symmetry states of the open bosonic string.
Actually, the currents that can be built in terms of the tachyon field are
totally symmetric, so that the tachyon can only couple directly to totally
symmetric fields, albeit arising from \emph{any} Regge trajectory.

\vskip 24pt


\scss{1--1--$s$ Couplings}\label{sec:11s}


This case exhibits a richer structure, since the generating function
\eqref{Curr} does not lead directly to a result that is gauge invariant
off--shell for both the spin--$s$ and spin--$1$ external legs. In this case the
on--shell expression \eqref{expl} reads
\begin{equation}
\begin{split}
\cA^{\,\pm}_{\,1-1-s}\,=\,&\left(\!\pm\sqrt{\frac{\a^{\,\prime}\!\!}{2}}\,\right)^{s-2}\,s(s-1)\, A_{1\,\m}\,A_{2\,\n}\,\phi^{\,\m\n\ldots}\,p_{\,12}^{\,s-2}\\
+&\left(\!\pm\sqrt{\frac{\a^{\,\prime}\!\!}{2}}\,\right)^s \Big[A_{1} \cdot A_{2}\ \phi\cdot p_{\,12}^{\,s}+s\,A_{1}\cdot p_{\,23}\,A_{2\,\n}\,\phi^{\,\n\ldots\,}p_{\,12}^{\,s-1}\\&\qquad\qquad\quad+s\,A_{2}\cdot p_{\,31}\, A_{1\,\n}\,\phi^{\,\n\ldots\,}p_{\,12}^{\,s-1}\Big]\\+&\left(\!\pm\sqrt{\frac{\a^{\,\prime}\!\!}{2}}\,\right)^{s+2}A_{1} \cdot p_{\,23}\,A_{2}\cdot p_{\,31}\, \phi\cdot p_{\,12}^{\,s}\ , \label{11s}
\end{split}
\end{equation}
and comprises three groups of terms, that we have ordered according to their
overall power of $\sqrt{\a^{\,\prime}}$ in order to stress their relative
weights in the $\a^{\,\prime}\,\ra\,\infty$ limit. Notice that only their sum is gauge invariant
on shell, as it should, insofar as the two \emph{massless} spin--$1$ external
legs are concerned. On the other hand, under the spin--$s$ gauge transformation
\be
\d\cA^{\,\pm}_{\,1-1-s}\,=\,\left(\!\pm\sqrt{\frac{\a^{\,\prime}\!\!}{2}}\,\right)^{s-2}\,\frac{s(s-1)}{2}\, \Big[A_{2}\cdot p_{\,31}\, A_{1\,\m}\,\Lambda^{\m\,\ldots\,}\cdot p_{\,12}^{\,s-2}-A_{1}\cdot p_{\,23}\, A_{2\,\m}\,\Lambda^{\m\,\ldots\,}\cdot p_{\,12}^{\,s-2}\Big]\ ,\label{lack}
\ee
that clearly does not vanish for generic values of $s$. While this type of
result is expected for massive excitations, \eqref{lack} has nonetheless a
peculiar feature: it does not contain any terms where the gauge parameter
$\Lambda$ is entirely contracted with momenta, but only contributions of the
type\footnote{Gauge variations of this type are proportional to the mass of the HS field, which in its turn is proportional to $s-1$. As a result, they can be compensated by terms involving divergences, that we ignoring at this point. The reader will find more details on this matter in Section \ref{sec:massgauginv}.} $A_\m\,\Lambda^{\m\ldots\,}$. Spin--1 curvatures give rise indeed to terms
of this form, together with overall factors $p^{\,2}_{\,i}$ that the massive
field equations have turned into powers of $1/\alpha^{\,\prime}$. Let us stress
that in the high--energy limit terms of this type are subleading, an important
feature that we shall soon elaborate upon.

As we have seen, the amplitudes \eqref{11s} are invariant under spin--1 gauge
transformations, and therefore it should not come as a surprise to the reader
that they can be recast in the form
\begin{equation}
\cA^{\,\pm}_{\,1-1-s}\,=\,-\, 2\left(\!\pm\sqrt{\frac{\a^{\,\prime}\!\!}{2}}\,\right)^{s+2} \left(F^{\,2}\right) \, \phi\cdot p_{\,12}^{\,s}\,+\, 4s\left(\!\pm\sqrt{\frac{\a^{\,\prime}\!\!}{2}}\,\right)^s\, \left(F^{\,2}\right)_{\mu\nu} \,\phi^{\,\m\n\ldots\,}p_{\,12}^{\,s-2}\ ,\label{onshell}
\end{equation}
that only involves the spin--1 curvatures
\begin{equation}
F_2^{\,\m\n}\,=\,-\,i\, \left( p_{\,2}^{\,\m}\, A_{\,2}^\n-p_{\,2}^{\,\n}\, A_{\,2}^\m\right) \ .
\end{equation}
This expression has actually an additional virtue: it becomes nicely invariant \emph{on
shell} under spin--$s$ gauge transformations in the limit where the spin--$s$ mass
disappears, so that all scalar products of the three momenta vanish, in striking
correspondence with the results of Metsaev \cite{Metsaev:2007rn}, that also include two
groups of terms carrying along different powers of the momenta. This is true although our
starting point was not gauge invariant with respect to the spin--$s$ field, and can be
regarded as a manifestation of the fact that, in String Theory, even the 1--1--$s$
couplings draw their origin somehow from conserved currents, exactly as was the case for
the 0--0--$s$ couplings of eq.~\eqref{scalar}. Notice that the first term in
\eqref{onshell} is a higher--derivative coupling proportional to the product of the
curvatures of the two spin--$1$ fields, while the second involves the portion of the
spin--1 energy--momentum tensor that can couple to the traceless $\phi_{\,\m_1\ldots
\m_s}$.

For $s\,=\,2$ one can actually go further, associating to eq.~\eqref{onshell}
the fully covariant form
\begin{equation}
\cL\,=\,\frac{\a^{\,\prime}\!\!}{8} \, \sqrt{g} \, F_{\m\n} F_{\r\s} R^{\,\m\n\r\s}\, -\, \frac{1}{4} \, \sqrt{g} \, g^{\,\m\r}g^{\,\n\s}F_{\m\n}F_{\r\s}\ ,
\end{equation}
while in general one can write for the two types of terms the current generating functions
\begin{equation}
j_{\,1}^{\,\pm}(\xi)\,=\,-\, \a^{\,\prime}\exp\left(\pm\sqrt{\frac{\a^{\,\prime}\!\!}{2}}\ p_{\,12}\cdot \xi\right)F_{\m\n}F^{\m\n}\label{J10}
\end{equation}
and
\begin{equation}
j_{\,2}^{\, \pm}(\xi)\,=\,-\,4\,\partial_\l\left(\partial_\chi\cdot\partial_\chi\right)\ \exp\left(\pm\,(1+\l)\,
\sqrt{\frac{\a^{\,\prime}\!\!}{2}}\ [\,(p_{\,12})_{\,\m}\,+\,i\,F_{\m\a}\,\chi^\a\,]\,\xi^{\m}\right)\Bigg|_{\chi\,=\,\l\,=\,0}\ ,\label{J20}
\end{equation}
where the role of the variable $\l$ is merely to produce an overall factor proportional to $s$, as in eq.~\eqref{onshell}. The coupling to the energy--momentum tensor that is present in \eqref{onshell}
is the one discussed by Berends, Burgers and van Dam \cite{bbvd}. The emergence, in the massless limit, of additional gauge invariant
structures beyond the ones that enter somewhat accidentally the 0--0--$s$ case,
resonates with the long--held suspicion that masses in String Theory draw their
origin from some form of spontaneous breaking.

A closer look at this simple example can shed some light on the structure that
is manifesting itself in the massless limit. We can actually use it to simulate
some situations that are to be faced in the general case, and to this end let
us give momentarily identical non--vanishing masses to the pair of spin--1
fields that are present in eq.~\eqref{onshell}, letting
\be
-\, p_{\,1}^{\,2}\, =\, -\, p_{\,2}^{\,2}\, = \, \frac{1}{\a^{\,\prime}}\ (n-1)\ ,
\ee
before expanding the spin--1 curvatures. This leads to the massive amplitude
\begin{equation}
\begin{split}
\cA^{\,\pm}_{\,1-1-s}\,&=\,\left(\!\pm\sqrt{\frac{\a^{\,\prime}\!\!}{2}}\,\right)^{s-2}\,s(s-2n+1)\,A_{1\,\m}\,A_{2\,\n} \,\phi^{\,\m\n\ldots\,}\cdot p_{\,12}^{\,s-2}\\
&+\,\left(\!\pm\sqrt{\frac{\a^{\,\prime}\!\!}{2}}\,\right)^s \Big[(2n-1)\,A_1\cdot A_2\ \phi\cdot p_{\,12}^{\,s}+s\,A_{1}\cdot p_{\,23}\,A_{2\,\n}\,\phi^{\,\n\ldots\,}\cdot p_{\,12}^{\,s-1}\\&\qquad\qquad\qquad\
+\,s\,A_{2\,\m}\cdot p_{\,31}\, A_{1\,\n}\,\phi^{\,\n\ldots\,}\cdot p_{\,12}^{\,s-1}\Big]\\&+\, \left(\!\pm\sqrt{\frac{\a^{\,\prime}\!\!}{2}}\,\right)^{s+2}A_{1}\cdot p_{\,23}\,A_{2}\cdot p_{\,31}\, \phi\cdot p_{\,12}^{\,s}\ ,\label{mass}
\end{split}
\end{equation}
that is still defined for momentum--space fields satisfying the transversality
constraints
\be p_{\,i}\cdot\phi_{\,i}(p_{\,i})\,=\,0\ . \label{diverg} \ee
Of course, now eq.~\eqref{mass} need not be gauge invariant, since it involves
massive excitations, but more importantly this way of presenting it is by no
means unique, once terms of the type \eqref{diverg} are ignored. Various choices obtained from one another making use of
transversality and momentum conservation, such as
\be
A_{1}\cdot p_{\,23}=2\, A_{1}\cdot p_{\,2}=-2\, A_{1}\cdot p_{\,3}\ , \label{ambiguity}
\ee
behave differently when one tries to perform gauge variations.
In this case there is actually a unique choice for the amplitude,
\begin{equation}
\begin{split}
\cA^{\,\pm}_{\,1-1-s}\,&=\,\left(\!\pm\sqrt{\frac{\a^{\,\prime}\!\!}{2}}\,\right)^{s-2}\,2\a^{\,\prime} s\, p_{\,1}\cdot p_{\,2}\,A_{1\,\m}A_{2\,\n}\, \phi^{\,\m\n\ldots\,}\cdot p_{\,12}^{\,s-2}\\
&+\,\left(\!\pm\sqrt{\frac{\a^{\,\prime}\!\!}{2}}\,\right)^s \Big[\left(2\a^{\,\prime}\,p_{\,1}\cdot p_{\,2}+s\right)A_{1}\cdot A_{2}\, \phi\cdot p_{\,12}^{\,s}+s\,A_{1}\cdot p_{\,23}\,A_{2\,\n}\,\phi^{\,\n\ldots\,}\cdot p_{\,12}^{\,s-1}\\&\qquad\qquad\qquad +s\,A_{2}\cdot p_{\,31}\, A_{1\,\n}\,\phi^{\,\n\ldots\,}\cdot p_{\,12}^{\,s-1}\Big]\\ &+\,\left(\!\pm\sqrt{\frac{\a^{\,\prime}\!\!}{2}}\,\right)^{s+2}A_{1}\cdot p_{\,23}\,A_{2}\cdot p_{\,31}\, \phi\cdot p_{\,12}^{\,s}\ , \label{11ssimple}
\end{split}
\end{equation}
where any explicit reference to the masses disappears, leaving way to scalar
products of the external momenta. Moreover, gauge variations give rise to
additional scalar products of the external momenta, which vanish kinematically
in the massless limit. This 1--1--$s$ example is actually a bit too simple, since
the amplitude \eqref{11ssimple} is invariant under gauge transformations of the
two spin--1 fields, despite their masses: after all, it is merely a proper
translation of our starting point, eq.~\eqref{onshell}. On the other hand, the
spin--$s$ gauge symmetry is violated, but precisely by terms involving scalar
products of the external momenta, that as we have stressed vanish for purely
kinematic reasons in the massless limit. The general lesson is that one can
present both this and other amplitudes involving massive states in such a way
that masses only enter via scalar products of the external momenta, which
vanish kinematically when, in the massless limit, the gauge symmetry is
restored.

It is interesting to exhibit what type of pattern accompanies the gauge
invariant results that emerge in the massless limit. Up to an overall
coefficient, the resulting tensorial structure can be ascribed to the action of
the differential operator
\be
\cG\,=\,\left(\sqrt{\frac{\a^{\,\prime}\!\!}{2}}\,\right)\Big[(\partial_{\xi_{\,1}}\cdot\partial_{\xi_{\,2}})(\partial_{\xi_{\,3}}\cdot p_{\,12})\,+\,(\partial_{\xi_{\,2}}\cdot\partial_{\xi_{\,3}})(\partial_{\xi_{\,1}}\cdot p_{\,23})\,+\,(\partial_{\xi_{\,3}}\cdot\partial_{\xi_{\,1}})(\partial_{\xi_{\,2}}\cdot
p_{\,31})\Big]\ ,\label{G}
\ee
acting on the symbols. This is actually the only operator containing both
traces and divergences that in the massless limit satisfies on--shell the commutation relations
\be
\left[\,\cG\,,\,\xi_{\,i}\,\cdot\,p_{\,i}\,\right]\,=\,0\ ,
\ee
and in terms of $\cG$ the amplitude takes the form
\be
\cA_{\,1-1-s}^{[0]\,\pm}\,=\,\left[1\,\pm\,\cG\,\right]\,A_{1}\cdot \left(\xi_{\,1}\pm\sqrt{\frac{\a^{\,\prime}\!\!}{2}}\,p_{\,23}\right) A_{2}\cdot\left(\xi_{\,2}\pm\sqrt{\frac{\a^{\,\prime}\!\!}{2}}\,p_{\,31}\right)
\phi_{\,3}\cdot\left(\xi_{\,3}\pm\sqrt{\frac{\a^{\,\prime}\!\!}{2}}\,p_{\,12}\right)^s\Bigg|_{\xi_{\,i}=0} ,
\ee
where the notation, that we shall use repeatedly in the following, is rather compact but
nonetheless suffices to indicate that all vector indices carried by the fields are
contracted with symbols shifted by momentum dependent terms as above. Moreover, the
superscript $[0]$ in $\cA_{\,1-1-s}^{[0]\,\pm}$ is meant to stress that all terms
involving scalar products of momenta have been eliminated, so that one is focussing on a
portion of a massless amplitude that suffices to determine it completely. Notice that
$\cA_{\,1-1-s}^{[0]\,\pm}$ compute precisely the terms in eq.~\eqref{11ssimple} not
involving $p_{\,1}\cdot p_{\,2}$.

We can now describe briefly how it is possible to translate these findings in field theory
language. To begin with, as we have stated in the Introduction, vertices that
are gauge invariant up to the mass--shell condition indicate that linearized
gauge transformations must generally be deformed. Two more steps would then be
needed in order to recover complete off--shell couplings. As in
\cite{Manvelyan}, one should first reinstate terms proportional to traces and
de Donder conditions, which were ignored so far in compliance with the Virasoro
conditions, but these are uniquely fixed by the structure of $\cG$, up to
partial integrations. The result would be a cubic vertex in Fronsdal's
constrained formulation \cite{Fronsdal}, but taking into account the
compensators $\a$ one would recover in a similar fashion the unconstrained
gauge symmetry of the given HS fields in the minimal setting of \cite{minimal} or even, with more additional fields, the unconstrained forms of Pashnev, Tsulaia, Buchbinder and others \cite{Buchbinder:2001bs,fotopoulos}.

In the following we shall apply similar arguments to exhibit the types of
structures that emerge from more general string amplitudes in the massless
limit. To begin with, we shall confine our attention to terms not involving traces or de Donder conditions, and then in the Appendix we shall return to the issue of completing them into fully gauge invariant cubic couplings.

\vskip 24pt


\scss{2--2--$s$ Couplings}\label{sec:22s}


We can now turn to the spin--$s$ current constructed from a pair of spin--$2$
fields that is embodied in eq.~\eqref{Apiumeno}. Collecting terms that contain
identical numbers of momenta, the result can be cast in the form
{\allowdisplaybreaks
\begin{eqnarray}
\cA_{\,2-2-s}^{\,\pm}&=&\left(\!\pm\,\sqrt{\frac{\a^{\,\prime}\!\!}{2}}\,\, \right)^{s-4}\,s(s-1)(s-2)(s-3)\, h_{\,1\,\a_1\a_2}\,h_{\,2\,\b_1\b_2}\,\phi^{\,\a_1\a_2\b_1\b_2\,\ldots\,}\cdot p_{\,12}^{\,s-4}\nonumber \\
&+&\left(\!\pm\,\sqrt{\frac{\a^{\,\prime}\!\!}{2}}\,\right)^{s-2}\left[\,4s(s-1) \,h_{\,1\,\a_1\a_2}\,{h_{\,2}^{\,\a_1}}_{\b_2}\,\phi^{\,\a_2\b_2\,\ldots\,}\cdot p_{\,12}^{\,s-2}\nonumber \right.\\
\phantom{\Big(\sqrt{\frac{\a^{\,\prime}\!\!}{2}}\Big)^{s}}&&\qquad\qquad\qquad+\,2s(s-1)(s-2)\,h_{\,1\,\a_1\a_2}\,{h_{\,2\,\b_2}}\cdot p_{\,31}\,\phi^{\,\a_1\a_2\b_2\,\ldots\,}\cdot p_{\,12}^{\,s-3}\nonumber\\
\phantom{\Big(\sqrt{\frac{\a^{\,\prime}\!\!}{2}}\Big)^{s}} &&\qquad\qquad\qquad+\left. 2s(s-1)(s-2)\,{h_{\,1\,\a_2}}\cdot p_{\,23}\,h_{\,2\,\b_1\b_2}\,\phi^{\,\a_2\b_1\b_2\,\ldots\,}\cdot p_{\,12}^{\,s-3}\right]\nonumber\\
&+&\left(\!\pm\,\sqrt{\frac{\a^{\,\prime}\!\!}{2}}\,\right)^{s\phantom{+2}} \left[\,2h_{\,1\,\a_1\a_2}\,h_2^{\,\a_1\a_2}\,\phi\cdot p_{\,12}^{\,s\vphantom{2\b}}+4s\,\left(\,h_{\,1\,\a_1\a_2}\,{h_2^{\,\a_2}}\cdot p_{\,31}\,\phi^{\,\a_1\,\ldots\,\vphantom{\b}}\cdot p_{\,12}^{\,s-1}\nonumber\right.\right.\\
\phantom{\Big(\sqrt{\frac{\a^{\,\prime}\!\!}{2}}\Big)}&&\qquad\qquad\qquad\left.+\,h_{\,1\,\a_1}\cdot p_{\,23}\, {h_2^{\,\a_1}}_{\b_2}\,\phi^{\,\b_2\,\ldots\,}\cdot p_{\,12}^{\,s-1}\right)\nonumber\\
\phantom{\Big(\sqrt{\frac{\a^{\,\prime}\!\!}{2}}\Big)}&&\qquad\qquad\qquad+s(s-1)\,\left(\,h_{\,1\,\a_1\a_2}\,h_{\,2}\cdot p_{\,31}^{\,2}\,\phi^{\,\a_1\a_2\,\ldots\,\vphantom{\b}}\cdot p_{\,12}^{\,s-2} \nonumber \right.\\
\phantom{\Big(\sqrt{\frac{\a^{\,\prime}\!\!}{2}}\Big)}&&\qquad\qquad\qquad+h_{\,1}\cdot p_{\,23}^{\,2}\,h_{\,2\,\b_1\b_2}\,\phi^{\,\b_1\b_2\,\ldots\,}\cdot p_{\,12}^{\,s-2} \nonumber\\
\phantom{\Big(\sqrt{\frac{\a^{\,\prime}\!\!}{2}}\Big)}&&\qquad\qquad\qquad +\left.\left. 4\,h_{\,1\,\a_1}\cdot p_{\,23}\, h_{\,2\,\b_1}\,\cdot p_{\,31}\, \phi^{\,\a_1\b_1\,\ldots\,}\cdot p_{\,12}^{\,s-2}\right)\right]\nonumber\\
&+&\left(\!\pm\,\sqrt{\frac{\a^{\,\prime}\!\!}{2}}\,\right)^{s+2}\left[\,2s\,h_{\,1\,\a_1}\cdot p_{\,23}\,h_{\,2}\cdot p_{\,31}^{\,2}\,\phi^{\,\a_1\,\ldots\,\vphantom{\b}}\cdot p_{\,12}^{\,s-1}\right.\nonumber\\\phantom{\Big(\sqrt{\frac{\a^{\,\prime}\!\!}{2}}\Big)}&&\qquad\qquad\qquad+\left.2s\,h_{\,1}\cdot p_{\,23}^{\,2}\,h_{\,2\,\b_1}\cdot p_{\,31}\,\phi^{\,\b_1\ldots\,}\cdot p_{\,12}^{\,s-1}+4h_{\,1\,\a_1}\cdot p_{\,23}\, h_2^{\,\a_1}\cdot p_{\,31}\, \phi\cdot p_{\,12}^{\,s}\right]\nonumber\\
&+&\left(\!\pm\,\sqrt{\frac{\a^{\,\prime}\!\!}{2}}\,\right)^{s+4}h_{\,1}\cdot p_{\,23}^{\,2}\,h_{\,2}\cdot p_{\,31}^{\,2}\,\phi\cdot p_{\,12}^{\,s}\ ,  \label{22s}
\end{eqnarray}}
\!\!where $h_{\,1}$, $h_{\,2}$ and $\phi$ are the two spin--$2$ and the
spin--$s$ momentum--space fields. The special form chosen for eq.~\eqref{22s}
guarantees, once more, that all terms where the gauge parameter is contracted
solely with momenta disappear from the gauge variation. In this fashion, in the
\emph{massless limit} one recovers the gauge invariant couplings
{\allowdisplaybreaks
\begin{eqnarray}
\cA^{\,[0]\, \pm}_{\, 2-2-s}&=& 2s(s-1)\,\left(\!\pm\,\sqrt{\frac{\a^{\,\prime}\!\!}{2}}\,\right)^{s}\left[ \,h_{\,1\,\a_1\a_2}\,h_{\,2}^{\,\a_1\a_2}\,\phi\cdot p_{\,12}^{\,s}+\ 2\,\left(\,h_{\,1\,\a_1\a_2}\,{h_2^{\,\a_2}}\cdot p_{\,31}\,\phi^{\,\a_1\,\ldots\,}\cdot p_{\,12}^{\,s-1}\vphantom{h_{\,1\,\a_1}\cdot p_{\,23}\, {h_2^{\,\a_1}}_{\b_2}\,\phi^{\,\b_2\,\ldots\,}\cdot p_{\,12}^{\,s-1}}\right.\right.\nonumber\\
&&\qquad\qquad\qquad\qquad\quad\ +\left.h_{\,1\,\a_1}\cdot p_{\,23}\, {h_2^{\,\a_1}}_{\b_2}\,\phi^{\,\b_2\,\ldots\,}\cdot p_{\,12}^{\,s-1}\right)\nonumber\\
&&\qquad\qquad\qquad\qquad\quad\ +\left(\,h_{\,1\,\a_1\a_2}\,h_{\,2}\cdot p_{\,31}^{\,2}\,\phi^{\,\a_1\a_2\,\ldots\,}\cdot p_{\,12}^{\,s-2}\vphantom{h_{\,1\,\a_1}\cdot p_{\,23}\, {h_2^{\,\a_1}}_{\b_2}\,\phi^{\,\b_2\,\ldots\,}\cdot p_{\,12}^{\,s-1}}\vphantom{h_{\,1\,\a_1}\cdot p_{\,23}\, {h_2^{\,\a_1}}_{\b_2}\,\phi^{\,\b_2\,\ldots\,}\cdot p_{\,12}^{\,s-1}}\right.\nonumber\\
&&\qquad\qquad\qquad\qquad\quad\ +\,h_{\,1}\cdot p_{\,23}^{\,2}\,h_{\,2\,\b_1\b_2}\,\phi^{\,\b_1\b_2\,\ldots\,}\cdot p_{\,12}^{\,s-2}\nonumber\\
&&\qquad\qquad\qquad\qquad\quad\ +\left.\left. 2\,h_{\,1\,\a_1}\cdot p_{\,23}\, h_{\,2\,\b_1}\cdot p_{\,31}\, \phi^{\,\a_1\b_1\,\ldots\,}\cdot p_{\,12}^{\,s-2}\vphantom{h_{\,1\,\a_1}\cdot p_{\,23}\, {h_2^{\,\a_1}}_{\b_2}\,\phi^{\,\b_2\,\ldots\,}\cdot p_{\,12}^{\,s-1}}\right)\right]\nonumber\\
&+&\!\!\!\!4s\left(\!\pm\,\sqrt{\frac{\a^{\,\prime}\!\!}{2}}\,\right)^{s+2}\!\!\!\!\!\left[\vphantom{h_{\,1\,\a_1}\cdot p_{\,23}\, {h_2^{\,\a_1}}_{\b_2}\,\phi^{\,\b_2\,\ldots\,}\cdot p_{\,12}^{\,s-1}}\,h_{\,1\,\a_1}\cdot p_{\,23}\,h_{\,2}\cdot p_{\,31}^{\,2}\,\phi^{\,\a_1\,\ldots\,}\cdot p_{\,12}^{\,s-1}\,+\, h_{\,1}\cdot p_{\,23}^{\,2}\,h_{\,2\,\b_1}\cdot p_{\,31}\,\phi^{\,\b_1\,\ldots\,}\cdot p_{\,12}^{\,s-1}\right.\nonumber\\
&&\qquad\qquad\quad\ \ +\left.\,h_{\,1\,\a_1}\cdot p_{\,23}\, h_2^{\,\a_1}\cdot p_{\,31}\, \phi\cdot p_{\,12}^{\,s}\vphantom{h_{\,1\,\a_1}\cdot p_{\,23}\, {h_2^{\,\a_1}}_{\b_2}\,\phi^{\,\b_2\,\ldots\,}\cdot p_{\,12}^{\,s-1}}\right]\nonumber\\
&+&\phantom{\a}\left(\!\pm\,\sqrt{\frac{\a^{\,\prime}\!\!}{2}}\,\right)^{s+4}h_{\,1}\cdot p_{\,23}^{2}\,h_{\,2}\cdot p_{\,31}^2\,\phi\cdot p_{\,12}^{s}\ , \label{22s_massless}
\end{eqnarray}}
\!\!\!Three types of contributions thus survive, with $s$, $s+2$ and $s+4$ momenta.
The first of these deforms the abelian gauge symmetry, since it rests on a
current that is only conserved on shell, while the others involve currents that
are \emph{identically} conserved and therefore do not deform the gauge algebra,
in agreement with Metsaev's results \cite{Metsaev:2007rn}. The actual massive
couplings include of course additional contributions, which can be regarded as
``decorations'' of these: they are subleading in the external momenta and
disappear in the massless limit.

As in the previous case, it is interesting to stress the type of pattern that
is emerging for the gauge invariant results that obtain in the massless limit.
The massless 2--2--$s$ coupling can be nicely expressed in the form
\begin{multline}
\cA^{\,[0]\, \pm}_{\, 2-2-s}\,=\,\left[1\,\pm\ \cG\,+\,\frac{1}{2}\ \cG^{\,2}\right]\\
\times\,h_{\,1}\cdot\left(\xi_{\,1}\,\pm\,\sqrt{\frac{\a^{\,\prime}\!\!}{2}}\,p_{\,23}\right)^2\,
h_{\,2}\cdot\left(\xi_{\,2}\,\pm\,\sqrt{\frac{\a^{\,\prime}\!\!}{2}}\,p_{\,31}\right)^2\,
\phi_{\,3}\cdot\left(\xi_{\,3}\,\pm\,\sqrt{\frac{\a^{\,\prime}\!\!}{2}}\,p_{\,12}\right)^s
\,\Bigg|_{\xi_{\,i}\,=\,0}\
,
\end{multline}
so that the three types of contributions involving different powers of the external momenta arise as iterations induced by the same operator $\cG$ that was presented in eq.~\eqref{G}.

\vskip 24pt


\scss{1--$s$--$s$ Couplings}\label{sec:1ss}


The 1--$s$--$s$ couplings are particularly interesting, since they codify the
electromagnetic interactions of spin--$s$ fields. As in the previous examples,
the starting point is eq.~\eqref{expl}, that leads to the expression
\begin{equation}
\begin{split}
\cA&^{\,\pm}_{\,1-s-s}\,=\,\\ \,=\,&\sum_{k\,=\,0}^{s-1}\left(\!\pm\sqrt{\frac{\a^{\,\prime}\!\!}{2}}\,\right)^{2s-2k-1}\!\!\!\!\!\!\!\!\!\!\!\!\!\!\!\!\!\frac{(s!)^{\,2}}{k!\,[(s-k-1)!\,]^{\,2}\!\!}\, \left[\frac{1}{(k+1)}\  p_{\,23}^{\,s-k-1}\cdot \phi_{\,1\,\a_{s-k}\ldots\,\a_s}\,{\phi_{\,3}^{\,\a_{s-k}\ldots\,\a_{s}}}\cdot p_{\,12}^{\,s-k-1}\,A_2\cdot p_{\,31}\right.\\
&\qquad\qquad\qquad\qquad\qquad\qquad\qquad\ +\left.\frac{2}{s-k}\ p_{\,23}^{\,s-k-1}\cdot \phi_{\,1\,\a_{s-k}\ldots\,\a_{s-1}\m}\,{\phi_{\,3}^{\,\a_{s-k}\ldots\,\a_{s-1}}}_{\n}\cdot p_{\,12}^{\,s-k-1}\ i\, F_2^{\,\m\n}\right]\\
+&\quad\ \ \,\left(\!\pm\sqrt{\frac{\a^{\,\prime}\!\!}{2}}\,\right)^{2s+1}p_{\,23}^{\,s}\cdot \phi_{\,1}\, \phi_{\,3}\cdot p_{\,12}^{\,s}\, A\cdot p_{\,31}\ ,\label{1ss}
\end{split}
\end{equation}
where $\phi_{\,1}$ and $\phi_{\,2}$ are totally symmetric spin--$s$ irreducible
momentum--space fields, $A_\m$ is the spin--1 momentum--space vector and we have grouped
together terms containing identical numbers of momenta. Again,
\begin{equation}
F_2^{\,\m\n}\,=\,-\,i\, \left( p_{\,2}^{\,\m}\, A_{\,2}^\n-p_{\,2}^{\,\n}\, A_{\,2}^\m\right) \ ,
\end{equation}
is a spin--1 field strength, defined for convenience up to a factor $i$. From
\eqref{1ss} one can extract, in particular, the 1--3--3 coupling, that in our
notation and up to an overall factor, reads
{\allowdisplaybreaks
\begin{eqnarray}
\cA^{\,\pm}_{\,1-3-3}&=&\left(\!\pm\sqrt{\frac{\a^{\,\prime}\!\!}{2}}\,\right) \ \left[i\, \phi_{\,1\,\a_1\a_2\m}\,{\phi_{\,3}^{\,\a_1\a_2}}_\n\ F_2^{\m\n}+\frac{1}{6}\ \phi_{\,1}\cdot \phi_{\,3}\  A_2\cdot p_{\,31}\right]\nonumber\\
&+&\left(\!\pm\sqrt{\frac{\a^{\,\prime}\!\!}{2}}\,\right)^3\left[i\, p_{\,23}\cdot \phi_{\,1\,\a_1\m}\,{\phi_{\,3}^{\,\a_1}}_\n\cdot p_{\,12}\ F_2^{\,\m\n}+\frac{1}{2}\ p_{\,23}\cdot \phi_{\,1\,\a_1\a_2}\, \phi_{\,3}^{\,\a_1\a_2}\cdot p_{\,12}\  A_2\cdot p_{\,31}\right]\nonumber\\
&+&\left(\!\pm\sqrt{\frac{\a^{\,\prime}\!\!}{2}}\,\right)^5\left[\frac{i}{6}\ p_{\,23}^{\,2}\cdot \phi_{\,1\,\m}\,\phi_{\,3\,n}\cdot p_{\,12}^{\,2}\ F_2^{\,\m\n}+\frac{1}{4}\ p_{\,23}^{\,2}\cdot \phi_{\,1\,\a_1}\, \phi_{\,3}^{\,\a_1}\cdot p_{\,12}^{\,2}\  A_2\cdot p_{\,31}\right]\nonumber\\
&+&\left(\!\pm\sqrt{\frac{\a^{\,\prime}\!\!}{2}}\,\right)^7\left[\frac{1}{36}\ p_{\,23}^{\,3}\cdot \phi_{\,1}\, \phi_{\,3}\cdot p_{\,12}^{\,3}\  A_2\cdot p_{\,31}\right]\ .\label{133}
\end{eqnarray}}
\!\!\!The relatively simple massive amplitude \eqref{1ss} thus involves a tower of
contributions, down to a term carrying no momenta. This is to be confronted
with the massless scaling limit of \cite{Boulanger:2008tg}, that on the
contrary contains only the terms with $2s-1$ momenta. The spin--$s$ modes are
massive in String Theory, and therefore all these amplitudes need only be
invariant under spin--$1$ gauge transformations. However, the choice made in
writing eq.~\eqref{1ss} guarantees again that, when attempting to perform
spin--$s$ gauge variations, all terms where the gauge parameter is contracted
solely with momenta cancel nonetheless, so that only scalar products of the
momenta, that vanish in the massless limit, are left over. This suggests that
off--shell couplings somehow invariant under spin--$s$ gauge transformations
underlie even these 1--$s$--$s$ interactions, and two types of off--shell couplings
dominate the massless limit. The first involves the highest number of momenta,
was identified in \cite{Gross} and is discussed in detail in the last paper in
\cite{west}. The next term, with $2s-1$ momenta,
is however more interesting since it deforms the abelian gauge symmetry.

These findings are nicely consistent with previous results of Metsaev
\cite{Metsaev:2007rn} and Boulanger, Leclercq and Sundell
\cite{Boulanger:2008tg}, that describe effectively a spin--$1$ current built
from a pair of massless spin--$s$ fields and involving $2s-1$ derivatives. This
current clearly lies behind the amplitude \eqref{133}, up to integrations by
parts, and in order to identify it explicitly one can recast eq.~\eqref{1ss} in
terms of the spin--1 field strength $F_{\m\n}$ and generalized de Wit--Freedman connections \cite{de Wit:1979pe}. The end result is
\begin{multline}
\cL^{(3)}\,=\,f_{[ab]}F^{\r\s}\left(\Gamma^{(s-1)a}_{\m_1\ldots\,\m_{s-1},\n_1\ldots\,\n_{s-1}\r}\, \Gamma^{(s-1)b}_{\m_1\ldots\,\m_{s-1},\n_1\ldots\,\n_{s-1}\sigma}\right.\\\left.-\, \Gamma^{(s-1)a}_{\m_1\ldots\,\r,\m_{s-1}\n_1\ldots\,\n_{s-1}}\,\Gamma^{(s-1)b}_{\m_1\ldots\, \s,\m_{s-1}\n_1\ldots\,\n_{s-1}}\right)\ ,
\end{multline}
and indeed this dipole coupling, after integrating by parts the derivative
acting on the spin--$1$ field, defines a current
\begin{equation}
J_\s\,=\,f_{[ab]}\partial_\r\left(\Gamma^{(s-1)a}_{\m_1\ldots\,\m_{s-1},\n_1\ldots\,\n_{s-1}\r}\, \Gamma^{(s-1)b}_{\m_1\ldots\,\m_{s-1},\n_1\ldots\,\n_{s-1}\sigma}\,- \,\Gamma^{(s-1)a}_{\m_1\ldots\,\r,\m_{s-1}\n_1\ldots\,\n_{s-1}}\,\Gamma^{(s-1)b}_{\m_1\ldots\, \s,\m_{s-1}\n_1\ldots\,\n_{s-1}}\right)\ ,
\end{equation}
that has clearly the form of an improvement and is thus trivially conserved. On the other
hand, the very same coupling generates a spin--$s$ current built from a spin--$2$ and a
spin--$s$ field that is only conserved on shell, and thus deforms the abelian spin--$s$
gauge transformation. This fact resonates with a number of no--go theorems that spell
trouble for minimal couplings in flat space for massless HS fields, and indeed this type
of multipolar couplings arise quite naturally in HS field theories \cite{Metsaev:2007rn}.
Before moving to the next example, it is interesting to stress that even in this case the
gauge invariant version of these 1--$s$--$s$ couplings can be also expressed in terms of
the operator $\cG$ defined in \eqref{G}:
\be
\cA_{\,1-s-s}^{\,[0]\, \pm}\,=\,\left[1\,\pm\,\cG\,\right]\,A_{1}\cdot\left(\xi_{\,1}
\pm\sqrt{\frac{\a^{\,\prime}\!\!}{2}}\,p_{\,23}\right) \phi_{\,2}\cdot\left(\xi_{\,2}\pm\sqrt{\frac{\a^{\,\prime}\!\!}{2}}\,p_{\,31}\right)^s
\phi_{\,3}\cdot\left(\xi_{\,3}+\sqrt{\frac{\a^{\,\prime}\!\!}{2}}\,p_{\,12}\right)^s \ .
\ee

\vskip 24pt


\scss{2--$s$--$s$ Couplings}\label{sec:2ss}


Let us now turn to the 2--$s$--$s$ amplitudes, that codify the spin--2 couplings
of a pair of spin--$s$ fields. The starting point is again eq.~\eqref{expl},
that in this case leads to
{\allowdisplaybreaks
\begin{eqnarray}
\cA^{\,\pm}_{\,2-s-s}&=&\vphantom{\left({\frac{\a^{\,\prime}\!\!}{2}}\right)^{s-k-1}} 2s\, s!\ \phi_{\,1\,\a_{1}\ldots\,\a_{s-1}\m_1}\,{\phi_{\,3}^{\,\a_{1}\ldots\,\a_{s-1}}}_{\n_1}\,h_2^{\m_1\n_1}\nonumber\\
&+&\sum_{k\,=\,0}^{s-2}\left({\frac{\a^{\,\prime}\!}{2}}\,\right)^{s-k-1}\frac{(s!)^2}{k!(s-k-2)!^2} \nonumber\\
&&\times\,\vphantom{\left({\frac{\a^{\,\prime}\!}{2}}\right)^{s-k-1}} \left\{\vphantom{\left({\frac{\a^{\,\prime}\!}{2}}\right)^{s-k-1}}\frac{1}{(k+1)(k+2)}\ p_{\,23}^{\,s-k-2}\cdot \phi_{\,1\,\a_{s-k-1}\ldots\,\a_{s}}\,{\phi_{\,3}^{\,\a_{s-k-1}\ldots\,\a_{s}}}\cdot p_{\,12}^{\,s-k-2}\,p_{\,31}^{\,2}\cdot h_2\right.\nonumber\\
&&\vphantom{\left(\!\pm\sqrt{\frac{\a^{\,\prime}\!}{2}}\,\right)^{2s-2k-2}}\ \ \ +\frac{2}{(k+1)(s-k-1)}\ \left(p_{\,31}\cdot h_2^{\,\n_1}\,p_{\,23}^{\,\m_1}+p_{\,31}\cdot h_2^{\,\m_1}\,p_{\,12}^{\,\n_1}\right) \nonumber\\
&&\vphantom{\left(\!\pm\sqrt{\frac{\a^{\,\prime}\!}{2}}\,\right)^{2s-2k-2}}\phantom{aaaaaaaaaaaaaa} \times\,p_{\,23}^{\,s-k-2}\cdot \phi_{\,1\,\a_{s-k-1}\ldots\,\a_{s-1}\m_1}\,{\phi_{\,3}^{\,\a_{s-k-1}\ldots\,\a_{s-1}}}_{\n_1}\cdot p_{\,12}^{\,s-k-2}\nonumber\\
&&\vphantom{\left(\!\pm\sqrt{\frac{\a^{\,\prime}\!}{2}}\,\right)^{2s-2k-2}}\ \ \ +\frac{1}{s-k-1}\  \left[\,\frac{1}{s-k}\left(h_2^{\n_1\n_2}\,p_{\,23}^{\,\m_1}\,p_{\,23}^{\,\m_2} +h_2^{\m_1\m_2}\,p_{\,12}^{\,\n_1}\,p_{\,12}^{\,\n_2}\right)+ \frac{2}{s-k-1}\ h_2^{\m_1\n_1}\,p_{\,23}^{\,\m_2}\,p_{\,12}^{\,\n_2}\right]\nonumber\\
&&\vphantom{\left(\!\pm\sqrt{\frac{\a^{\,\prime}\!}{2}}\,\right)^{2s-2k-2}}\left.\phantom{aaaaaaaaaaaaaaaaaa}\times p_{\,23}^{\,s-k-2}\cdot \phi_{\,1\,\a_{s-k-1}\ldots\,\a_{s-2}\m_1\m_2}\,{\phi_{\,3}^{\,\a_{s-k-1}\ldots\,\a_{s-2}}}_{\n_1\n_2}\cdot p_{\,12}^{\,s-k-2}\vphantom{\left(\!\pm\sqrt{\frac{\a^{\,\prime}\!}{2}}\,\right)^{2s-2k-2}}\right\}\nonumber\\
&+&\left({\frac{\a^{\,\prime}\!}{2}}\,\right)^{s} \Big[\ s^2\,p_{\,23}^{\,s-1}\cdot \phi_{\,1\,\a_s}\,\phi_{\,3}^{\,\a_s}\cdot p_{\,12}^{\,s-1}\, p_{\,31}^{\,2}\cdot h_2\nonumber\\
&&\vphantom{\left(\!\pm\sqrt{\frac{\a^{\,\prime}\!}{2}}\,\right)^{2s-2k-2}}\qquad\ \ + 2s\ p_{\,23}^{\,s-1}\cdot \phi_{\,1\,\m_1}\,\phi_{\,3\,\n_1}\cdot p_{\,12}^{\,s-1}\,\left(p_{\,31}\cdot h_2^{\,\n_1}\,p_{\,23}^{\,\m_1}+ p_{\,31}\cdot h_2^{\,\m_1}\,p_{\,12}^{\,\n_1}\right)\Big]\nonumber\\
\vphantom{\left(\!\pm\sqrt{\frac{\a^{\,\prime}\!}{2}}\,\right)^{2s-2k-2}} &+&\left({\frac{\a^{\,\prime}\!}{2}}\,\right)^{s+1} p_{\,23}^{\,s}\cdot \phi_{\,1}\,\phi_{\,3}\cdot p_{\,12}^{\,s}\, p_{\,31}^{\,2}\cdot h_2\ ,\label{2ss}
\end{eqnarray}}
\!\!\!where we have grouped together terms with the same number of momenta and
$\phi_{\,1\,\m_1\ldots\,\m_s}$, $\phi_{\,3\,\n_1\ldots\,\n_s}$ and $h_{2\,
\m\n}$ are the two spin--$s$ symmetric and traceless momentum--space fields and
the spin--$2$ symmetric and traceless momentum--space field. It is instructive
to take a closer look at the 2--3--3 example, that was analyzed in detail in
\cite{Boulanger:2008tg} in the massless case. Up to an overall factor,
eq.~\eqref{2ss} becomes in this case
{\allowdisplaybreaks
\begin{eqnarray}
\cA^{\pm}_{\,2-3-3}&=&\vphantom{\left(\!\pm\sqrt{\frac{\a^{\,\prime}\!}{2}}\,\right)} \phi_{\,1\,\a_1\a_2\m_1}\,{\phi_{\,3}^{\,\a_1\a_2}}_{\n_1}\ h^{\,\m_1\n_1}\nonumber\\
&+&\vphantom{\left(\sqrt{\frac{\a^{\,\prime}\!}{2}}\right)} {\frac{\a^{\,\prime}\!}{2}}\ \ \left[\frac{1}{2}\ \phi_{\,1\,\a_1\m_1\m_2}\,{\phi_{\,3}^{\,\a_1}}_{\,\n_1\n_2}\ \left(h^{\,\n_1\n_2}\,p_{\,23}^{\,\m_1}\,p_{\,23}^{\,\m_2}+h^{\,\m_1\m_2}\,p_{\,12}^{\,\n_1}\,p_{\,12}^{\,\n_2} \right)\right.\nonumber\\
&&\qquad+\vphantom{\left(\sqrt{\frac{\a^{\,\prime}\!}{2}}\right)}\phi_{\,1\,\a_1\a_2\m_1}\, {\phi_{\,3}^{\,\a_1\a_2}}_{\,\n_1}\ \left(p_{\,31}\cdot h^{\,\n_1}p_{\,23}^{\,\m_1}+p_{\,31}\cdot h^{\,\m_1}p_{\,12}^{\,\n_1}\right)+\frac{1}{6}\ \phi_{\,1}\cdot \phi_{\,3}\ p_{\,31}^2\cdot h\nonumber\\
&&\qquad+\vphantom{\left(\sqrt{\frac{\a^{\,\prime}\!}{2}}\right)}\left.2\ p_{\,23}\cdot \phi_{\,1\,\a_1\m_1}\,{\phi_{\,3}^{\,\a_1}}_{\,\n_1}\cdot p_{\,12}\,h^{\,\m_1\n_1}\right]\nonumber\\
&+&\!\!\!\!\! \left({\frac{\a^{\,\prime}\!}{2}}\right)^2 \left[\frac{1}{6}\ p_{\,23}\cdot \phi_{\,1\,\m_1\m_2}\ {\phi_{\,3\,\n_1\n_2}}\cdot p_{\,12}\ \left(h^{\,\n_1\n_2}\,p_{\,23}^{\,\m_1}\,p_{\,23}^{\,\m_2}+h^{\,\m_1\m_2}\,p_{\,12}^{\,\n_1}p_{\,12}^{\,\n_2} \right)\right.\nonumber\\
&&\qquad+\vphantom{\left(\sqrt{\frac{\a^{\,\prime}\!}{2}}\right)}p_{\,23}\cdot \phi_{\,1\,\a_1\m_1}\, {\phi_{\,3}^{\,\a_1}}_{\n_1}\cdot p_{\,12}\ \left(p_{\,31}\cdot h^{\,\n_1}\,p_{\,23}^{\,\m_1}+p_{\,31}\cdot h^{\,\m_1}\,p_{\,12}^{\,\n_1}\right)\nonumber\\
&&\qquad+\vphantom{\left(\sqrt{\frac{\a^{\,\prime}\!}{2}}\right)}\left.\frac{1}{2}\ p_{\,23}\cdot \phi_{\,1\,\a_1\a_2}\,\phi_{\,3}^{\,\a_1\a_2}\cdot p_{\,12}\ p_{\,31}^{\,2}\cdot h+\frac{1}{2}\ p_{\,23}^{\,2}\cdot \phi_{\,1\,\m_1}\,\phi_{\,3\,\n_1}\cdot p_{\,12}^{\,2}\ h^{\,\m_1\n_1}\right]\nonumber\\
&+&\!\!\!\!\! \left({\frac{\a^{\,\prime}\!}{2}}\right)^3 \left[\frac{1}{6}\ p_{\,23}^{\,2}\cdot \phi_{\,1\,\m_1}\,\phi_{\,3\,\n_1}\cdot p_{\,12}^{\,2}\ \left(p_{\,31}\cdot h^{\,\n_1}\,p_{\,23}^{\,\m_1}+p_{\,31}\cdot h^{\,\m_1}p_{\,12}^{\,\n_1}\right)\right.\nonumber\\
&&\qquad+\vphantom{\left(\sqrt{\frac{\a^{\,\prime}\!}{2}}\right)}\left.\frac{1}{4}\ p_{\,23}^{\,2}\cdot \phi_{\,1\,\a_1}\,\phi_{\,3}^{\,\a_1}\cdot p_{\,12}^{\,2}\ p_{\,31}^{\,2}\cdot h\right]\nonumber\\
&+&\!\!\!\!\! \left({\frac{\a^{\,\prime}\!}{2}}\right)^4 p_{\,23}^{\,3}\cdot \phi_{\,1}\ \phi_{\,3}\cdot p_{\,12}^{\,3}\ p_{\,31}^{\,2}\cdot h\ .\label{233}
\end{eqnarray}}
\!\!\!Eqs.~\eqref{2ss} and \eqref{233} reveal an interesting pattern for the
couplings of this type, with several groups of terms that are accompanied by
different numbers of momenta. A gauge symmetry is not present, but nonetheless
the terms where the gauge parameter is contracted solely with momenta are not
present in the gauge variation, as in all previous examples. The gauge
invariant terms that are left over in the limit of vanishing tension comply to
the general pattern of the previous examples: they involve $2s-2$ momenta, four
in the last example, and only three different types of tensor structures, that
are distinguished by the number of indices carried by the spin--$2$ field that
are contracted with spin--3 fields.

For the actual massive string spectrum, the types of off--shell terms giving rise to
these couplings include a chain of lower--derivative ``decorations''. But even in this
class of examples the pattern is consistent with the presence of a ``seed'' involving
$2s-2$ derivatives, in agreement with previous results of Metsaev \cite{Metsaev:2007rn}
and Boulanger, Leclercq and Sundell \cite{Boulanger:2008tg}. More in detail, the gauge
invariant off--shell coupling was shown to comprise \cite{Boulanger:2008tg} a coupling
with $2s+2$ derivatives, a second type of coupling with $2s$ derivatives that is only
present in space--time dimensions $d>4$, and a third type of coupling with $2s-2$
derivatives, which is most interesting from our vantage point since it deforms the
abelian gauge symmetry of the free theory. These types of structures are visible in the
relatively simple 2--3--3 example.

Let us stress that the couplings that survive in the massless limit are again
induced by a spin--2 current that is algebraically conserved, and thus of
multipolar type. Moreover, even in this case one can recast our gauge invariant
limiting result in the form
{\allowdisplaybreaks
\begin{multline}
\cA^{\,[0]\,\pm}_{\, 2-2-s}\,=\,\left[1\,\pm\,\cG\,+\,\frac{1}{2}\,\cG^{\,2}\right]\\ \times\,h_{\,1}\cdot\left(\xi_{\,1}\,\pm\,\sqrt{\frac{\a^{\,\prime}\!\!}{2}}\,p_{\,23}\right)^2\,
\phi_{\,2}\cdot\left(\xi_{\,2}\,\pm\,\sqrt{\frac{\a^{\,\prime}\!\!}{2}}\,p_{\,31}\right)^s\, \phi_{\,3}\cdot\left(\xi_{\,3}\,\pm\,\sqrt{\frac{\a^{\,\prime}\!\!}{2}}\,p_{\,12}\right)^s\,\Bigg|_{\xi_{\,i}\,=\,0}\ ,
\end{multline}}
\!\!in terms of the operator $\cG$ defined in eq.~\eqref{G}, so that all couplings are somehow induced by the highest--derivative one.

\vskip 24pt


\scss{The general case of $s_1$--$s_2$--$s_3$ couplings}\label{sec:mnl}


We can now conclude our examples of string couplings by showing that the terms
where the gauge parameter is contracted solely with momenta disappear in the
gauge variation of all three--point amplitudes. To this end, starting from the
general expression given in eq.~\eqref{Apiumeno}, or alternatively in
eq.~\eqref{expl}, let us consider the effect of the spin--$s_1$ gauge
variations. The result,
{\allowdisplaybreaks
\begin{eqnarray}
\d\cA^{\,\pm}&\!\!\!\!\!=&\!\!\!\sum_{\cI}\,\vphantom{\int} \frac{1}{2}\frac{s_1!\,s_2!\,s_3!\ \d^{\a}_{23}\ \d^{\,\b}_{\,31}\ \d^{\,\g}_{\,12} \ p_{\,23}^{\,s_1-\b-\g-1}\ p_{\,31}^{\,s_2-\a-\g}\ p_{\,12}^{\,s_3-\a-\b}}{\a!\,\b!\,\g!\,(s_1-\b-\g-1)!\,(s_2-\a-\g)!\,(s_3-\a-\b)!} \,\Big[s_3-s_2+\g-\b+\a'\,p_{\,1}\,\cdot\,p_{\,23}\Big]\nonumber\\
\vphantom{{\phi}_{\,1}
\left(p_{\,1},\,\xi_{\,1}\,\pm\,\sqrt{\frac{\a^{\,\prime}\!\!}{2}}\,p_{\,23}\right)}
&&\quad\quad\vphantom{\int}\ \times\,\L_{\,1}(p_{\,1})\ \phi_{\,2}(p_{\,2})\ \phi_{\,3}(p_{\,3})\\
&\!\!\!=&\!\!\! \frac{1}{2}\,\Big(\partial_{\xi_{\,1}}\cdot \partial_{\xi_{\,2}}\,-\,\partial_{\xi_{\,1}}\cdot \partial_{\xi_{\,3}}\Big)\,\exp{\Big[\,\partial_{\xi_{\,1}}\cdot \partial_{\xi_{\,2}}+\partial_{\xi_{\,2}}\cdot \partial_{\xi_{\,3}}+\partial_{\xi_{\,3}}\cdot \partial_{\xi_{\,1}}\Big]} \vphantom{{\phi}_{\,1}\left(p_{\,1},\,\xi_{\,1}\,\pm\,\sqrt{\frac{\a^{\,\prime}\!\!}{2}}\,p_{\,23}\right)}\nonumber\\
\vphantom{\int} && \times\,{\Lambda}_{\,1}\left(p_{\,1},\,\xi_{\,1}\,\pm\,\sqrt{\frac{\a^{\,\prime}\!\!}{2}}\,p_{\,23}\right)\,{\phi}_{\,2}\left( p_{\,2},\,\xi_{\,2}\,\pm\,\sqrt{\frac{\a^{\,\prime}\!\!}{2}}\,p_{\,31}\right)\, {\phi}_{\,3}\left(p_{\,3},\,\xi_{\,3}\,\pm\,\sqrt{\frac{\a^{\,\prime}\!\!}{2}}\,p_{\,12}\right)\Bigg|_{\xi_{\,i}\,=\,0}\ ,\label{cancelations}\nonumber
\end{eqnarray}}
\!\!shows explicitly that a large number of cancelations takes place, in particular
among contributions containing different numbers of momenta, in such a way that
all leftover terms involve some contractions between gauge parameters and
residual gauge fields. As in previous sections, for brevity in the first part
of this equation we are leaving space--time indices implicit, and as a result we
are not indicating that each of the $p_{\,ij}$ is contracted with a
corresponding ${\phi}_{\,k}$, with $k$ different from $i$ and $j$. Moreover,
the powers of the $\d_{\,ij}$ indicate the numbers of contractions between
${\phi}_{\,i}$ and ${\phi}_{\,j}$. On the other hand, vector indices are
indicated more explicitly in the second part, via the corresponding (shifted)
symbols.

Taking into account, as in the previous special cases, the terms proportional
to the masses and fixing the ambiguity in the gauge variation,  in the
\emph{massless} limit one is naturally led to the expression
\begin{multline}
\cA^{\,[0]\,\pm}\,=\,\vphantom{\int}\exp{\left\{\pm\sqrt{\frac{\a^{\,\prime}\!\!}{2}}\ \Big[(\partial_{\xi_{\,1}}\cdot\partial_{\xi_{\,2}})
(\partial_{\xi_{\,3}}\cdot p_{\,12})\,+\,(\partial_{\xi_{\,2}}\cdot\partial_{\xi_{\,3}})(\partial_{\xi_{\,1}}\cdot p_{\,23})\,+\,(\partial_{\xi_{\,3}}\cdot\partial_{\xi_{\,1}})(\partial_{\xi_{\,2}}\cdot
p_{\,31})\Big]\right\}}\\ \vphantom{\int} \times\, \phi_{\,1}\left(p_{\,1}\,,\,\xi_{\,1}\,\pm\,\sqrt{\frac{\a^{\,\prime}\!\!}{2}}\,p_{\,23}\right)\,
\phi_{\,2}\left(p_{\,2}\,,\,\xi_{\,2}\,\pm\,\sqrt{\frac{\a^{\,\prime}\!\!}{2}}\,p_{\,31}\right)\, \phi_{\,3}\left(p_{\,3}\,,\,\xi_{\,3}\,\pm\,\sqrt{\frac{\a^{\,\prime}\!\!}{2}}\,p_{\,12}\right)\,\Bigg|_{\xi_{\,i}\,=\,0}\ ,\label{masslessGenfunc}
\end{multline}
that involves again the differential operator $\cG$ defined in eq.~\eqref{G}. In other
words, this off--shell cubic vertex lies behind the low--tension limit of the
three--point string amplitudes that we have described so far. As we have already
stressed, $\cG$ commutes \emph{on shell} with gauge transformations, so that once this
vertex is added to the free theory, the linearized gauge transformations must be
complemented in general by contributions that deform the gauge algebra. In the Appendix
we shall see how complete off--shell vertices can be reconstructed from these expressions
reinstating the terms proportional to multiple traces and de Donder conditions that were
ignored so far.

String Theory makes use of exponentials of $\cG$, but from a field theory perspective we can not exclude at present that HS couplings of more general types
\begin{multline}
\cA^{\,[0]\,\pm}\,=\,a\left(\pm\,\cG\right)\\ \times\,\phi_{\,1}\left(p_{\,1}\,,\,\xi_{\,1}\,\pm\,\sqrt{\frac{\a^{\,\prime}\!\!}{2}}\,p_{\,23}\right)\,
\phi_{\,2}\left(p_{\,2}\,,\,\xi_{\,2}\,\pm\,\sqrt{\frac{\a^{\,\prime}\!\!}{2}}\,p_{\,31}\right)\, \phi_{\,3}\left(p_{\,3}\,,\,\xi_{\,3}\,\pm\,\sqrt{\frac{\a^{\,\prime}\!\!}{2}}\,p_{\,12}\right)\,\Bigg|_{\xi_{\,i}\,=\,0}\ ,
\end{multline}
where the exponential of eq.~\eqref{masslessGenfunc} is replaced by another
operator--valued function $a(\cG)$, be equally consistent. The explicit
analysis of the quartic order will be instrumental to identify the options that
are actually available. At any rate, the emergence of the iterative structure
suggested by String Theory and based on a single operator $\cG$ is very
interesting and puts somehow all abelian and non abelian couplings on the same
footing. Moreover, in this fashion the highest--derivative coupling
\be
\cA^{\,[0]} \ \sim \ \phi_{\,1}\left( p_{\,1}\right)\cdot \left(p_{\,23}\right)^{s_1} \ \phi_{\,2}\left( p_{\,2}\right)\cdot \left(p_{\,31}\right)^{s_2}\ \phi_{\,3}\left( p_{\,3}\right)\cdot \left(p_{\,12}\right)^{s_3} \ ,
\ee
generates somehow all other contributions via $\cG$, and in this respect it plays a special role\footnote{Although this coupling is gauge invariant on--shell, it can completed by other terms, as discussed in \cite{cubic}, in such a way that the result \emph{does not} induce deformations of the gauge algebra but merely field redefinitions.}.

\vskip 24pt


\scss{Massive on--shell gauge invariance} \label{sec:massgauginv}


We can conclude this Section with an instructive exercise related to the Stueckelberg gauge symmetries of the cubic couplings that we have analyzed. As we shall see, the procedure has the additional virtue of exhibiting some features of generic, mixed--symmetry couplings that will allow us to make, at the end, an educated guess for their actual form.

The first observation concerns the form of the physical states that enter the scattering amplitudes. As we have stressed in the Introduction, resorting to momentum--space fields of the form
\begin{equation}
\phi(p\,,\xi)\,=\,\phi_{\m_1\ldots\,\m_s}(p)\ \xi^{\,\m_1}\ldots\, \xi^{\,\m_s}\ ,\label{normalf}
\end{equation}
is somehow on the par with fixing a gauge for a Yang--Mills theory.

At this point, one has two seemingly distinct ways to reinstate a gauge symmetry, that are actually related to one another by the string model. The first is to perform in the vertex
\be\label{normalfs=2}
\cV_{\,2}\,=\,\oint\,h_{\m_1\m_2}(p)\,\partial X^{\,\m_1}\,\partial X^{\,\m_2}\,e^{ip\cdot X}\ ,
\ee
a Stueckelberg shift, thus obtaining
\be
\cV^{\ \text{shift}}_{\,2}\,=\,\oint\,\Big[\vphantom{\partial X^{\,\m_s}} h_{\m_1\m_2}(p)\,-\,i p_{\,\m} \, b_\n (p)\Big] \,\partial X^{\,\m_1}\,\partial X^{\,\m_2}\,e^{ip\cdot X}\ .\label{shift}
\ee
On this expression, the physical state conditions amount to
\be
\begin{split}
  h^{\,\prime}(p)\,-\,\,i\,p\cdot b(p)\,&=\,0\ ,\\
  p\cdot h_{\,\m}(p)\,-\,\frac{i}{2}\,p_{\,\m}\,p\cdot b(p)\,&=\,-\,\frac{i}{2\a^{\,\prime}\!\!}\ b_{\,\m}(p)\ ,\label{phys}
\end{split}
\ee
while the Stueckelberg symmetry
\be
\delta\, h_{\,\m\n}(p)\,=\,-\,\frac{i}{2}\,\left[p_{\,\m}\,\L_{\n}(p)\,+\,p_{\,\n}\,\L_{\,\m}(p)\right]\ ,\qquad
\delta\, b_{\,\m}(p)\,=\,-\,\L_{\,\m}(p)\ ,\label{Stueckelberg}
\ee
where the mass--shell condition $p^{\,2}\,=\,-1/\a^{\,\prime}$ has been used,
is clearly manifest, and does not require any additional constraints on the
gauge parameter $\L_{\,\m}$. On the other hand, the second possibility rests on
the existence of the trivial \emph{integrated} vertex
\be
\cV_{null}\,=\,\oint\,\partial\left[b_{\,\m_1}(p)\,\partial X^{\,\m_1}\,e^{ip \cdot X}\right]\,=\,\oint\,\left[ip_{\,\m_1}\,b_{\,\m_2}(p)\,\partial X^{\,\m_1}\,\partial X^{\,\m_2}\,+\,b_{\,\m_1}(p)\,\partial^{\,2} X^{\,\m_1}\right]\,e^{ip\cdot X}\ ,\label{null}
\ee
that, when added to the vertex operator \eqref{shift}, can turn it into the expression
\be
\cV_{\,2}\,=\,\oint\,\Big[\,h_{\m_1\m_2}(p)\,\partial X^{\,\m_1}\,\partial X^{\,\m_2}\,+b_{\,\m_1}(p)\,\partial^{\,2} X^{\,\m_1}\Big]\,e^{ip\cdot X}\ ,\label{UnintVertexGauge}
\ee
that also involves the oscillator $\a_{\,-2}$. In order to use this vertex
operator, rather than \eqref{normalfs=2}, to compute string amplitudes, one
should also consider its unintegrated form, since the $SL(2,R)$ symmetry is
usually accounted for keeping some vertices at fixed positions. For
unintegrated vertices the two choices of eqs.~\eqref{shift} and
\eqref{UnintVertexGauge} cease to be equivalent and therefore, when applied to
the former, the Virasoro constraints translate into a different set conditions:
\be
b_{\,\m}(p)\,-\,{2\a^{\,\prime}}\,i\,\ p\cdot h_{\,\m}(p)\,=\,0\ ,\qquad h^{\,\prime}(p)\,+\,4\a^{\,\prime}\, p\cdot h(p)\cdot p\,=\,0\ .
\ee
The new constraints, however, are compatible with a Stueckelberg--like symmetry only if the gauge parameter is also subject to the differential constraint
\be
p\cdot\L\,=\,0\ .
\ee

This seemingly different behavior reflects the interplay between the two oscillators $\a_{-1}$ and $\a_{-2}$. While they seem to mix different Regge trajectories, in this case they are actually not independent, since as we have stressed their contributions correspond to integrated vertices that differ by a total derivative. These two types of options continue to exist for generic symmetric tensors of the first Regge trajectory, where seemingly different options for the integrated vertices differ by the trivial integrated vertices
\be
\begin{split}
\cV_{null}\,&=\,\oint\,\partial\Big[b_{\,\m_1\ldots\,\m_{s-1}}(p)\,\partial X^{\,\m_1}\!\ldots\,\partial X^{\,\m_{s-1}}\,e^{\,ip\, \cdot X}\Big]\label{nullgen}\\&=\, \oint\ \ \Big[ip_{\,\m_1}\,b_{\,\m_2\ldots\,\m_s}(p)\,\partial X^{\,\m_1}\!\ldots\,\partial X^{\,\m_s}\\&+\,(s-1)\,b_{\,\m_1\ldots\,\m_{s-1}}(p)\,\partial^{\,2} X^{\,\m_1}\,\partial X^{\,\m_2}\!\ldots\,\partial X^{\,\m_{s-1}}\Big]\,e^{\,ip\,\cdot X}\ .
\end{split}
\ee
These relate in fact the Stueckelberg shifted vertices
\be
\cV^{\ \text{shift}}_{\,s}\,=\,\oint\,\Big[\vphantom{\partial X^{\,\m_s}} \phi_{\m_1\ldots\m_s}(p)\,-\,\frac{i}{s-1}\  p_{\m_1} \, b_{\m_{2}\ldots\m_s} (p)\Big] \,\partial X^{\,\m_1}\,\partial X^{\,\m_2}\ldots\,\partial X^{\,\m_s}\,e^{ip\cdot X}\ ,\label{shift_s}
\ee
whose integrand is to be traceless and transverse, that by construction possess the Stueckelberg symmetry
\be
\delta\,\phi_{\,\m_1\ldots\,\m_s}(p)\,=\,-\,\frac{1}{s}\left[ip_{\,\m_1}\,\L_{\,\m_2\ldots\,\m_s}(p)\,+\,\text{cyclic}\right]\ ,\qquad \delta\, b_{\,\m_1\ldots\,\m_{s-1}}(p)\,=\,-\,(s-1)\,\L_{\,\m_1\ldots\,\m_{s-1}}(p)\ ,\label{gaugesym}
\ee
to generalizations of eq.~\eqref{UnintVertexGauge},
\be
\cV_{\,s}\,=\,\oint\,\Big[\phi_{\,\m_1\ldots\,\m_s}(p)\,\partial X^{\,\m_1}\ldots\,\partial X^{\,\m_s}\,+\,b_{\,\m_1\ldots\,\m_{s-1}}(p)\,\partial^2 X^{\,\m_1}\ldots\,\partial X^{\,\m_{s-1}}\Big]\,e^{\,ip\,\cdot X}\ .\label{physgen}
\ee
When applied to the integrands of eqs.~\eqref{nullgen} and \eqref{physgen}, the Virasoro constraints imply the mass--shell condition
\be
-p^{\,2}\,=\,\frac{s-1}{\a^{\,\prime}}\ ,
\ee
together with the relations
\be
\begin{split}
b_{\,\m_1\ldots\,\m_{s-1}}(p)\,-\,i\, s\,\a^{\,\prime}\ p\,\cdot\phi_{\,\m_1\ldots\,\m_{s-1}}(p)\,&=\,0\ ,\\
p\cdot b_{\,\m_1\ldots\,\m_{s-2}}(p)\,&=\,0 \ ,\vphantom{\sqrt{\frac{\a^{\,\prime}\!\!}{2}}}\\
\eta^{\,\m_1\m_2}\,\phi_{\,\m_1\ldots\,\m_s}(p)\vphantom{\sqrt{\frac{\a^{\,\prime}\!\!}{2}}}\,&=\,0 \ ,\vphantom{\sqrt{\frac{\a^{\,\prime}\!\!}{2}}}\\
\eta^{\,\m_1\m_2}\,b_{\m_1\ldots\,\m_{s-1}}(p)\vphantom{\sqrt{\frac{\a^{\,\prime}\!\!}{2}}}\,&=\,0\vphantom{\sqrt{\frac{\a^{\,\prime}\!\!}{2}}}\ ,\label{ViraField}
\end{split}
\ee
where the previous arguments show that the last two conditions are only present for $s\,>\,2$. Moreover, the gauge parameter is to satisfy the constraints
\be
\begin{split}
\eta^{\,\m_1\m_2}\,\L_{\,\m_1\ldots\,\m_{s-1}}(p)\,&=\,0 \ ,\\
p\cdot \L_{\,\m_1\ldots\,\m_{s-2}}(p)\,&=\,0\ .\label{ViraGauge}
\end{split}
\ee
Hence, after imposing the Virasoro conditions one would seem to end up with a
\emph{constrained} gauge symmetry, at least insofar as the unintegrated
vertices are concerned. This, however, is merely an artifact of the mechanical
string model: the dependence on the $y_i$ is bound to disappear eventually from the amplitudes, together with the distinction between integrated
and unintegrated vertices, leaving way to the complete Stueckelberg symmetry
expected in Field Theory. Having anticipated the nature of the answer, we can
now profit from a closer look at the vertex \eqref{physgen}.

It is now convenient to work with the vertex \eqref{UnintVertexGauge}, albeit with $b_{\,\m_1\ldots \m_{s-1}}$ expressed in terms of $\phi_{\,\m_1\ldots \m_{s}}$ via the first of the Virasoro constraints \eqref{ViraField}. In terms of the symbols of the first two oscillators this amounts to working with
\be
\bar{\phi}\,\left(p\,,\,\xi^{\,(1)},\,\xi^{\,(2)}\right)\,=\,\left[1-\frac{1}{\hat{s}-1}\,\sqrt{\frac{\a^{\,\prime}\!\!}{2}}\, \left(p\cdot\partial_{\xi^{\,(1)}}\right)\left(\xi^{\,(2)}\cdot\partial_{\xi^{\,(1)}}\right)\right] \,\phi\,\left(p\,,\,\xi^{\,(1)}\right)\ ,  \label{gaugeregge}
\ee
where now $\phi\left(p\,,\,\xi^{\,(1)}\right)$ is to satisfy the constraints
\begin{gather}
\partial_{\xi^{\,(1)}}\cdot \partial_{\xi^{\,(1)}}\ \phi\,\left(p\,,\,\xi^{\,(1)}\right)\,=\,0\ \nonumber ,\\
\left(p\cdot \partial_{\xi^{\,(1)}}\right)^2\,\phi\,\left(p\,,\,\xi^{\,(1)}\right)\,=\,0\ ,\label{Viraphi1}
\end{gather}
that for $s\,>\,2$ are implied by \eqref{ViraField}, and where $\hat{s}$ computes the spin of the generating functions to which it is applied.
In this fashion, one is led to generating functions for three--point amplitudes that also involve the second symbol $\xi^{(2)}$. These are special, and yet non--trivial, cases of the general three--point generating function
{\allowdisplaybreaks
\begin{eqnarray}
&&\!\!\!\!\!\!\!\!\!\!\mathbf{Z}\,=\,i\,g_o\,\frac{(2\pi)^{\,d}}{\a^{\,\prime}}\ \delta^{\,(d)}(p_{\,1}+p_{\,2}+p_{\,3})\,\nonumber\\
&&\!\!\!\!\!\!\!\!\!\!\times\exp\left[\sum_{\ n,\,m=1}^{\infty}\,\frac{(-1)^{n+1} (n+m-1)!}{(n-1)!(m-1)!}\, \left(\xi_{\,1}^{(n)}\cdot \xi_{\,2}^{\,(m)}\,\bra y_{12}\ket^{n+m}\left|\frac{y_{31}}{y_{23}}\right|^{n-m}\right.\right.\nonumber\\
&&\left.\qquad\qquad\qquad+\,\xi_{\,1}^{(n)}\cdot \xi_{\,3}^{\,(m)}\,\bra y_{13}\ket^{n+m}\left|\frac{y_{12}}{y_{23}}\right|^{n-m}\,+\,\xi_{\,2}^{(n)}\cdot \xi_{\,3}^{\,(m)}\bra y_{23}\ket^{n+m}\left|\frac{y_{12}}{y_{31}}\right|^{n-m}\right)\nonumber\\
&&\!\!\!\!-\,\sqrt{\frac{\a^{\prime}\!\!}{2}}\ \sum_{n=1}^{\infty}\left(\,\xi_{\,1}^{(n)}\cdot p_{\,23}\left|\frac{y_{12}\,y_{13}}{y_{23}}\right|^n\,\frac{y_{31}^n-y_{21}^n}{y_{21}^n\,y_{31}^n}
\,-\,\xi_{\,1}^{(n)}\cdot p_{\,1}\ \left|\frac{y_{12}\, y_{13}}{y_{23}}\right|^n\,\frac{y_{31}^n+y_{21}^n}{y_{31}^n\, y_{21}^n}\nonumber\right.\\\vphantom{\sqrt{\frac{\a^{\prime}\!\!}{2}}}
&&\qquad\qquad+\ \xi_{\,2}^{(n)}\cdot p_{\,31}\left|\frac{y_{12}\, y_{23}}{y_{13}}\right|^n\,\frac{y_{12}^n-y_{32}^n}{y_{12}^n\,y_{32}^n}\,-\,\xi_{\,2}^{(n)}\cdot p_{\,2}\ \left|\frac{y_{12}\, y_{23}}{y_{13}}\right|^n\,\frac{y_{12}^n+y_{32}^n}{y_{32}^n\, y_{12}^n}\nonumber\\\vphantom{\sqrt{\frac{\a^{\prime}\!\!}{2}}}
&&\left.\left.\!\qquad\qquad+\ \xi_{\,3}^{(n)}\cdot p_{\,12}\left|\frac{y_{13}\, y_{23}}{y_{12}}\right|^n\,\frac{y_{23}^n-y_{13}^n}{y_{13}^n\,y_{23}^n}\,
-\,\xi_{\,3}^{(3)}\cdot p_{\,3}\ \left|\frac{y_{13}\, y_{23}}{y_{12}}\right|^n\,\frac{y_{23}^n+y_{13}^n}{y_{23}^n\, y_{13}^n}\right)\right]\ ,\label{GeneralZ}
\end{eqnarray}}
\!\!where the dependence on the $y_i$ would not seem to disappear in any way
as in eq.~\eqref{Gen3}. The problem is due to the symbols of the lower
trajectories, and indeed in its present form \eqref{GeneralZ} contains much
spurious information, just as was the case for the original expression \eqref{Gen2} for the
first Regge trajectory. Only correlation functions involving states that
satisfy the Virasoro constraints, like those of eq.~\eqref{gaugeregge}, are to
be independent of the $y_i$, and we can now see that this is again the case.
For definiteness, let us thus fix the order of the vertices according to
\be
y_{\,1}\,>\,y_{\,2}\,>\,y_{\,3}\ ,
\ee
and let us concentrate on the portion of \eqref{GeneralZ} that depends only on $\xi^{\,(1)}$ and $\xi^{\,(2)}$, which suffices for our present purposes and reads
{\allowdisplaybreaks
\begin{eqnarray}
\mathbf{Z}&\!\!=&\!\!i\,g_o\,\frac{(2\pi)^{\,d}}{\a^{\,\prime}}\ \delta^{\,(d)}(p_{\,1}+p_{\,2}+p_{\,3})\nonumber\\
\times &\!\!\!\!\exp&\!\!\!\!\!\left\{\
\xi_{\,1}^{(1)}\cdot\xi_{\,2}^{(1)}+\xi_{\,2}^{(1)}\cdot\xi_{\,3}^{(1)}+\xi_{\,3}^{(1)}\cdot\xi_{\,1}^{(1)}-
2\,\xi_{\,1}^{(1)}\cdot\xi_{\,2}^{(2)}\ \frac{y_{23}}{y_{31}}-
2\,\xi_{\,2}^{(1)}\cdot\xi_{\,3}^{(2)}\ \frac{y_{31}}{y_{12}}-
2\,\xi_{\,3}^{(1)}\cdot\xi_{\,1}^{(2)}\ \frac{y_{12}}{y_{23}}\right.\nonumber\\
&\!\!+&\!\!
2\,\xi_{\,1}^{(2)}\cdot\xi_{\,2}^{(1)}\ \frac{y_{31}}{y_{23}}+
2\,\xi_{\,2}^{(2)}\cdot\xi_{\,3}^{(1)}\ \frac{y_{12}}{y_{31}}+
2\,\xi_{\,3}^{(2)}\cdot\xi_{\,1}^{(1)}\ \frac{y_{23}}{y_{12}}+
6\,\xi_{\,1}^{(2)}\cdot\xi_{\,2}^{(2)}+6\,\xi_{\,2}^{(2)}\cdot\xi_{\,3}^{(2)}+6\,\xi_{\,3}^{(2)}\cdot\xi_{\,1}^{(2)}\nonumber\\
&\!\!+&\!\!\sqrt{\frac{\a^{\,\prime}\!\!}{2}}\ \left[
\,\xi_{\,1}^{(1)}\cdot\left(p_{\,23}+p_{\,1}\,\frac{y_{31}+y_{21}}{y_{23}}\right)\right.\!\!+
\,\xi_{\,2}^{(1)}\cdot \left(p_{\,31}+p_{\,2}\,\frac{y_{12}+y_{32}}{y_{13}}\right)\nonumber\\
&&\quad\ \ +\
\xi_{\,3}^{(1)}\cdot \left(p_{\,12}+p_{\,3}\,\frac{y_{23}+y_{13}}{y_{12}}\right)+
\,\xi_{\,1}^{(2)}\cdot\left(p_{\,23}\,\frac{y_{31}+y_{21}}{y_{23}}+p_{\,1}\,\frac{y_{31}^2+y_{21}^2}{y_{23}^2}\right)\\
&&\quad\ \ +\
\xi_{\,2}^{(2)}\cdot \left(p_{\,31}\,\frac{y_{12}+y_{32}}{y_{13}}+p_{\,2}\,\frac{y_{12}^2+y_{32}^2}{y_{13}^2}\right)+\left.\left.\!
\xi_{\,3}^{(2)}\cdot \left(p_{\,12}\,\frac{y_{23}+y_{13}}{y_{12}}+p_{\,3}\,\frac{y_{23}^2+y_{13}^2}{y_{12}^2}\right)\right]\right\}\ .\nonumber
\end{eqnarray}}
\!\!The three--point amplitudes compatible with the massive gauge symmetry that we have identified are then encoded in
\begin{multline}
\mathbf{\cA}\,=\,\left\{\mathbf{Z}\,\star_{\,\xi_{\,i}^{(2)}}\,\left[1-\frac{1}{\hat{s}_1-1}\,\sqrt{\frac{\a^{\,\prime}\!\!}{2}}\, \left(p_{\,1}\cdot\partial_{\xi_{\,1}^{(1)}}\right)\left(\xi_{\,1}^{(2)}\cdot\partial_{\xi_{\,1}^{(1)}}\right)\right]\right.
\\\left.
\times\,\left[1-\frac{1}{\hat{s}_2-1}\,\sqrt{\frac{\a^{\,\prime}\!\!}{2}}\, \left(p_{\,2}\cdot\partial_{\xi_{\,2}^{(1)}}\right)\left(\xi_{\,2}^{(2)}\cdot\partial_{\xi_{\,2}^{(1)}}\right)\right]
\left[1-\frac{1}{\hat{s}_3-1}\,\sqrt{\frac{\a^{\,\prime}\!\!}{2}}\, \left(p_{\,3}\cdot\partial_{\xi_{\,3}^{(1)}}\right)\left(\xi_{\,3}^{(2)}\cdot\partial_{\xi_{\,3}^{(1)}}
\right)\right]\right\}\\\star_{\,\xi_{\,i}^{(1)}}\,\phi_{\,1}\,\left(p_{\,1}\,,\,\xi_{\,1}^{(1)}\right)
\phi_{\,2}\,\left(p_{\,2}\,,\,\xi_{\,2}^{(1)}\right)\phi_3\,\left(p_{\,3}\,,\,\xi_{\,3}^{(1)}\right)\ ,\label{Agaugeinv}
\end{multline}
where the inner $\star$--product is on $\xi_{\,i}^{(2)}$ and, for brevity, we refer
to \eqref{ampl} for the complete expression involving also Chan--Paton factors.
This generating function can be shown \emph{not to} depend on the $y_i$ if
eqs.~\eqref{Viraphi1} are satisfied. Moreover, one can notice that
\eqref{Agaugeinv} differs from \eqref{Apiumeno} only by terms proportional to
$p\cdot \phi$, which were previously ignored.

In this respect, it is interesting to compute explicitly from \eqref{Agaugeinv} some relatively simple scattering amplitudes where one of the three legs is of the form \eqref{gaugeregge} while the others have been gauge fixed in the unitary gauge. The simplest example is the 1--1--$s$ amplitude,
\begin{equation}
\begin{split}
\cA^{\pm}_{1-1-s}\,&=\,\left(\!\pm\sqrt{\frac{\a^{\,\prime}\!\!}{2}}\,\right)^{s-2}\,s(s-1)\, A_{1\,\m}\,A_{2\,\n}\,\phi_{\,3}^{\,\m\n\ldots}\cdot p_{\,12}^{\,s-2}\\
&+\,\left(\!\pm\sqrt{\frac{\a^{\,\prime}\!\!}{2}}\,\right)^s \Big[A_{1} \cdot A_{2}\, \phi_{\,3}\cdot p_{\,12}^{\,s}+s\,A_{1}\cdot p_{\,23}\,A_{2\,\n}\,\phi_{\,3}^{\,\n\ldots\,}\cdot(p_{\,12}\,-\,p_{\,3})\,p_{\,12}^{\,s-2}\\
&\,\qquad\qquad\quad\ \ +s\,A_{2}\cdot p_{\,31}\, A_{1\,\n}\,\phi_{\,3}^{\,\n\ldots\,}\cdot(p_{\,12}\,+\,p_{\,3})\,p_{\,12}^{\,s-2}\Big]\\
&+\,\left(\!\pm\sqrt{\frac{\a^{\,\prime}\!\!}{2}}\,\right)^{s+2}A_{1} \cdot p_{\,23}\,A_{2}\cdot p_{\,31}\, \phi_{\,3}\cdot p_{\,12}^{\,s}\ , \label{11sg}
\end{split}
\end{equation}
where one can see that the terms proportional to the divergence of $\phi$ produce a gauge variation that cancels exactly the contribution \eqref{lack}.

Actually, eq.~\eqref{Agaugeinv} simplifies considerably when one external excitation is described by \eqref{gaugeregge} while the other two are of the form \eqref{normalf}. The result is completely independent of the $y_i$ and is given by
\begin{multline}
\cA\,=\,\exp{\Big(\partial_{\xi_{\,1}}\cdot\partial_{\xi_{\,2}}\,+\,\partial_{\xi_{\,2}}\cdot\partial_{\xi_{\,3}}\,
+\,\partial_{\xi_{\,3}} \cdot\partial_{\xi_{\,1}}\Big)}\\\times \left[1\,+\,\frac{1}{\hat{s}_1-1}\ \left(\sqrt{\frac{\a^{\,\prime}\!\!}{2}}\ p_{\,1}\cdot\partial_{\xi_{\,1}}\right)\Bigg(\partial_{ \xi_{\,2}}\cdot\partial_{\xi_{\,1}}\,- \,\partial_{\xi_{\,3}}\cdot\partial_{\xi_{\,1}}\Bigg)\right]\\
\times\,{\phi}_{\,1}\left(p_{\,1},\,\xi_{\,1}\,\pm\,\sqrt{\frac{\a^{\,\prime}\!\!}{2}}\,p_{\,23}\right) \,{\phi}_{\,2}\left( p_{\,2},\,\xi_{\,2}\,\pm\,\sqrt{\frac{\a^{\,\prime}\!\!}{2}}\,p_{\,31}\right) \\\times\,{\phi}_{\,3}\left(p_{\,3},\,\xi_{\,3}\,\pm\,\sqrt{\frac{\a^{\,\prime}\!\!}{2}}\,p_{\,12}\right) \Bigg|_{\xi_{\,i}\,=\,0}\ .
\end{multline}
In this relatively simple example it is instructive to use the equations of motion implied by the Virasoro constraints backwards, referring to \eqref{ViraField}, since the result,
{\allowdisplaybreaks
\begin{eqnarray}
\cA &=&\vphantom{\Bigg|_p}
\exp{\Big(\partial_{\xi_{\,1}}\cdot\partial_{\xi_{\,2}}\,+\,\partial_{\xi_{\,2}}\cdot\partial_{ \xi_{\,3}}\,+\,\partial_{\xi_{\,3}} \cdot\partial_{\xi_{\,1}}\Big)}\nonumber\\\vphantom{\Bigg|_p}
&&\times \left[
{\phi}_{\,1}\left(p_{\,1},\,\xi_{\,1}\,\pm\,\sqrt{\frac{\a^{\,\prime}\!\!}{2}}\,p_{\,23}\right) \,{\phi}_{\,2}\left( p_{\,2},\,\xi_{\,2}\,\pm\,\sqrt{\frac{\a^{\,\prime}\!\!}{2}}\,p_{\,31}\right)
{\phi}_{\,3}\left(p_{\,3},\,\xi_{\,3}\,\pm\,\sqrt{\frac{\a^{\,\prime}\!\!}{2}}\,p_{\,12}\right)\right. \nonumber\\\vphantom{\Bigg|_p}&&\quad-\Big(\partial_{\xi_{\,2}}\cdot\partial_{\xi_{\,1}}\,- \,\partial_{ \xi_{\,3}}\cdot\partial_{\xi_{\,1}}\Big)\ {b}_{\,1}\left(p_{\,1},\,\xi_{\,1}\,\pm\, \sqrt{\frac{\a^{\,\prime}\!\!}{2}}\,p_{\,23}\right) \,{\phi}_{\,2}\left( p_{\,2},\,\xi_{\,2}\,\pm\,\sqrt{\frac{\a^{\,\prime}\!\!}{2}}\,p_{\,31}\right) \nonumber\\&&\left.\qquad\times\,{\phi}_{\,3}\left(p_{\,3},\,\xi_{\,3}\,\pm\,\sqrt{\frac{\a^{\,\prime}\!\!}{2}}\,p_{\,12}\right) \right] \Bigg|_{\xi_{\,i}\,=\,0}\ .
\end{eqnarray}
}
\!\!is invariant under the Stueckelberg transformations
\be
\delta\, \phi_{\,1}(p_{\,1},\xi_{\,1})\,=\,-\,i\,p_{\,1}\cdot\xi_1\,\L_{\,1}(p_{\,1},\xi_{\,1})\ , \qquad \delta\, b_{\,1}(p_{\,1},\xi_{\,1})\,=\,-\,\L_{\,1}(p_{\,1},\xi_{\,1})\ .
\ee
All in all, the version of the symmetry related to the Stueckelberg shift of the vertex operator is finally recovered. One can actually go further along the lines of the previous computation, obtaining a gauge invariant amplitude for three states of the form \eqref{ViraField}. The end result is again completely independent of the $y_i$ and reads
{\allowdisplaybreaks
\begin{eqnarray}
\cA&=&\exp{\Big(\partial_{\xi_{\,1}}\cdot\partial_{\xi_{\,2}}\,+\,\partial_{\xi_{\,2}}\cdot\partial_{ \xi_{\,3}}\,+\,\partial_{\xi_{\,3}} \cdot\partial_{\xi_{\,1}}\Big)}\nonumber\\
&\times& \left\{\prod_{i\,=\,1}^3\left[1\,+\,\frac{1}{\hat{s}_i-1}\ \left(\sqrt{\frac{\a^{\,\prime}\!\!}{2}}\ p_{\,i}\cdot\partial_{\xi_{\,i}}\right)\Bigg(\partial_{\xi_{\,i+1}}\cdot\partial_{\xi_{\,i}}\,- \,\partial_{ \xi_{\,i-1}}\cdot\partial_{\xi_{\,i}}\Big)\right]\right.\\
&&-\left.\,3\,\left(\frac{1}{\hat{s}_1-1}\,\frac{1}{\hat{s}_2-1}\right)\left(\sqrt{\frac{\a^{\,\prime}\!\!}{2}}\ p_{\,1}\cdot\partial_{\xi_{\,1}}\right)\left(\sqrt{\frac{\a^{\,\prime}\!\!}{2}}\ p_{\,2}\cdot\partial_{ \xi_{\,2}}\right)\,\partial_{\xi_{\,1}}\cdot\partial_{\xi_{\,2}}\,+\,\text{cyclic}
\vphantom{\prod_{i\,=\,1}^3}\right\}\nonumber\\
&\times&{\phi}_{\,1}\left(p_{\,1},\,\xi_{\,1}\,\pm\,\sqrt{\frac{\a^{\,\prime}\!\!}{2}}\,p_{\,23}\right) \,{\phi}_{\,2}\left( p_{\,2},\,\xi_{\,2}\,\pm\,\,\sqrt{\frac{\a^{\,\prime}\!\!}{2}}\,p_{\,31}\right)
\,{\phi}_{\,3}\left(p_{\,3},\,\xi_{\,3}\,\pm\,\sqrt{\frac{\a^{\,\prime}\!\!}{2}}\,p_{\,12}\right) \Bigg|_{\xi_{\,i}\,=\,0}\!\!\!\!,\nonumber
\end{eqnarray}}
\!\!\!where, $1\,\leq\,i\,\leq \,3$ is an integer defined modulo three, so that for instance $4 \sim 1$.

All reference to the string world sheet has thus disappeared from the final result,
despite the fact that the two unintegrated vertices of eqs.~\eqref{shift} and
\eqref{UnintVertexGauge} are apparently inequivalent. Of course, the catch is that, for
purely technical reasons that reflect the infinite group volume, one is dealing
asymmetrically with the $SL(2,R)$ symmetry, but nonetheless the final amplitude is fully
compatible with it. As a result, the gauge symmetry is precisely what the Stueckelberg
shift would induce in \eqref{Apiumeno} after imposing the Virasoro constraints. Of
course, we see this fact only to some extent here, since for spin--$s$ fields one should
consider more complicated shifts involving lower trajectories, but these arguments
vindicate the consistency of our strategy for dealing with the gauge variation of the
massive amplitudes \eqref{ampl}, and provides a rationale for the cancelations that were
noticed in previous sections. Moreover, on--shell the constraints on fields and gauge
parameters become manifest in the low--tension limit, since
\be
p\,\cdot\phi(p)\,\simeq\, p^{\,2}\,=\,0\ ,\qquad p\,\cdot\L(p)\,\simeq\, p^{\,2}\,=\,0\ ,
\ee
so that one recovers the usual gauge symmetry for massless fields
\be
\delta\,\phi\,=\,-\,i\,(p\,\cdot \xi)\,\L(p\,,\,\xi)\ ,
\ee
while longitudinal modes decouple.

As we have anticipated, in order to see the Stueckelberg symmetry in full--fledged form, the shifted vertex that we have here considered should be
generalized adding lower--spin fields down to spin zero. Let us consider here
for brevity the spin--2 case, where the complete shift would take the form
\be
h_{\,\m\n}(p)\,\ra\,\tilde{h}_{\,\m\n}(p)\,=\,h_{\,\m\n}(p)\,-
\,\frac{i}{2}\left[\,p_{\m}\,b_{\,\n}(p)\,-\,p_{\n}\,b_{\,\m}(p)\right]\,-\,p_{\,\m}\,p_{\,\n}\,c(p)\ ,
\ee
so that the integrated vertex operator would become
\be
\cV^{\ \text{shift}}_{\,2}\,=\,\oint\,\Big[\vphantom{\partial X^{\,\m_s}} h_{\m_1\m_2}(p)\,-\,i p_\m \, b_\n (p)\,-\,p_{\,\m}\,p_{\,\n}\,c(p)\Big] \,\partial X^{\,\m_1}\,\partial X^{\,\m_2}\,e^{ip\cdot X}\ .\label{shift_0}
\ee
In this fashion one would end up with the two constraints
\be
\begin{split}
p\cdot h_{\,\m}(p)\,-\,\frac{i}{2}\,p_{\,\m}\,p\cdot b(p)\,+\,\frac{1}{\a^{\,\prime}}\ p_{\,\m}\,c(p)\,&=\,-\,\frac{i}{2\a^{\,\prime}}\ b_{\,\m}(p)\ ,\\
h^{\,\prime}(p)\,-\,i\,p\cdot b(p)\,+\,\frac{1}{\a^{\,\prime}}\ c(p)\,&=\,0\ ,
\end{split}
\ee
that are actually related to the de Donder--like constraints considered by
Metsaev in~\cite{Metsaev:2009hp}, up to a redefinition of the form
\be
h_{\,\m\n}(p)\,\ra\,h_{\,\m\n}(p)\,+\,\frac{1}{d-2}\ \eta_{\,\m\n}\,\frac{c(p)}{\a^{\,\prime}\!\!}\ ,
\ee
where $d$ denotes the space--time dimension, and a rescaling of the other two fields.
Repeating this exercise for symmetric tensors of arbitrary spin would be interesting,
relatively simple and potentially quite instructive, since it would shed light on the
roles of constrained and unconstrained symmetries in this context. We leave a more
detailed discussion of this issue, and more generally of mixed--symmetry fields, for
future work.

We can now conclude this section with a comment. In the unitary gauge the cubic
effective Lagrangian originating from String Theory would take the schematic form
\begin{multline}
\cL\,=\,\frac{1}{2}\
\phi(p\,,\,\xi)\,\star\,\left(-\,p^{\,2}-\frac{1}{\a^{\,\prime}\!\!}\
(\xi\cdot\partial_\xi-1)\right)\phi(-\,p\,,\,\xi)+\frac{1}{3}\
\phi\,(-p_{\,1}-p_{\,2}\,,\,\xi)\\ \star\
\underbrace{e^{\partial_{\zeta_1}\cdot\partial_{\zeta_2}+{\xi}\cdot\left(\partial_{\zeta_1}
+\partial_{\zeta_2}+\sqrt{\frac{\a^{\,\prime}\!\!}{2}}\ p_{12}\right)}\,
\phi\left(p_{\,1},\,\zeta_1\,+\,\sqrt{2\a^{\,\prime}}\,p_{\,2}\right)
\,\phi\left(p_{\,2},\,\zeta_2\,-\,\sqrt{2\a^{\,\prime}}\,p_{\,1}\right)\Bigg|_{\zeta_i\,=\,0}}_{j(p_{\,1}+p_{\,2}\,,\,\xi)}\
,
\end{multline}
where, for brevity, we have displayed explicitly the generating function \eqref{Curr}
leaving out Chan--Paton factors. As we have seen, the underlying gauge symmetry that we
have elaborated upon defines a smooth massless limit under the key assumption of current
conservation. Sorting out from the gauge variation of the amplitudes all terms that are
proportional to the masses, one can present the result as a gauge invariant coupling,
together with a term proportional to $p^{\,2}$, that on--shell is therefore proportional
to a mass,
\begin{equation}
\begin{split}
\cL\,&=\,\frac{1}{2}\ \phi(p\,,\,\xi)\,\star\,\left(-\,p^{\,2}-\frac{1}{\a^{\,\prime}\!\!}\
\left(\xi\cdot\partial_\xi-1\right)\right)\phi(-\,p\,,\,\xi)\,+\,\delta\phi(p\,,\,\xi)\star \left(-p^{\,2}\right)\phi(p\,,\,\xi)\\
&+\,\frac{1}{3}\ \phi(-p_{\,1}-p_{\,2}\,,\,\xi)\,\star\,
e^{\sqrt{\frac{\a^{\,\prime}\!\!}{2}}\ \xi\cdot p_{\,12}}\ \exp\left(\sqrt{\frac{\a^{\,\prime}\!\!}{2}}\ \xi_{\,\a}
\left[\partial_{\zeta_1}\cdot\partial_{\zeta_2}\,p^{\,\a}_{\,12}-2\,\partial^{\,\a}_{\zeta_1}\,
\partial_{\zeta_2}\cdot p_{\,1}
+\,2\,\partial^{\,\a}_{\zeta_2}\,\partial_{\zeta_1}\cdot p_{\,2}\right]\right)\\
&\qquad\qquad\times\,\phi\left(p_{\,1},\,\zeta_1\,+\,\sqrt{2\a^{\,\prime}}\,p_{\,2}\right)
\,\phi\left(p_{\,2},\,\zeta_2\,-\,\sqrt{2\a^{\,\prime}}\,p_{\,1}\right)\Bigg|_{\zeta_i\,=\,0}\ ,\label{masslessLagr}
\end{split}
\end{equation}
where, again, Chan--Paton factors are not displayed explicitly. In the massless limit
terms proportional to the masses are subleading, and only the gauge invariant part of the
coupling is left. Eq.~\eqref{masslessLagr} thus contains the explicit generating function
of all on--shell N\"oeter currents in momentum space,
\begin{multline}
j^{\, [0]\,\pm} (p_{\,1}\,,\,
p_{\,2}\,;\,\xi)\,=\,e^{\pm\,\sqrt{\frac{\a^{\,\prime}\!\!}{2}}\ \xi\cdot p_{\,12}}\
\exp\left(\pm\, \sqrt{\frac{\a^{\,\prime}\!\!}{2}}\ \xi_{\,\a}
\left[\partial_{\zeta_1}\cdot\partial_{\zeta_2}\,p^{\,\a}_{\,12}-2\,\partial^{\,\a}_{\zeta_1}\,
\partial_{\zeta_2}\cdot p_{\,1}
+\,2\,\partial^{\,\a}_{\zeta_2}\,\partial_{\zeta_1}\cdot p_{\,2}\right]\right)\\\times\,\phi_{\,1}\left(p_{\,1},\,\zeta_1\,\pm\,\sqrt{2\a^{\,\prime}}\,p_{\,2}\right)\,
\phi_{\,2}\left(p_{\,2},\,\zeta_2\,\mp\,\sqrt{2\a^{\,\prime}}\,p_{\,1}\right)\Bigg|_{\zeta_i\,=\,0}\ ,
\end{multline}
where for brevity we have left out all trace and/or de Donder terms. After a
Fourier transform, one is then led to the corresponding coordinate--space
currents
\begin{multline} \label{cons_current}
J^{\, [0] \,\pm} (x\,;\,\xi)\,=\, \exp\left(\mp\ i\,\sqrt{\frac{\a^{\,\prime}\!\!}{2}}\
\xi_{\,\a}
\left[\partial_{\zeta_1}\cdot\partial_{\zeta_2}\,\partial^{\,\a}_{\,12}-2\,\partial^{\,\a}_{\zeta_1}\,
\partial_{\zeta_2}\cdot \partial_{\,1}
+\,2\,\partial^{\,\a}_{\zeta_2}\,\partial_{\zeta_1}\cdot \partial_{\,2}\right]\right)\\\times\,\Phi_{\,1}\left(x\,\mp\,i\sqrt{\frac{\a^{\,\prime}\!\!}{2}}\ \xi\, ,\,\zeta_1\,\mp\,i\,\sqrt{2\a^{\,\prime}}\,\partial_{\,2}\right)\, \Phi_{\,2}\left(x\,\pm\,i\sqrt{\frac{\a^{\,\prime}\!\!}{2}}\ \xi \,,\,\zeta_2\,\pm\,i\,\sqrt{2\a^{\,\prime}}\,\partial_{\,1}\right)\Bigg|_{\zeta_i\,=\,0}\ ,
\end{multline}
that are conserved up to the massless Klein--Gordon equation. Again, one could
reinstate de Donder and/or trace terms, arriving at off--shell currents that are conserved up to the compensator equations of
\cite{minimal}
\be
\cF(\phi_{\,i}) \ - \ \frac{1}{2} \ \left(\xi \cdot \partial\right)^{\, 3}\, \a_i \ = \ 0 \ ,
\ee
where $\cF(\phi_{\,i})$ are the Fronsdal operators for $\phi_{\,i}$ and $\a_{\,i}$ are the corresponding compensators.

Interestingly, the formalism suggested by String Theory for totally symmetric
fields affords a natural extension to mixed--symmetry fields. Indeed, the
following gauge invariant generalization of \eqref{masslessGenfunc},
\be
\begin{split}
\cA^{\,\pm}\,=\,\exp&\left\{\pm\sqrt{\frac{\a^{\,\prime}\!\!}{2}}\,\left[(\partial_{\xi^1_1}+\ldots+\partial_{\xi^n_1})
\cdot(\partial_{\xi^1_2}+\ldots+\partial_{\xi^n_2})\right]
\left[(\partial_{\xi^1_3}+\ldots+\partial_{\xi^n_3})\cdot p_{\,12}\right]\,+\text{cyclic}\right\}\\&\times\,\phi_{\,1}
\left(p_{\,1},\,\xi^k_1\,\pm\,\sqrt{\frac{\a^{\,\prime}\!\!}{2}}\,p_{\,23}\right)
\phi_{\,2}\left(p_{\,2},\,\xi^k_2\,\pm\,\sqrt{\frac{\a^{\,\prime}\!\!}{2}}\,p_{\,31}\right)
\phi_{\,3}\left(p_{\,3},\,\xi^k_3\,\pm\,\sqrt{\frac{\a^{\,\prime}\!\!}{2}}\,p_{\,12}\right)\Bigg|_{\xi_i=0}\ ,
\end{split} \label{guess}
\ee
appears a natural generating function of massless cubic amplitudes for mixed--symmetry
fields, where the $\xi^k_i$ are the symbols associated to the $n$--th index family of the
$i$--th field and where Chan--Paton factors are left implicit, referring for brevity to
\eqref{ampl}. Again, up to mixed--symmetry de Donder--like and/or trace terms.

\vskip 36pt


\scs{Current Exchanges}\label{sec:exchanges}


In this section we elaborate upon the current exchange
amplitudes of \cite{minimal}, constructing generating functions for some instructive
cases. We also discuss how the results of
\cite{Bekaert:2009ud} can be extended to the case of mixed--symmetry fields, that is
particularly important for String Theory. In the next section these results
will be used to compute directly some interesting four--point amplitudes along the lines of \cite{taronna09}.

\vskip 24pt


\scss{Higher--Spin Current Exchanges}\label{sec:HSexchanges}


Our first task is obtaining a compact form for the generating
function of current exchange amplitudes for totally symmetric HS
Bose fields. In our mostly--positive convention for the space--time signature,
these propagators take in general the form
\begin{equation}
\cP^{(s)}_{\m_1\ldots\,\m_s;\n_1\ldots\,\n_s}\,=\,-\, \frac{1}{p^{\,2}+M^{2}}\
P_{\m_1\ldots\,\m_s\, ;\,\n_1\ldots\,\n_s} \ ,
\end{equation}
where the two sets of $s$ indices are associated to the incoming and outgoing
spin--$s$ currents. These propagators are uniquely determined by the
requirement that they lead to the proper current exchange amplitudes. This
requires that $P_{\m_1\ldots\,\m_s\,;\,\n_1\ldots\,\n_s}$ be a projector onto
traceless symmetric Lorentz tensors that is also symmetric under the
interchange of its two sets of indices. For massive particles, the irreducible
representations are associated to Young projections that are traceless and
transverse with respect to time--like momenta. On the other hand, for massless
particles the irreducible representations are traceless on the
$(d-2)$--dimensional space that is left over by the gauge symmetry.

In order to describe the propagators for totally symmetric spin--$s$ fields, let us
therefore introduce two sets of symbols, $\{\xi_\m\}$ and $\{\lambda_\m\}$, that are to
be associated to the Lorentz labels of the ingoing and outgoing modes. To begin with, let
us consider conserved currents, but as we shall see shortly one can extend relatively
simply the results to the more general case of external currents that are not conserved.
Moreover, relatively simple and natural modifications make it possible to adapt the
formalism to mixed--symmetry fields.

In the totally symmetric case, the propagator polynomial simplifies drastically
when it is sandwiched between a pair of transverse currents. Only traces can in
fact survive, and taking into account the in--out symmetry these give rise to
only two types of terms, $\xi^{\,2} \lambda^{\,2}$ and $\xi\cdot \lambda$.
Moreover, one is only to account for trace conditions in the transverse
subspace, since the symmetrization is automatically enforced by the commuting
nature of the $\xi$ and $\lambda$ symbols. In this formalism the trace
operators in the two symbol spaces coincide with the corresponding Laplace
operators, so that the trace conditions translate into the conditions that
$\widehat{\cP}^{\,(s)}$ be harmonic in $\xi$ and $\lambda$:
\begin{equation}
(\partial_\xi\cdot\partial_\xi)\ \widehat{\,\!\cP}^{\,(s)}(\xi,\lambda)\,=\,0\ ,
\qquad (\partial_\lambda\cdot\partial_\lambda)\ \widehat{\,\!\cP}^{\,(s)}(\xi,\lambda)\,=\,0\ . \label{trace_cond}
\end{equation}
All in all, $\widehat{\cP}^{\,(s)}$ is thus a harmonic polynomial that is homogeneous of
degree $s$ in both the $\xi$ and $\lambda$ symbols, as pertains to the nature of the
external currents. Moreover, these currents are conserved, as we have stressed,
and hence their mere presence enforces on the $\xi$ and $\lambda$ symbols a projection
onto the transverse $(d-2)$--dimensional subspace, that translates into the
conditions
\begin{equation}
(\partial_\xi\cdot\partial_\xi)\ \xi^{\,2}\,\equiv \, (\partial_\xi\cdot\partial_\xi)\ \xi_\m \, \eta_{\,T}^{\,\m\n}\, \xi_\m \, =\, 2\,(d-2) \ ,
\end{equation}
as pertains to the implicit presence of the transversely projected metric $\eta_{\,T}$ in their contractions.

One can naturally construct from the two basic monomials $\xi^{\,2}\, \lambda^{\,2}$ and $\xi\cdot \lambda$ the quantity
\begin{equation}
x\,=\,\cos\theta\,=\,\frac{\xi\cdot \lambda}{\sqrt{\xi^{\,2} \, \lambda^{\,2}}}\ , \label{y}
\end{equation}
that is homogeneous of degree zero and can be taken to define an angle $\theta$ in the
transverse space, so that a natural ansatz for a massless spin--$s$ propagator takes the
form
\begin{equation}
\widehat{\,\!\cP}^{\,(s)}(\xi,\lambda)\,=\,\frac{K}{p^{\,2}}\ \left(\xi^{\,2}\, \lambda^{\,2}\right)^{s/2}\ f\left(\frac{\xi\cdot \lambda}{\sqrt{\xi^{\,2} \, \lambda^{\,2}}}\right)\ , \label{eq_radial}
\end{equation}
where the exponent $s/2$ reflects the degree of homogeneity of $\widehat{\cP}^{(s)}$ and
$K$ is an overall normalization constant.

When applied to eq.~\eqref{eq_radial}, the Laplace operator results in the
ordinary differential equation
\begin{equation}
(1-x^2)\ f''(x)\, -\, (2\b+1)\,x\ f'(x)\,+\, s(s+2\b)\ f(x)\,=\,0\
\label{Laplace}
\end{equation}
for $f(x)$, where
\begin{equation}
\b\,=\,\frac{d}{2}-2\ .
\end{equation}
For $\b>0$ eq.~\eqref{Laplace} admits the polynomial solution
\begin{equation}
f^{[\b]}_s(x)\,=\,C_s^{\,[\b]}(x)\ ,
\end{equation}
where $C_s^{\,[\b]}(x)$ is a Gegenbauer polynomial, while for $\b\,=\,0$, that
corresponds to $d\,=\,4$, it admits the polynomial solution
\begin{equation}
f^{[0]}_s(x)\,=\,T_s(x)\ ,
\end{equation}
where $T_s(x)$ a Chebyshev polynomial of the first kind. It is now convenient
to fix the value of $K$ in such a way that the coefficient of the monomial
$(\xi\,\cdot\, \l)^{\,s}$ in the propagator be $1/s!$, since this simplifies the
$\star$--products. As a result, for $\b>0$
\begin{equation}
K\,=\,\frac{1}{2^s}\ \frac{\Gamma(\b)}{\Gamma(\b+s)}\ ,
\end{equation}
while for $\b\,=\,0$
\begin{align}
K&\,=\,\frac{1}{s!}\ 2^{\,-s+1}\ ,& s&\,\geq\, 1\ ,\\
K&\,=\,1\ ,& s&\,=\,0\ .
\end{align}

For totally symmetric spin--$s$ fields the portion of the massless propagator
that contributes to the exchanges for conserved currents can thus be cast in
the form
\begin{equation}
\widehat{\,\!\cP}^{\,(s)}(\xi,\lambda)\,=\,-\, \frac{K}{p^{\,2}}\ \left(\xi^{\,2}\,\lambda^{\,2}\right)^{s/2}f^{[\b]}_s\left(\frac{\xi\cdot \lambda}{\sqrt{\xi^{\,2}\, \lambda^{\,2}}}\right)\ ,
\end{equation}
where $f_s^{[\b]}(x)$ is the polynomial solution of \eqref{Laplace}. To
reiterate, within a spin--$s$ propagator one can distinguish a radial part,
that simply reflects the degree of homogeneity and thus the spin, and an
angular part that for totally symmetric fields in $d$ dimensions is a
Gegenbauer or Chebyshev polynomial.

For mixed--symmetry \emph{reducible} fields of the type $\phi_{\m_1 \ldots
\m_{s_1};\, \n_1 \ldots \n_{s_2};\ldots}$, fully symmetric under interchanges
of indices belonging to the various families but with no further symmetries
relating the various families, the trace conditions \eqref{trace_cond} are
naturally replaced by the constraints
\begin{equation}
\partial_{\xi_{\,i}}\cdot\partial_{\xi_{\,j}}\ \widehat{\,\!\cP}\left(\{\xi_m,\lambda_{\,n}\}\right)\,=\,0\ ,\qquad \partial_{\lambda_{\,i}}\cdot\partial_{\lambda_{\,j}}\ \widehat{\,\!\cP}\left(\{\xi_m,\lambda_{\,n}\}\right)\,=\,0\ ,\ \ \forall\ i,j\ , \label{laplace_n}
\end{equation}
that eliminate all traces within the various families and all mixed traces
involving pairs of them. Remarkably, the generating functions for the
propagators of mixed--symmetry fields bearing a total of $s$ indices,
\begin{multline}
\widehat{\,\!\cP}_s(\{\xi_m,\lambda_{\,n}\})\,=\,-\,\frac{K^{\b}_{\{s\}}}{p^{\,2}}\
\left[(\xi_{\,1}+\ldots+\xi_{\,N})^{\,2}(\lambda_{\,1}+\ldots
+\lambda_{\,N})^{\,2}\right]^{s/2}\\ \times f^{[\b]}_s\,
\left(\frac{(\xi_{\,1}+\ldots+\xi_{\,N})\cdot (\lambda_{\,1}+\ldots
+\lambda_{\,N})}{\sqrt{(\xi_{\,1}+\ldots+\xi_{\,N})^{\,2} (\lambda_{\,1}+\ldots
+\lambda_{\,N})^{\,2}}}\right)\ ,\label{Mixed}
\end{multline}
are then relatively simple polynomial solutions of eq.~\eqref{laplace_n}. The monomials
$\xi_{\,1}^{s_1}\ldots \xi_{\,N}^{s_N}\,\lambda_{\,1}^{s_1}\ldots \lambda_{\,N}^{s_N}$
with $s\,=\,s_1+\ldots +s_N$ that are present in this expression encode the more
complicated fully traceless projectors for all types of transverse tensors with $N$ index
families. For instance, for two--family tensors of type $(s_1,s_2)$ the propagator
polynomial that extracts from eq.~\eqref{Mixed} the current--exchange amplitude for the
corresponding conserved currents is
\begin{multline}
\widehat{\,\!\cP}^{\,(s_1,s_2)}(\xi_{\,1},\xi_{\,2},\lambda_1,\lambda_2)\,=\,-\, \frac{K^{\b}_{s_1,s_2}}{p^{\,2}}\ (\partial_{\rho_{\,1}}\,\partial_{\sigma_{\,1}})^{\,s_1}(\partial_{\rho_{\,2}}\,\partial_{\sigma_{\,2}})^{\,s_2} \\ \times\,\left\{\left[(\rho_{\,1}\,\xi_{\,1}+ \rho_{\,2}\,\xi_{\,2})^{\,2}(\sigma_{\,1}\,\lambda_{\,1}+\sigma_{\,2}\,\lambda_{\,2})^{\,2}\right]^{\frac{s_1+s_2}{2}}\right.\\ \left.\times f^{[\b]}_{s_1+s_2}\left(\frac{(\rho_{\,1}\,\xi_{\,1}+\rho_{\,2}\, \xi_{\,2})\cdot (\sigma_{\,1}\,\lambda_{\,1}+\sigma_{\,2}\,\lambda_{\,2})}{\sqrt{(\rho_{\,1}\,\xi_{\,1}+\rho_{\,2}\,\xi_{\,2})^{\,2} (\sigma_{\,1}\,\lambda_{\,1}+\sigma_{\,2}\,\lambda_{\,2})^{\,2}}}\right)\right\}\ ,
\end{multline}
where it is actually not necessary to let $\rho_{\,i}\,=\,\sigma_{\,i}\,=\,0$,
since the result is automatically independent of $\rho$ and $\sigma$. Similar
relations would hold for the propagators of fields bearing arbitrary numbers of
index families.

The results obtained so far are only valid for massless particles, but they can
be simply extended to the massive case provided the currents are still
conserved. This can be attained via the formal substitution $d\ra d+1$ or
$\b\ra\b+\frac{1}{2}\,$, that amounts to referring to the massive little group
$SO(d-1)$ rather than to its massless counterpart $SO(d-2)$. In this fashion
one can also obtain the massive totally symmetric HS propagators
\begin{equation}
\widehat{\cP}^{(s)}\,=\,-\, \frac{1}{p^{\,2}+M_s^{\,2}}\ \left\{K\ \left(\xi^{\,2}\, \lambda^{\,2}\right)^{s/2}f^{[\b+\frac{1}{2}]}_s\left(\frac{\xi\cdot \lambda}{\sqrt{\xi^{\,2}\, \lambda^{\,2}}}\right)\right\}\ ,
\end{equation}
and corresponding expressions that apply to the mixed--symmetry case.

If the external currents are not conserved, or equivalently are not transverse with respect to the exchanged momentum, the propagators must be modified in such a way that they effectively project them onto their transverse parts. One can account very conveniently for this complication in the present formalism: in order to project the currents, it suffices to project the symbols $\xi_{\,i}$ and $\lambda_{\,i}$ according to
\begin{equation}
\xi_{\,i}\ra \hat{\xi}_{\,i}\,=\,\xi_{\,i}\, +\, p\ \frac{\xi_{\,i}\cdot p}{M^2}\ ,\qquad \lambda_{\,i}\ra \hat{\lambda}_{\,i}\,=\,\lambda_{\,i}\, +\, p\ \frac{\lambda_{\,i}\cdot p}{M^2}\ ,
\end{equation}
and performing the substitutions $\xi_{\,i}\ra \hat{\xi}_{\,i}$ and $\lambda_{\,i}\ra \hat{\lambda}_{\,i}$ in the previous expressions one can recover the massive propagators. For totally symmetric fields the general result is
\begin{equation}
\widehat{\!\,\cP}^{(s)}\,=\,-\, \frac{1}{p^{\,2}+M_s^{\,2}}\left\{K\left(\hat{\xi}^{\,2}\,\hat{\lambda}^{\,2}\right)^{s/2} f^{[\b+\frac{1}{2}]}_s\left(\frac{\hat{\xi}\cdot
\hat{\lambda}}{\sqrt{\hat{\xi}^{\,2}\, \hat{\lambda}^{\,2}}}\right)\right\}\ ,
\end{equation}
while similar relations, with $\xi_{\,i}\ra\, \hat{\xi\!}_{\,i}$ and
$\lambda_{\,i}\ra \hat{\lambda}_{\,i}$, hold in the mixed--symmetry case.

Finally, one can consider the generating function of totally symmetric HS propagators
\begin{equation}
\widehat{\,\!\cP}\,=\,-\,\sum_{s\,=\,0}^{\infty}\ \frac{1}{p^{\,2}+M_s^{\,2}}\left\{K^{\,\b(s)}_{s} \left(\hat{\xi}^{\,2}\,\hat{\lambda}^{\,2}\right)^{s/2}f^{[\b(s)]}_s\left(\frac{\hat{\xi}\cdot
\hat{\lambda}}{\sqrt{\hat{\xi}^{\,2}\, \hat{\lambda}^{\,2}}}\right)\right\}\ ,\label{propSum1}
\end{equation}
where
\begin{equation}
\b(s)\,=\,\frac{d}{2}\, -\, 2
\end{equation}
for the massless spin--$s$ particles, and
\begin{equation}
\b(s)\,=\,\frac{d+1}{2}\, -\, 2
\end{equation}
for the massive ones, while $f^{[\b]}_s(x)$ and $K_s^{\,\b}$ are defined as
before. For the application to the string currents computed in the previous
section, or to currents whose definitions include the coupling constants, one
can confine the attention to eq.~\eqref{propSum1}. In general, however, it is
possible to consider, as in \cite{Bekaert:2009ud}, arbitrary spin--dependent
coupling constants $a_s>0$, that can be conveniently grouped in a generating
function
\begin{equation}
a(z)\,=\,\sum_{s\,=\,0}^{\infty}\frac{a_s}{s!} \  z^s\ ,\label{cau}
\end{equation}
so that
\begin{equation}
\widehat{\cP}(a)\,=\,-\,\sum_{s\,=\,0}^{\infty}\frac{a_s}{p^{\,2}+M_s^{\,2}}\left\{K^{\b(s)}_{s}\,
\left(\hat{\xi}^{\,2}\,\hat{\lambda}^{\,2}\right)^{s/2}f^{[\b(s)]}_s\left(\frac{\hat{\xi}
\cdot \hat{\lambda}}{\sqrt{\hat{\xi}^{\,2}\, \hat{\lambda}^{\,2}}}\right)\right\}\ .
\end{equation}

\vskip 24pt


\scss{Generating Function for Massless and Massive Exchanges}\label{sec:ExchangeGenFunc}


In this section we begin by considering the case in which
only massless HS modes take part in the current exchanges, so that the generating function for totally symmetric fields reduces to
\begin{equation}
\widehat{\,\!\cP}(a)\,=\,-\, \sum_{s\,=\,0}^{\infty}\frac{a_{s}}{p^{\,2}}\left\{K^{\b}_{s}\left(\xi^{\,2}\,\lambda^{\,2}\right)^{s/2}f^{[\b]}_s
\left(\frac{\xi\cdot\lambda}{\sqrt{\xi^{\,2}\, \lambda^{\,2}}}\right)\right\}\ .\label{Propsum}
\end{equation}
We then turn our attention to an infinite set of massive modes corresponding to the first Regge trajectory of the open bosonic string. In both cases we begin with two special choices, $d\,=\,4$ for massless particles and $d\,=\,3$ for massive ones. Both lead in fact to $\b\,=\,0$, so that \eqref{Propsum} simplifies considerably and takes the form
\begin{equation}
\widehat{\,\!\cP}(a)\,=\,-\, \sum_{s\,=\,0}^{\infty}\frac{1}{p^{\,2}+M_s^{\,2}}\left\{\frac{2 a_s}{s!}\left(\frac{\xi^{\,2}\, \lambda^{\,2}}{4}\right)^{s/2}T_s\left(\frac{\xi\cdot \lambda}{\sqrt{\xi^{\,2} \, \lambda^{\,2}}}\right)-a_0\right\}\ ,\label{D4}
\end{equation}
with vanishing masses in the first case. We then illustrate how this result
changes if $\b>0$, i.e. when additional space--time dimensions are present.


\scsss{$\b\,=\,0$} \label{sec:alpha0}


In this case one can sum the series resorting to the generating function of the Chebyshev polynomials
\begin{equation}
\cT(y,t)\,=\,\sum_{s\,=\,0}^{\infty}T_s(y)\,t^{\,s}\,=\,\frac{1-y\,t}{1-2\,y\,t+t^{\,2}}\ ,\label{g1}
\end{equation}
with $y$ as in eq.~\eqref{y}, observing that the sum in \eqref{D4}, with $a_s\,=\,1$ for all $s$, is precisely of the form
\begin{equation}
f(y,t)\,=\,\sum_{s\,=\,0}^{\infty}\frac{1}{s!}\ T_s(y)\,t^{\,s}\ .\label{g2}
\end{equation}
Inserting in \eqref{g2} the Hankel representation for the reciprocal of the Euler $\G$ function
\begin{equation}
\frac{1}{\Gamma(q)}\,=\,\frac{1}{2\pi i}\oint_C e^z z^{-q}\, dz\ ,\label{Hankel}
\end{equation}
where the contour $C$ starts and ends at $-\infty$ in the $z$--plane and leaves the origin to its left, gives
\be
\sum_{s\,=\,0}^{\infty}\frac{1}{\Gamma(s+1)}\ T_s(y)\,t^{\,s}\,=\,\frac{1}{2\pi i}\oint_C dz\ e^z\frac{z-y\,t}{z^{\,2}-2\,y\,z\,t+t^{\,2}}\ .
\ee
This integral is simply determined by the residues at
\begin{equation}
z_{\pm}\,=\,y\,t\,\pm\, t\,\sqrt{y^2-1}\ ,
\end{equation}
and the final result is thus
\begin{equation}
f(y,t)\,=\,\sum_{s\,=\,0}^{\infty}\frac{1}{s!}\ T_s(y)\,t^{\,s}\,=\,\frac{1}{2}\left(e^{(y+\sqrt{y^2-1})t}+e^{(y-\sqrt{y^2-1})t}\right)\ .\label{eq}
\end{equation}

One can also address along these lines the case of generic $a_s$ couplings, as in eq.~\eqref{cau}, considering the generating function
\begin{equation}
g(y,t)\,=\,\sum_{s\,=\,0}^{\infty}\frac{a_s}{s!}\ T_s(y)\,t^{\,s}\ .
\end{equation}
The Fourier Transform of $f(y,t)$ with respect to $t$ is now
\begin{equation}
\tilde{f}(y,q)\,=\,\frac{1}{2}\left[e^{i\left(y+\sqrt{y^2-1}\, \right)\partial_q}+e^{i\left(y-\sqrt{y^2-1}
\, \right)\partial_q}\right] \, \sqrt{2\pi}\ \delta(q)\ ,
\end{equation}
and resorting the identity
\begin{equation}
\sum_s\,\frac{a_s}{s!}\ T_s(y)\,t^{\,s}\,=\,\int \frac{dz}{\sqrt{2\pi}}\
a(i\,z\,t)\ \tilde{f}(y,z)\
\end{equation}
finally leads to
\begin{equation}
g(y,t)\,=\,\frac{1}{2}\left[a(t\,y+t\sqrt{y^2-1})+a(t\,y-t\sqrt{y^2-1})\right]\ ,
\end{equation}
where $a(z)$ is defined in eq.~\eqref{cau}. Hence, the generating function for
massless totally symmetric exchanges \eqref{D4} takes the form
\begin{equation}
\widehat{\cP}\,=\,-\frac{1}{p^{\,2}}\left[a\left(\frac{1}{2}\ \xi\cdot \lambda\,+\,\frac{1}{2}\ \sqrt{(\xi\cdot \lambda)^{\,2}\,-\,\xi^{\,2}\,\lambda^{\,2}}\right)+a\left(\frac{1}{2}\ \xi\cdot \lambda\, -\, \frac{1}{2}\ \sqrt{(\xi\cdot \lambda)^{\,2}-\xi^{\,2}\,\lambda^{\,2}}\right)-a_0\right]\ ,\label{massless4}
\end{equation}
and for the special choice $a(z)\,=\,e^z$ reduces to \eqref{eq}. This result
was recently presented by Bekaert, Joung and Mourad in \cite{Bekaert:2009ud},
with reference to the scalar currents \eqref{scalar} that were identified by
Berends, Burgers and van Dam in \cite{bbvd}. Here we have recovered it starting
from the expression \eqref{Propsum} for the propagator in terms of the symbols,
that can be directly of use in more general cases. To begin with, in three
dimensions and with an infinite sequence of HS modes whose masses are spaced as
in the first Regge trajectory of the bosonic string, so that
\be
M_s^{\,2} \, =\, \frac{s-1}{\a^{\,\prime}} \, ,
\ee
one is led to a relatively handy result that can be obtained starting
from \eqref{massless4},
\be
\begin{split}
\widehat{\,\!\cP}\,=\,\a^{\,\prime}\int_0^1 dy\,y^{-\a^{\,\prime}
s-2}&\left[\,a\left(\frac{y}{2}\ \xi\cdot \lambda+\frac{y}{2}\
\sqrt{(\xi\cdot \lambda)^{\,2}-\xi^{\,2}\
\lambda^{\,2}}\right)\right.\\&+\left.a\left(\frac{y}{2}\ \xi\cdot
\lambda-\frac{y}{2}\ \sqrt{(\xi\cdot \lambda)^{\,2}-\xi^{\,2}\
\lambda^{\,2}}\right)-a_0\right]\ .
\end{split}
\ee
%


\scsss{$\b>0$}


The cases with $\b>0$ can be similarly addressed, but the end result is more cumbersome, since the generating function of all HS exchanges is given by
\begin{equation}
\widehat{\,\!\cP}\,=\,\sum_{s\,=\,0}^{\infty}\,-\frac{a_s}{p^{\,2}+M_s^{\,2}}\left\{\frac{\Gamma(\b)}{\Gamma(\b+s)} \left(\frac{\xi^2\,\lambda^2}{4}\right)^{s/2}\ G^{\,[\b]}_s\left(\frac{\xi\cdot \lambda}{\sqrt{\xi^2\, \lambda^2}}\right)\right\}\ ,\label{Propsum2}
\end{equation}
where the $G_s^{\,[\b]}(x)$ are Gegenbauer polynomials. As before, one can start from their generating function,
\begin{equation}
\cG^{\,[\b]}(x,t)\,=\,\sum_{s\,=\,0}^{\infty}\ G_s^{\,[\b]}(x)\, t^{\,s}\,=\,\frac{1}{(1-2\,x\,t+t^{\,2})^{\,\b}}\ ,\label{cG}
\end{equation}
and in fact if all $a_s\,=\,1$ for massless exchanges the sum over $s$
in \eqref{Propsum2} is exactly of the form
\begin{equation}
k^{[\b]}(x,t)\,=\,\sum_{s\,=\,0}^{\infty}\frac{1}{\Gamma(\b+s)}\ G_s^{\,[\b]}(x)\, t^{\,s}\ .
\end{equation}
The $k^{[\b]}(x,t)$ can be computed resorting as before to the Hankel contour
integral \eqref{Hankel}, with the end result
\be
k^{[\b]}(x,t)\,=\,\frac{1}{2\pi i}\oint_C dz\ \frac{e^{z}\, z^{\b}}{(z^2-2\,x\,t\,z+t^2)^{\,\b}}\ .
\ee
The massless propagator function is thus
\begin{equation}
\widehat{\,\!\cP}\,=\,-\ \frac{1}{p^{\,2}}\ \frac{\Gamma(\b)}{2\pi i}\oint_C dz\frac{e^z z^\b}{\left(z^2-\xi\cdot \lambda\, z+\frac{1}{4}\ \xi^{\,2}\, \lambda^{\,2} \right)^\b}\ ,\label{SP1}
\end{equation}
while proceeding as in Section \ref{sec:alpha0} the massive propagator
function for an infinite set of states lying along the first Regge trajectory
of the open bosonic string is
\begin{equation}
\widehat{\cP}\,=\,\a^{\,\prime}\int_0^1 dy\,
y^{-\a^{\,\prime}s-2}\,\frac{\Gamma(\b)}{2\pi i}\oint_C dz\, \frac{e^z \,
z^\b}{\left(z^2-y\,\xi\cdot \lambda\, z\, +\,\frac{1}{4}\ y^{2}\, \xi^{\,2}\, \lambda^{\,2}
\right)^\b}\ , \label{SP2}
\end{equation}
where the value $\b\,=\,23/2$ corresponds to $d=26$. In order to address the
general case with arbitrary coupling constants $a_s$, one can proceed again as
in Section \ref{sec:alpha0}. However, the non--exponential terms do not have a
simple operator meaning, and the end result
\begin{equation}
\begin{split}
K^{[\b]}(x,t)&\,=\,\sum_{s\,=\,0}^{\infty} a_s \ \frac{\Gamma(\b)}{\Gamma(\b+s)}\ G_s^{[\b]}(x)\, t^s\,=\,e^{(t\,\partial_{u_1}\partial_{u_2})}a(u_1)\ \cG^{[\b]}(x,u_2)\Big|_{u_1\,=\,u_2\,=\,0}\\&\,=\,
\int\frac{du}{\sqrt{2\pi}}\ \tilde{a}(u)\ {\cG}^{[\b]}(x,\,i\,u\,t)\\
&\,=\,\int \,\frac{du_1\, du_2}{{2\pi}}\,\frac{{b}(u_1)\,e^{-iu_1u_2}}{(1-2\,i\,x\,u_2\,t-u_2^2\,t^2)^\b}\ ,
\end{split}
\end{equation}
where $\cG^{[\b]}(x,u_2)$ is defined in eq.~\eqref{cG},
is rather involved, since $b(z)$ is defined via the series
\be
b(z)\,=\,\sum_{s\,=\,0}^{\infty}\frac{\Gamma(\b)}{\Gamma(\b+s)}\,\frac{a_s}{s!}\ z^{\,s}\ .
\ee
All in all, the propagator generating function for massless exchanges reads
\begin{equation}
\widehat{\,\!\cP}\,=\,-\, \frac{1}{p^{\,2}}\int\frac{du_1\,du_2}{2\pi}\,\frac{{b}(u_1)\,
e^{-iu_1u_2}}{\left(1-i\,\xi\cdot \lambda\ u_2- \frac{1}{4}\ \xi^{\,2}\  \lambda^{\,2}\, u_2^2\right)^\b}\ ,
\end{equation}
while in the massive case with the spectrum of the first Regge trajectory one
obtains
\begin{equation}
\widehat{\,\!\cP}\,=\,\a^{\,\prime}\int_0^1 dy\,y^{-\a^{\,\prime}s-2}\int\frac{du_1 \, du_2}{{2\pi}}\, \frac{{b}(u_1)\,e^{-iu_1u_2}}{\left(1-iy\, \xi\cdot \lambda\ u_2-\frac{1}{4} \ y^2\,\xi^2\, \lambda^2\, u_2^2\right)^\b}\ .
\end{equation}

Actually, in \eqref{SP1} for massless particles in even space--time dimensions the $z$--integral has only poles, so that one can evaluate it summing residues, with the end result
\begin{equation}
k^{[\b]}(x,t)\,=\,\frac{1}{\Gamma(\b)}\left[\partial_z^{\,\b-1}\left(\frac{e^z z^\b}{(z-z_-)^\b}\right)\Big|_{z\,=\,z_+}\, +\, \partial_z^{\,\b-1}\left(\frac{e^z z^\b}{(z-z_+)^\b}\right)\Big|_{z\,=\,z_-}\right]\ ,\label{resul}
\end{equation}
where
\begin{equation}
z_{\pm}\,=\,x\,t\,\pm\, t\,\sqrt{x^2-1}\ .
\end{equation}
One can also simplify eq.~\eqref{resul} resorting to the identity
\begin{equation}
\frac{d^{\,n}}{dz^n}\ \left[e^zg(z)\right] \,=\,e^z\sum_{i\,=\,0}^{n}\binom{n}{i}\,g^{\,(i)}(z)\ ,
\end{equation}
with
\be
g^{\,(i)}\,=\,\frac{d^{\,i}}{dz^i}\,g(z)\ ,
\ee
thus obtaining
\be
\begin{split}
k^{[\b]}(x,t)\,=\,\frac{1}{\Gamma(\b)}&\left[\ e^{\,x\,t\,+\, t\,\sqrt{x^2-1}}\ \sum_{i\,=\,0}^{\b-1}\binom{\b-1}{i}\, g_{\,-}^{(i)}\left(x\,t+t\sqrt{x^2-1}\right)\right.\\&\!\!\!+\left.\,e^{\,x\,t \,-\, t\,\sqrt{x^2-1}}\ \sum_{i\,=\,0}^{\b-1}\binom{\b-1}{i}\, g_{\,+}^{(i)}\left(x\,t \,-\, t\sqrt{x^2-1}\right)\right]\ ,
\end{split}
\ee
where we have defined
\begin{equation}
g_{\,\pm}(z)\,=\,\left(\frac{z}{z-z_\pm}\right)^\b\ ,
\end{equation}
and we have pulled out an exponential factor that dominates the asymptotic behavior of HS amplitudes, so that in this case the propagator function takes finally the compact form
\be
\begin{split}
\widehat{\,\!\cP}\,=\,-\frac{1}{p^{\,2}}&\left[\,e^{\,\frac{1}{2}\, \xi\,\cdot\, \lambda+\frac{1}{2}\,\sqrt{(\xi\,\cdot\, \lambda)^2-\xi^2\,\lambda^2}}\ \sum_{i\,=\,0}^{\b-1}\binom{\b-1}{i}\,g_{\,-}^{\,(i)}\left(\frac{1}{2}\ \xi\cdot \lambda+\frac{1}{2}\sqrt{(\xi\cdot\lambda)^{\,2}-\xi^{\,2}\,\lambda^{\,2}}\right) \right.\\&\!\!\!\!+\left.\, e^{\,\frac{1}{2}\,\xi\,\cdot\, \lambda-\frac{1}{2}\,\sqrt{(\xi\,\cdot \,\lambda)^{2}-\xi^{2}\,\lambda^{2}}}\ \sum_{i\,=\,0}^{\b-1}\binom{\b-1}{i}\,g_{\,+}^{\,(i)}\left(\frac{1}{2}\ \xi\cdot \lambda-\frac{1}{2}\sqrt{(\xi\cdot \lambda)^{\,2}-\xi^{\,2}\,\lambda^{\,2}}\right)\right]\ .
\end{split}
\ee

We can conclude this section with some generating functions for $d>4$, where
special choices for the coupling constants lead to instructive and yet
relatively simple results. The simplest choice is
\begin{equation}
a(t)\,=\,\frac{1}{(1-t)^{\,\b}}\,=\,\sum_{s\,=\,0}^\infty \frac{1}{s!}\frac{\Gamma(\b+s)}{\Gamma(\b)}\ t^s\ , \label{hyper_a}
\end{equation}
that leads to
\begin{equation}
\widehat{\,\!\cP}\,=\,-\frac{1}{p^{\,2}}\left(1-\xi\cdot \lambda+\frac{\xi^2\, \lambda^2}{4}\right)^{-\b}\ ,\label{P1}
\end{equation}
in the massless case, and to
\begin{equation}
\widehat{\,\!\cP}\,=\,\a^{\,\prime}\int_0^1dy\,y^{-\a^{\,\prime}s-2}\left(1-\xi\cdot
\lambda\,y+\frac{\xi^2\, \lambda^2\,y^2}{4}\right)^{-\b}\ \label{P2}
\end{equation}
for massive exchanges corresponding to the first Regge trajectory of the open
bosonic string. Here $s$ is one of the Mandelstam variables, and the results
obtained are valid in any space--time dimension $d$, with the only proviso that
$d\,=\,2\b+4$ in the massless case and $d\,=\,2\b+3$ in the massive one.

Another instructive choice is
\begin{equation}
a(t)\,=\,\,_1 F_1(\b;1;t)\,=\,\sum_{s\,=\,0}^\infty \frac{1}{s!}\frac{\Gamma(\b+s)}{s!\,\Gamma(\b)}\,t^{\,s}\ ,
\end{equation}
with $_1 F_1$ a confluent hypergeometric function, that simplifies a bit for
massive particles in four dimensions or for massless particles in five
dimensions. In fact in these cases
\begin{equation}
a(t)\,=\,e^{\frac{t}{2}}\ I_0\left(\frac{t}{2}\right)\ ,
\end{equation}
with $I_0$ a modified Bessel function. It is then possible to obtain the rather
compact expression
\begin{equation}
\widehat{\,\!\cP}\,=\,-\frac{1}{p^{\,2}}\,e^{\frac{1}{2}\,\xi\,\cdot\, \lambda}\ I_0\left(\frac{1}{2}\sqrt{(\xi\cdot \lambda)^{\,2}-\xi^{\,2}\,\lambda^{\,2}}\right)\label{P3}
\end{equation}
for the propagator function for massless particles in five dimensions, or
\begin{equation}
\widehat{\,\!\cP}\,=\,\a^{\,\prime}\int_0^1dy\,y^{-\a^{\,\prime}s-2}\ e^{\frac{y}{2}\,\xi\,\cdot\, \lambda}\ I_0\left(\frac{y}{2}\sqrt{(\xi\cdot \lambda)^{\,2}-\xi^{\,2}\,\lambda^{\,2}}\right)\label{P4}
\end{equation}
for an infinite set of four--dimensional particles whose masses correspond to the first Regge trajectory of the open bosonic string.

\vskip 36pt


\scs{Some Field Theory Amplitudes}\label{sec:QFTamplitudes}


In this section we analyze in more detail the string couplings and the
corresponding currents that we have identified. In particular, we compute on
the field theory side HS current exchanges and tree--level four--point scattering
amplitudes in which infinitely many higher--spin particles are interchanged,
with special emphasis on their high--energy behavior.

Let us begin by considering the relatively simple case of massless HS particles in four dimensions, before turning to some simple examples involving massive fields or dimensions larger than four. Our starting point, as before, is the propagator function
\begin{equation}
\widehat{\,\!\cP}\,=\,-\,\frac{1}{p^{\,2}}\left[a\left(\frac{1}{2}\ \xi\cdot \lambda+\frac{1}{2}\,\sqrt{(\xi\cdot \lambda)^{\,2}-\xi^{\,2}\,\lambda^{\,2}}\right)+a\left(\frac{1}{2}\, \xi\cdot \lambda-\frac{1}{2}\sqrt{(\xi\cdot \lambda)^{\,2}-\xi^{\,2}\,\lambda^{\,2}}\right)-a_0\right]\ ,
\end{equation}
while the conserved currents of interest are given in eq.~\eqref{J20} for the
case of a pair of spin--1 fields, and more generally
\begin{equation}
j_{\,1}^{\,\pm}(\xi)\,=\,\exp\left(\pm\sqrt{\frac{\a^{\,\prime}\!\!}{2}}\ p_{\,12}\cdot \xi\right)\phi_{\,1}\left(p_{\,1},\pm\,\sqrt{2\a^{\,\prime}}\ p_{\,2}\right)\,\phi_{\,2}\left(p_{\,2},\mp\,\sqrt{2\a^{\,\prime}}\ p_{\,1}\right)\ ,\label{J30}
\end{equation}
if we confine the attention to the highest derivative terms built from a pair of fields $\phi_{\,1}$ and $\phi_{\,2}$ of spins $s_1$ and $s_2$. As we have stressed, the highest derivative terms are conserved for any spin. Furthermore, they reduce to eq.~\eqref{J10} if $s_1$ and $s_2$ are both equal to one.

In terms of the scalar product \eqref{contra} the general form of the amplitudes is
\begin{equation}
\cA\,=\,\cJ(\xi)\star\widehat{\,\!\cP}(\xi,\lambda)\star\cJ(\lambda)\ ,
\end{equation}
and for simplicity we can confine our attention to the $s$--channel contribution, since the others can be obtained from it by simple redefinitions of the Mandelstam variables.

Considering the amplitude with a pair of $j_{\,1}^{\,\pm}$ one can compute the
$s$--channel contribution observing that the current has an exponential dependence on
$\xi$ and $\lambda$. As a result, the $\star$--product simply effects in the propagator
function the substitutions
\begin{equation}
\begin{split}
\xi\,\cdot\, \lambda\ \lra&\ \ \frac{\a^{\,\prime}\!\!}{2}\,p_{\,12}\cdot p_{\,34}\,=\,\frac{\a^{\,\prime}\!\!}{2}\,(u-t)\ ,\\
\xi^2\ \lra&\ \ \frac{\a^{\,\prime}\!\!}{2}\,p_{\,12}\cdot p_{\,12}\,=\,\frac{\a^{\,\prime}\!\!}{2}\,s\,=\,-\frac{\a^{\,\prime}\!\!}{2}\,(u+t)\ ,\\
\lambda^2\ \lra&\ \ \frac{\a^{\,\prime}\!\!}{2}\,p_{\,34}\cdot p_{\,34}\,=\,\frac{\a^{\,\prime}\!\!}{2}\,s\,=\,-\frac{\a^{\,\prime}\!\!}{2}\,(u+t)\ , \label{subst_prop}
\end{split}
\end{equation}
so that the $s$--channel amplitude becomes
\be
\begin{split}
\cA^{(s)}\,&=\,-\frac{1}{\a^{\,\prime}s}\left[a\left(\frac{\a^{\,\prime}}{4}\,(u-t)+\frac{\a^{\,\prime}\!\!}{2}\, \sqrt{-ut}\right)+ a\left(\frac{\a^{\,\prime}}{4}\,(u-t)-\frac{\a^{\,\prime}\!\!}{2}\,\sqrt{-ut}\right)-a_0\right]\\&\times \phi_{\,1}\left(p_{\,1},\pm\,\sqrt{2\a^{\,\prime}}\,p_{\,2}\right) \phi_{\,2}\left(p_{\,2},\mp\,\sqrt{2\a^{\,\prime}}\,p_{\,1}\right) \phi_{\,3}\left(p_{\,1},\pm\,\sqrt{2\a^{\,\prime}}\,p_{\,4}\right)
\phi_{\,4}\left(p_{\,2},\mp\,\sqrt{2\a^{\,\prime}}\,p_{\,3}\right)\ .
\end{split}
\ee
In this case the dominant behavior is the same for all types of external states, and is determined by the factor
\begin{equation}
a\left(\frac{\a^{\,\prime}}{4}(u-t)+\frac{\a^{\,\prime}\!\!}{2}\sqrt{-ut}\right)+a\left(\frac{\a^{\,\prime}}{4}(u-t)-\frac{\a^{\,\prime}\!\!}{2}\sqrt{-ut}\right)-a_0\ .
\end{equation}
If $a(z)\,=\,e^z$ this result maintains in even space--time dimensions $d>4$ the relatively simple form
\begin{equation}
\begin{split}
\cA^{(s)}\,=\,-\frac{1}{\a^{\,\prime} s}&\left[e^{\frac{\a^{\,\prime}\!\!}{4}\,(u-t)+\frac{\a^{\,\prime}\!\!}{2}\,\sqrt{-ut}}\, \sum_{i\,=\,0}^{\b-1}\binom{\b-1}{i}\,g_{\,-}^{(i)} \left(\frac{\a^{\,\prime}\!}{4}\,(u-t)+\frac{\a^{\,\prime}\!\!}{2}\,\sqrt{-ut}\right)\right.\\
&\!\!\!\!\!\left.+\ e^{\frac{\a^{\,\prime}\!\!}{4}\,(u-t)-\frac{\a^{\,\prime}\!\!}{2}\,\sqrt{-ut}} \,\sum_{i\,=\,0}^{\b-1}\binom{\b-1}{i}\,g_{\,+}^{(i)} \left(\frac{\a^{\,\prime}\!}{4}\,(u-t) -\frac{\a^{\,\prime}\!\!}{2}\,\sqrt{-ut}\right)\right]\\
&\times \phi_{\,1}\left(p_{\,1},\pm\,\sqrt{2\a^{\,\prime}}\,p_{\,2}\right)
\phi_{\,2} \left(p_{\,2},\mp\,\sqrt{2\a^{\,\prime}}\,p_{\,1}\right)\\ &\times\phi_{\,3}\left(p_{\,1},\pm\,\sqrt{2\a^{\,\prime}}\,p_{\,4}\right)
\phi_{\,4}\left(p_{\,2},\mp\,\sqrt{2\a^{\,\prime}}\,p_{\,3}\right)\ ,
\end{split}
\end{equation}
where
\begin{equation}
g_{\,\pm}(z)\,=\,\left(\frac{z}{z-\frac{\a^{\,\prime}\!\!}{4}\ (u-t)\mp\frac{\a^{\,\prime}\!\!}{2}\,\sqrt{-ut}}\right)^\b\ .
\end{equation}
Similar results obtain for massive particles filling the first Regge trajectory of the open bosonic string in three dimensions, where making use again of the substitutions \eqref{subst_prop} gives
\be
\begin{split}
\!\!\cA^{(s)}\,&=\,\a^{\,\prime}\!\!\int_0^1 dy\, y^{-\a^{\,\prime}s-2}\left[a\left(\frac{\a^{\,\prime}y}{4}\,(u-t)+\frac{\a^{\,\prime}y}{2}\,\sqrt{-ut}\right) +a\left(\frac{\a^{\,\prime}y}{4}\,(u-t)-\frac{\a^{\,\prime}y}{2}\,\sqrt{-ut}\right)-a_0\right]\\&\times \phi_{\,1}\left(p_{\,1},\pm\,\sqrt{2\a^{\,\prime}}\,p_{\,2}\right)
\phi_{\,2}\left(p_{\,2},\mp\,\sqrt{2\a^{\,\prime}}\,p_{\,1}\right)
\phi_{\,3}\left(p_{\,1},\pm\,\sqrt{2\a^{\,\prime}}\,p_{\,4}\right) \phi_{\,4}\left(p_{\,2},\mp\,\sqrt{2\a^{\,\prime}}\,p_{\,3}\right)\ .
\end{split}
\ee

Other interesting examples can be obtained starting from the confluent hypergeometric form of the coupling function \eqref{hyper_a}, which leads to
\be
\begin{split}
\cA^{(s)}\,&=\,-\frac{1}{\a^{\,\prime}s} \left(1-\frac{\a^{\,\prime}\!\!}{2}\,(u-t)+\frac{\a^{\,\prime\, 2}\!\!\!}{16}\,(u+t)^2\right)^{-\b}\\ &\times
\phi_{\,1}\left(p_{\,1},\pm\,\sqrt{2\a^{\,\prime}}\,p_{\,2}\right)
\phi_{\,2}\left(p_{\,2},\mp\,\sqrt{2\a^{\,\prime}}\,p_{\,1}\right) \phi_{\,3}\left(p_{\,1},\pm\,\sqrt{2\a^{\,\prime}}\,p_{\,4}\right)
\phi_{\,4}\left(p_{\,2},\mp\,\sqrt{2\a^{\,\prime}}\,p_{\,3}\right)\ ,
\end{split}
\ee
in the massless case in any number of space--time dimensions, or to
\be
\begin{split}
\cA^{(s)}\,&=\,\a^{\,\prime}\int_0^1 dy\, y^{-\a^{\,\prime}s-2}\left(1-\frac{\a^{\,\prime}y}{2}\,(u-t)+
\frac{\a^{\,\prime\,2}\,y^2}{16}\, (u+t)^2\right)^{-\b}\\&\times \phi_{\,1}\left(p_{\,1},\pm\,\sqrt{2\a^{\,\prime}}\,p_{\,2}\right)
\phi_{\,2}\left(p_{\,2},\mp\,\sqrt{2\a^{\,\prime}}\,p_{\,1}\right) \phi_{\,3}\left(p_{\,1},\pm\,\sqrt{2\a^{\,\prime}}\,p_{\,4}\right) \phi_{\,4}\left(p_{\,2},\mp\,\sqrt{2\a^{\,\prime}}\,p_{\,3}\right)\ ,
\end{split}
\ee
in the massive case for a multiplet filling the first Regge trajectory of the open bosonic string. In particular, starting from eqs.~\eqref{P3} and \eqref{P4} leads to
\be
\begin{split}
\cA^{(s)}\,&=\,-\frac{1}{\a^{\,\prime}s}\ e^{\frac{\a^{\,\prime}\!\!\!}{4}(u-t)}\,I_0\left(\frac{\a^{\,\prime}\!\!}{2}\sqrt{-tu}\right)\\ &\times \phi_{\,1}\left(p_{\,1},\pm\,\sqrt{2\a^{\,\prime}}\,p_{\,2}\right)
\phi_{\,2}\left(p_{\,2},\mp\,\sqrt{2\a^{\,\prime}}\,p_{\,1}\right) \phi_{\,3}\left(p_{\,1},\pm\,\sqrt{2\a^{\,\prime}}\,p_{\,4}\right)
\phi_{\,4}\left(p_{\,2},\mp\,\sqrt{2\a^{\,\prime}}\,p_{\,3}\right)\ ,
\end{split}
\ee
for the massless case in five dimensions, and
\be
\begin{split}
\cA^{(s)}\,&=\,\a^{\,\prime}\int_0^1 dy\, y^{-\a^{\,\prime}s-2}\ e^{\frac{\a^{\,\prime}y}{4}(u-t)}\ I_0\left(\frac{\a^{\,\prime}y}{2}\sqrt{-tu}\right)\\&\times \phi_{\,1}\left(p_{\,1},\pm\,\sqrt{2\a^{\,\prime}}\,p_{\,2}\right)
\phi_{\,2}\left(p_{\,2},\mp\,\sqrt{2\a^{\,\prime}}\,p_{\,1}\right) \phi_{\,3}\left(p_{\,1},\pm\,\sqrt{2\a^{\,\prime}}\,p_{\,4}\right)
\phi_{\,4}\left(p_{\,2},\mp\,\sqrt{2\a^{\,\prime}}\,p_{\,3}\right)\ ,
\end{split}
\ee
for the massive case in four dimensions. Notice that a soft behavior in the high--energy limit in every channel is only possible if the couplings grow with the spin.

Similar manipulations can be made in the amplitude with two $j_2^{\,\pm}$, and the end
result are the following substitutions in the propagator function:
\begin{equation}
\begin{split}
\xi\,\cdot\, \lambda\ \lra&\ \ \frac{\a^{\,\prime}\!\!}{2}(1+\r_1)\,(1+\r_2)\,[\,p_{\,12}^{\,\m}+{i\,F^{\,\m}}_{\a}\,\chi_{\,1}^\a\,]\,
[\,p_{\,34\m}\,+\,i\,F_{\m\a}\,\chi_{\,2}^\a\,]\ ,\\
\xi^2\ \lra&\ \ \frac{\a^{\,\prime}\!\!}{2}(1+\r_1)^2[\,p_{\,12\m}\,+\,i\,F_{\m\a}\,\chi_{\,1}^\a\,]^{\,2}\ ,\\
\lambda^2\ \lra&\ \ \frac{\a^{\,\prime}\!\!}{2}(1+\r_2)^2[\,p_{\,34\m}\,+\,i\,F_{\m\a}\,\chi_{\,2}^\a\,]^{\,2}\ .
\end{split}
\end{equation}
For brevity, let us consider massless particles, since the other cases can be easily
reproduced making the same type of substitutions in the corresponding generating
functions of current exchanges. In four dimensions one thus arrives at the following
generalization of the result of \cite{Bekaert:2009ud}:

\begin{equation}
\begin{split}
{\cA}\,=\,-\frac{\hat{A}}{\a^{\,\prime} s} \left[\vphantom{\Big(\frac{\a^{\,\prime}}{4}\Big)}\right.
&\,a\left(\frac{\a^{\,\prime}}{4}\,(1+\rho_1)\,(1+\rho_2)\,[\,p_{\,12}^{\,\m}+{i\, F^{\,\m}}_{\a}\,\chi_{\,1}^{\,\a}\,]\, [\,p_{\,34\m}\,+\,i\,F_{\m\a}\,\chi_{\,2}^{\,\a}\,]\right.\\ +&\frac{\a^{\,\prime}}{4}\left[\vphantom{\Big|}\left((1+\rho_1)\,(1+\rho_2)\, [\,p_{\,12}^{\,\m}+{i\,F^{\,\m}}_{\a}\,\chi_{\,1}^{\,\a}\,]\,[\,p_{\,34\m}\,+\,i\, F_{\m\a}\,\chi_{\,2}^{\,\a}\,]\right)^2\right.\\ -&\left.\left.(1+\rho_1)^2\,[\,p_{\,12\m}\,+\,i\,F_{\m\a}\,\chi_{\,1}^{\,\a}\,]^2\,(1+\rho_2)^2\, [\,p_{\,34\m}\,+\,i\,F_{\m\a}\,\chi_{\,2}^{\,\a}\,]^2\vphantom{\Big|}\right]^{1/2}\right)\\ +&\,a\left(\frac{\a^{\,\prime}}{4}\,(1+\rho_1)\,(1+\rho_2)\,[\,p_{\,12}^{\,\m}+{i\,F^{\,\m}}_{\a}\,\chi_{\,1}^{\,\a}\,]\, [\,p_{\,34\m}\,+\,i\,F_{\m\a}\,\chi_{\,2}^{\,\a}\,]\right.\\ -&\frac{\a^{\,\prime}}{4}\left[\vphantom{\Big|}\left((1+\rho_1)\,(1+\rho_2)\,[\,p_{\,12}^{\,\m}+{i\,F^{\,\m}}_{\a}\,\chi_{\,1}^{\,\a}\,]\, [\,p_{\,34\m}\,+\,i\,F_{\m\a}\,\chi_{\,2}^{\,\a}\,]\right)^2\right.\\ -&\left.\left.\left.(1+\rho_1)^2\,[\,p_{\,12\m}\,+\,i\,F_{\m\a}\,\chi_{\,1}^{\,\a}\,]^2\,(1+\rho_2)^2\, [\,p_{\,34\m}\,+\,i\,F_{\m\a}\,\chi_{\,2}^{\,\a}\,]^2\vphantom{\Big|}\right]^{1/2}\right)-a_0\right]\ ,
\end{split}
\end{equation}
where $\hat{A}$ is the differential operator
\begin{equation}
\hat{A}\,=\,\partial_{\rho_1}\,\partial_{\rho_2}\ (\partial_{\chi_{\,1}}\cdot\partial_{\chi_{\,1}})\, (\partial_{\chi_{\,2}}\cdot\partial_{\chi_{\,2}}) \Big|_{\rho_i\,=\,\chi_{\,i}\,=\,0}\ .
\end{equation}

As another possibility, one can consider the mixed amplitude, with the current $j_{\,1}^{\,\pm}$ \eqref{J10} at one end and the current $j_{\,2}^{\,\pm}$ \eqref{J30} at the other. In this case the final result can be obtained making the substitutions
\begin{equation}
\begin{split}
\xi\,\cdot\, \lambda\ \lra&\ \ \frac{\a^{\,\prime}\!\!}{2}\ (1+\rho_2)\ p_{\,12}^{\,\m}\cdot[\,p_{\,34\m}\,+\,i\,F_{\m\a}\,\chi_{\,2}^\a\,]\ ,\\
\xi^2\ \lra&\ \ \frac{\a^{\,\prime}\!\!}{2}\ p_{\,12}\cdot p_{\,12}\,=\,-\frac{\a^{\,\prime}\!\!}{2}\ (u+t)\ ,\\
\lambda^2\ \lra&\ \ \frac{\a^{\,\prime}\!\!}{2}\ (1+\rho_2)^2\,[\,p_{\,34\m}\,+\,i\,F_{\m\a}\,\chi_{\,2}^\a\,]^{\,2}\ ,
\end{split}
\end{equation}
so that the amplitude is finally
\begin{equation}
\begin{split}
\cA\,&=\,-\frac{\hat{B}}{\a^{\,\prime} s}\left[\vphantom{\Big(\frac{\a^{\,\prime}\!\!}{4} \,(1+\rho_2)\,p_{\,12\m}}\right.
a \left(\frac{\a^{\,\prime}\!\!}{4}\,(1+\rho_2)\,p_{\,12}^{\,\m}\,[\,p_{\,34\m}\,+\,i\,F_{\m\a}\,\chi_{\,2}^{\,\a}\,] +\frac{\a^{\,\prime}\!\!}{4}\,\left[\vphantom{\Big|}\left((1+\rho_2)\, p_{\,12}^{\,\m}\, [\, p_{\,34\m}
\,+\,i\,F_{\m\a}\,\chi_{\,2}^{\,\a}\,]\right)^2\right.\right.\\ &-\left.\left.(u+t)^2\,(1+\rho_2)^2\,[\,p_{\,34\m}\,+\,i\,F_{\m\a}\,\chi_{\,2}^{\,\a}]^2\vphantom{\Big|}\right]^{1/2}\right)+ a\left(\frac{\a^{\,\prime}\!\!}{4}\,(1+\rho_2)\,p_{\,12}^{\,\m}\,[\,p_{\,34\m}
\,+\,i\,F_{\m\a}\,\chi_{\,2}^{\,\a}]\right.\\ &-\left.\left.\frac{\a^{\,\prime}\!\!}{4}\,\left[\vphantom{\Big|}\left((1+\rho_2)\,p_{\,12}^{\,\m}\, [\,p_{\,34\m}
\,+\,i\,F_{\m\a}\,\chi_{\,2}^{\,\a}\,]\right)^2 -(u+t)^2\,(1+\rho_2)^2\,[\,p_{\,34\m}\,+\,i\,F_{\m\a}\,\chi_{\,2}^{\,\a}\,]^2\right]^{1/2}\right)-a_0\right]\\&\qquad\ \times \ \phi_{\,1}\left(p_{\,1},\pm\,\sqrt{2\a^{\,\prime}}\,p_{\,2}\right)
\phi_{\,2}\left(p_{\,2},\mp\,\sqrt{2\a^{\,\prime}}\,p_{\,1}\right)\ ,
\end{split}
\end{equation}
with $\hat{B}$ defined as
\begin{equation}
\hat{B}\,=\,\partial_{\rho_2}\ (\partial_{\chi_{\,2}}\cdot\partial_{\chi_{\,2}})\Big|_{\rho_2\,=\,\chi_{\,2}\,=\,0}\ .
\end{equation}

Similar results can be obtained starting from the generating function of all conserved currents involving massless totally symmetric HS states
\begin{multline}
j^{\,\pm}(\xi)\,=\,e^{\,\pm\,\sqrt{\frac{\a^{\,\prime}\!\!}{2}}\ \xi\cdot p_{\,12}}\ \exp\left(\!\pm\,\sqrt{\frac{\a^{\,\prime}\!\!}{2}}\,\ \xi_{\,\a}
\left[\partial_{\zeta_1}\cdot\partial_{\zeta_2}\,p^{\,\a}_{\,12}-2\,\partial^{\,\a}_{\zeta_1}\,
\partial_{\zeta_2}\cdot p_{\,1}
+\,2\,\partial^{\,\a}_{\zeta_2}\,\partial_{\zeta_1}\cdot p_{\,2}\right]\right)\\
\ \ \ \times\,\phi_{\,1}\left(p_{\,1},\,\zeta_1\,\pm\,\sqrt{2\a^{\,\prime}}\,p_{\,2}\right)
\,\phi_{\,2}\left(p_{\,2},\,\zeta_2\,\mp\,\sqrt{2\a^{\,\prime}}\,p_{\,1}\right)\ ,
\end{multline}
where the dominant factor is given again by the exponential
$$e^{\,\pm\,\sqrt{\frac{\a^{\,\prime}\!\!}{2}}\ \xi\cdot p_{\,12}}\ ,$$
that gives rise to a behavior along the lines of that observed in the previous cases.

Let us conclude this section by stressing that, independently of the tensorial structure, for both massless particles in four dimensions and massive particles in three dimensions the high--energy behavior is determined by the zeros of $a(z)$ and its derivatives. As a result, any of these amplitudes can be extremely soft at high energies: the available propagator functions clearly suggest that towers of HS fields can give rise to a soft high--energy behavior. As stressed in \cite{Bekaert:2009ud}, one can soften the behavior at high energies with a coupling function $a(z)$ whose power series has a finite radius of convergence. The examples resting on eqs.~\eqref{P1}, \eqref{P2} comply to this expectation. With strong couplings of this type, however, the $S$ matrix typically develops kinematical singularities that arise at some given impact parameter or angle parameter. This effect resonates with the expectation that theories of this type should be describing extended objects, and these singularities appear to reflect corresponding form factors.

\vskip 36pt


\scs{Conclusions}\label{sec:conclusions}


In this article we have studied three and four--point scattering amplitudes for
states belonging to the first Regge trajectory of the open bosonic string,
arriving at a handy explicit form for the cubic couplings of these massive HS
modes. Our key result is perhaps the simple expression for the three--point
amplitudes contained in eqs.~\eqref{Apiumeno} and \eqref{Apiumeno2}. The
derivation was made possible by a careful scrutiny of the physical state
conditions that lie behind the $SL(2,R)$ invariance of the amplitudes, and as
in \cite{Bekaert:2009ud} the Weyl--Wigner calculus provided a very convenient
tool.

One can extract from the massive amplitudes off--shell couplings that behave
nicely in the limit of vanishing external masses, where they display an
unbroken gauge symmetry. Lo and behold, one thus gains access to the types of
currents and couplings that emerge in the tensionless limit of String Theory.
We would like to stress that these couplings include the highest--derivative couplings first identified in \cite{Gross,AmatiCiafaloniVeneziano,Taiwan,west,fotopoulos}
but also additional terms that maintain a rich non--abelian structure even in
the limit. Interestingly, all these couplings, including the ``seeds'' of
\cite{Boulanger:2008tg}, are somehow induced by the highest--derivative ones via the
action of the operator $\cG$ defined in eq.~ \eqref{G}. In this respect, the
peculiarity of String Theory appears to lie in its use of $\exp(\cG)$, rather
than other seemingly viable operator--valued functions, to relate the various
contributions.

The results on cubic couplings led us to construct some HS four--point
scattering amplitudes directly on the Field Theory side, elaborating on the
beautiful results of Bekaert, Joung and Mourad \cite{Bekaert:2009ud} that may
be regarded as a first step toward the deconstruction of four--point string
amplitudes. As pointed out in \cite{Bekaert:2009ud}, the high--energy behavior
depends crucially on a coupling function $a(z)$ that collects the infinitely
many dimensionless couplings that size the interactions. Our corollary to
\cite{Bekaert:2009ud} confirms their basic conclusions: independently of the
types of external states, the amplitudes can be well behaved at high energies
only if the function $a(z)$ vanishes at infinity. In String Theory, the
coupling function is exponential, $a(z)=e^z$, and therefore has an essential
singularity at infinity, which makes the sub--leading Regge trajectories
necessary to grant a soft high--energy behavior. Moreover, non--local quartic
terms appear unavoidable, in general, in order to arrive at four--point
amplitudes with a high--energy behavior compatible with standard $S$--matrix
bounds. One is perhaps not used to see matters in this fashion, since the
Veneziano formula apparently does not suffer from these problems. However, it
is an $S$--matrix formula, that as such combines the contributions of the cubic
and quartic couplings that are needed to guarantee its familiar high--energy
behavior. Non--local quartic couplings are not at all unnatural in this
framework, insofar as they contribute to the definition of local and unitary
$S$--matrix amplitudes. Allowing non--local Lagrangians, a feat that has long
been regarded as inevitable for massless HS, opens up interesting new
possibilities. For instance, field equations containing infinitely many
derivatives may well loose their dynamical meaning altogether, in such a way
that their role be confined to the transfer of information encoded in boundary
conditions \cite{Moeller}. This state of affairs is actually reminiscent of
Vasiliev's construction \cite{Vasiliev:1988sa}, where the ``gauge function
method'' can be regarded as a reflection of this fact and has already led to
interesting classes of non--trivial classical solutions \cite{VasilievSoluz}.

\vskip 35pt


\section*{Acknowledgments}


We are grateful to N.~Boulanger, A.~Campoleoni, D.~Francia, C.~Iazeolla,
E.~Joung, R.~Manvelyan, J.~Mourad, R.~Rahman and P.~Sundell for several
stimulating discussions. We are also grateful to the GGI for the hospitality
while this work was in progress. The present research was supported in part by
Scuola Normale Superiore, by INFN, by the MIUR-PRIN contract 2007-5ATT78 and by
the ERC Advanced Investigator Grant no. 226455 ``Supersymmetry, Quantum Gravity
and Gauge Fields'' (SUPERFIELDS). Finally, we would like to thank the referee for his interest in this work and a careful reading of the manuscript.

\newpage

\begin{appendix}


\scs{On off--Shell Cubic and Quartic Couplings}


In the original version of the paper we refrained from displaying the off--shell completion of the couplings corresponding to eqs.~\eqref{116} and \eqref{117}, since R.~Manvelyan, K.~Mktrchyan and W.~Ruhl had communicated with us on the matter and we were considering the opportunity of writing a joint paper on the issue. We are now adding this appendix, on this and other matters, in order to abide to a request of the referee, who asked some details on the claims made in Section \ref{sec:couplings}.

The key message that we would like to convey is twofold. First, knowing all
terms without divergences and traces suffices to determine the
\emph{off--shell} vertices. This is true both in Fronsdal's constrained form of
\cite{Fronsdal} and in the minimal forms of \cite{FranciaSagnotti,minimal}.
Moreover, working with the reduced set of terms without divergences and traces
can provide a useful starting point for the analysis of higher--point vertices.

Let us illustrate this fact for Fronsdal's constrained version, with doubly-traceless fields and traceless gauge parameters. In this case one can exploit the relation
\be
p^{\,2}\,\phi(p\,,\,\xi)\,=\,-\,\cF(p\,,\,\xi)\,+\,i\,(p\cdot \xi)\,\cD(p\,,\,\xi)\,\phi(p\,,\,\xi)\ ,
\ee
for the generating function $\phi(p\,,\,\xi)$, together with the gauge variations
\be
\delta\,\cD(p\,,\,\xi)\,\phi(p\,,\,\xi)\,=\,-\,p^{\,2}\,\L(p\,,\,\xi)\ ,\qquad \delta\, [\partial_{\xi}\cdot\partial_{\xi}\,\phi(p\,,\,\xi)]\,=\,-\,2\,i\,(p\cdot \partial_{\,\xi})\,\L(p\,,\,\xi)\ ,
\ee
where the de Donder operator is here defined as
\be
\cD(p\,,\,\xi)\,=\,-\,i\,\left( p \cdot \partial_\xi \,  \, - \, \frac{1}{2} \ p \cdot \xi \ \partial_\xi \cdot \partial_\xi \right)\ .
\ee
Starting from eq.~\eqref{masslessGenfunc}, one can thus compute gauge
variations compensating them with appropriate counterterms and reconstructing
the complete cubic vertex. The whole procedure rests crucially on the
commutators
\begin{eqnarray}
[\cG\,,\,p_{\,1}\cdot\xi_{\,1}]\,\phi_{\,1}\,\phi_{\,2}\,\L_{\,3}&=&\left\{-\,p_{\,2}\cdot\xi_{\,2} \left[(\partial_{\xi_{\,2}}\cdot\partial_{\xi_{\,3}}\,+\,1) \,i\,\cD_{\,2}\,+\,\frac{1}{4}\,\partial_{\xi_{\,2}} \cdot\partial_{\xi_{\,2}}\,(p_{\,12}\cdot\partial_{\xi_{\,3}}\,+\,p_{\,3}\cdot\partial_{\xi_{\,3}})\right]\right.
\nonumber\\ \vphantom{\left(\!\pm\sqrt{\frac{\a^{\,\prime}\!\!}{2}}\,\right)^{\,3}}
&+& \left.\left[(\partial_{\xi_{\,2}}\cdot\partial_{\xi_{\,3}}\,+\,1)\ p^{\,2}_{\,3}\,-\,\frac{1}{2}\,p_{{\,3}} \cdot\partial_{\xi_{\,3}}\,(p_{\,31}\cdot\partial_{\xi_{\,2}}\,-\,p_{\,2}\cdot\partial_{\xi_{\,2}})\right]\right\}\, \phi_{\,1}\,\phi_{\,2}\,\L_{\,3}\ ,\nonumber\\
{[}\cG\,,\,p_{\,2}\cdot\xi_{\,2}{]}\,\phi_{\,1}\,\phi_{\,2}\,\L_{\,3}&=&\left\{+\,p_{\,1}\cdot\xi_{\,1}\left[(\partial_{\xi_{\,3}}\cdot\partial_{\xi_{\,1}}\,+\,1) \,i\,\cD_{\,1}\,-\,\frac{1}{4}\,\partial_{\xi_{\,1}} \cdot\partial_{\xi_{\,1}}\,(p_{\,12}\cdot\partial_{\xi_{\,3}}\,-\,p_{\,3}\cdot\partial_{\xi_{\,3}})\right]\nonumber\right.\\ \vphantom{\left(\!\pm\sqrt{\frac{\a^{\,\prime}\!\!}{2}}\,\right)^{\,3}}
&-&\left. \left[(\partial_{\xi_{\,3}}\cdot\partial_{\xi_{\,1}}\,+\,1)\ p^{\,2}_{\,3}\,+\,\frac{1}{2}\,p_{{\,3}} \cdot\partial_{\xi_{\,3}}\,(p_{\,23}\cdot\partial_{\xi_{\,1}}\,+\,p_{\,1}\cdot\partial_{\xi_{\,1}})\right]\right\}\, \phi_{\,1}\,\phi_{\,2}\,\L_{\,3}\ ,\nonumber\\
{[}\cG\,,\,p_{\,3}\cdot\xi_{\,3}{]}\,\phi_{\,1}\,\phi_{\,2}\,\L_{\,3}&=&\left\{p_{\,2}\cdot\xi_{\,2}\left[(\partial_{\xi_{\,1}}\cdot\partial_{\xi_{\,2}}\,+\,1) \,i\,\cD_{\,2}\,-\,\frac{1}{4}\,\partial_{\xi_{\,2}} \cdot\partial_{\xi_{\,2}}\,(p_{\,23}\cdot\partial_{\xi_{\,1}}\,-\,p_{\,1}\cdot\partial_{\xi_{\,1}})\right]\nonumber\right.\\ \vphantom{\left(\!\pm\sqrt{\frac{\a^{\,\prime}\!\!}{2}}\,\right)^{\,3}}
&-&\left.\ \,p_{\,1}\cdot\xi_{\,1}\left[(\partial_{\xi_{\,1}}\cdot\partial_{\xi_{\,2}}\,+\,1) \,i\,\cD_{\,1}\,+\,\frac{1}{4}\,\partial_{\xi_{\,1}} \cdot\partial_{\xi_{\,1}}\,(p_{\,31}\cdot\partial_{\xi_{\,2}}\,+\,p_{\,2}\cdot\partial_{\xi_{\,2}})\right]\right\}\nonumber\\ &&\times\,\phi_{\,1}\,\phi_{\,2}\,\L_{\,3}\ ,\label{commutators}
\end{eqnarray}
with ${\cal G}$ the operator of eq.~\eqref{G}, that hold modulo the Fronsdal
equation and that give rise to a handy iterative structure whose form permits
to build the counterterms order by order in $\sqrt{\a^{\,\prime}}$. Reinstating
the missing terms, one thus arrives at the final result for the complete cubic
vertices in Fronsdal's constrained form, that reads
{\allowdisplaybreaks
\begin{eqnarray}\label{offcubic}
\cA^{\, [0]\,\pm}&=&\exp\Bigg\{\pm\ \sqrt{\frac{\a^{\, \prime}\!\!}{2}}\ \Big[(\partial_{\xi_{\,1}}\cdot\partial_{\xi_{\,2}}+1)
(\partial_{\xi_{\,3}}\cdot p_{\,12})\,+\,(\partial_{\xi_{\,2}}\cdot\partial_{\xi_{\,3}}+1)(\partial_{\xi_{\,1}}\cdot p_{\,23})\nonumber\\&&+\,(\partial_{\xi_{\,3}}\cdot\partial_{\xi_{\,1}}+1)(\partial_{\xi_{\,2}}\cdot
p_{\,31})\Big]\Bigg\}\nonumber\\\vphantom{\left(\!\pm\sqrt{\frac{\a^{\,\prime}\!\!}{2}}\,\right)^{\,3}}
\times &&\!\!\!\!\!\!\!\!\!\!\!\!\!\!\left\{\vphantom{\left(\!\pm\sqrt{\frac{\a^{\,\prime}\!\!}{2}}\,\right)^{\,3}}
1\,+\,\frac{\a^{\,\prime}\!\!}{2}
\left[(\partial_{\xi_{\,1}}\cdot\partial_{\xi_{\,2}}\,+\,1)\,i\,\cD_{\,2}\,-\,\frac{1}{4}\,\partial_{\xi_{\,2}} \cdot\partial_{\xi_{\,2}}\,(p_{\,23}\cdot\partial_{\xi_{\,1}}\,-\,p_{\,1}\cdot\partial_{\xi_{\,1}})\right]\right.\nonumber\\ \vphantom{\left(\!\pm\sqrt{\frac{\a^{\,\prime}\!\!}{2}}\,\right)^{\,3}}&&\ \ \times \left[(\partial_{\xi_{\,3}}\cdot\partial_{\xi_{\,1}}\,+\,1)\,i\,\cD_{\,3}\,+\,\frac{1}{4}\,\partial_{\xi_{\,3}} \cdot\partial_{\xi_{\,3}}\,(p_{\,23}\cdot\partial_{\xi_{\,1}}\,+\,p_{\,1}\cdot\partial_{\xi_{\,1}})\right]\nonumber\\ \vphantom{\left(\!\pm\sqrt{\frac{\a^{\,\prime}\!\!}{2}}\,\right)^{\,3}}
&&+\qquad\text{cyclic}\nonumber\\
+ &&\!\!\!\!\!\!\!\!\!\!\!\!\!\!\left(\!\pm\,\sqrt{\frac{\a^{\,\prime}\!\!}{2}}\,\right)^{\,3}
\left[(\partial_{\xi_{\,1}}\cdot\partial_{\xi_{\,2}}\,+\,1)\,i\,\cD_{\,1}\,+\,\frac{1}{4}\,\partial_{\xi_{\,1}} \cdot\partial_{\xi_{\,1}}\,(p_{\,31}\cdot\partial_{\xi_{\,2}}\,+\,p_{\,2}\cdot\partial_{\xi_{\,2}})\right]\nonumber\\ \vphantom{\left(\!\pm\sqrt{\frac{\a^{\,\prime}\!\!}{2}}\,\right)^{\,3}}
&&\quad\ \ \times\,\left[(\partial_{\xi_{\,2}}\cdot\partial_{\xi_{\,3}}\,+\,1)\,i\,\cD_{\,2}\,+\,\frac{1}{4}\,\partial_{\xi_{\,2}} \cdot\partial_{\xi_{\,2}}\,(p_{\,12}\cdot\partial_{\xi_{\,3}}\,+\,p_{\,3}\cdot\partial_{\xi_{\,3}})\right]\nonumber\\ \vphantom{\left(\!\pm\sqrt{\frac{\a^{\,\prime}\!\!}{2}}\,\right)^{\,3}}
&&\quad\ \ \times\,\left[(\partial_{\xi_{\,3}}\cdot\partial_{\xi_{\,1}}\,+\,1)\,i\,\cD_{\,3}\,+\,\frac{1}{4}\,\partial_{\xi_{\,3}} \cdot\partial_{\xi_{\,3}}\,(p_{\,23}\cdot\partial_{\xi_{\,1}}\,+\,p_{\,1}\cdot\partial_{\xi_{\,1}})\right]\nonumber\\
\vphantom{\left(\!\pm\sqrt{\frac{\a^{\,\prime}\!\!}{2}}\,\right)^{\,3}}
- &&\!\!\!\!\!\!\!\!\!\!\!\!\!\!\left(\!\pm\,\sqrt{\frac{\a^{\,\prime}\!\!}{2}}\,\right)^{\,3}\ \frac{1}{2}\ \partial_{\xi_{\,1}}\cdot\partial_{\xi_{\,1}}\,(1\,+\,\partial_{\xi_{\,1}}\cdot\partial_{\xi_{\,2}})\nonumber\\
&&\quad\ \ \times\,\left[(\partial_{\xi_{\,2}}\cdot\partial_{\xi_{\,3}}\,+\,1)\,i\,\cD_{\,2}\,+\,\frac{1}{4}\,\partial_{\xi_{\,2}} \cdot\partial_{\xi_{\,2}}\,(p_{\,12}\cdot\partial_{\xi_{\,3}}\,+\,p_{\,3}\cdot\partial_{\xi_{\,3}})\right]\nonumber\\ \vphantom{\left(\!\pm\sqrt{\frac{\a^{\,\prime}\!\!}{2}}\,\right)^{\,3}}
&&\quad\ \ \times\,\left[\frac{1}{2}\ (p_{\,12}\cdot\partial_{\xi_{\,3}}\,-\,p_{\,3}\cdot\partial_{\xi_{\,3}})\,i\,\cD_{\,3}\,+\,\frac{1}{4}\,\partial_{\xi_{\,3}} \cdot\partial_{\xi_{\,3}}\,(p_{\,3}^{\,2}\,-\,p_{\,2}^{\,2}\,+\,p_{\,1}^2)\right]\nonumber\\
\vphantom{\left(\!\pm\sqrt{\frac{\a^{\,\prime}\!\!}{2}}\,\right)^{\,3}}
- &&\!\!\!\!\!\!\!\!\!\!\!\!\!\!\left(\!\pm\,\sqrt{\frac{\a^{\,\prime}\!\!}{2}}\,\right)^{\,3}
\left[(\partial_{\xi_{\,1}}\cdot\partial_{\xi_{\,2}}\,+\,1)\,i\,\cD_{\,2}\,-\,\frac{1}{4}\,\partial_{\xi_{\,2}} \cdot\partial_{\xi_{\,2}}\,(p_{\,23}\cdot\partial_{\xi_{\,1}}\,-\,p_{\,1}\cdot\partial_{\xi_{\,1}})\right]\nonumber\\ \vphantom{\left(\!\pm\sqrt{\frac{\a^{\,\prime}\!\!}{2}}\,\right)^{\,3}}
&&\quad\ \ \times\,\left[(\partial_{\xi_{\,3}}\cdot\partial_{\xi_{\,1}}\,+\,1)\,i\,\cD_{\,1}\,-\,\frac{1}{4}\,\partial_{\xi_{\,1}} \cdot\partial_{\xi_{\,1}}\,(p_{\,12}\cdot\partial_{\xi_{\,3}}\,-\,p_{\,3}\cdot\partial_{\xi_{\,3}})\right]\nonumber\\ \vphantom{\left(\!\pm\sqrt{\frac{\a^{\,\prime}\!\!}{2}}\,\right)^{\,3}}
&&\quad\ \ \times\,\left[(\partial_{\xi_{\,2}}\cdot\partial_{\xi_{\,3}}\,+\,1)\,i\,\cD_{\,3}\,-\,\frac{1}{4}\,\partial_{\xi_{\,3}} \cdot\partial_{\xi_{\,3}}\,(p_{\,31}\cdot\partial_{\xi_{\,2}}\,-\,p_{\,2}\cdot\partial_{\xi_{\,2}})\right]
\vphantom{\left(\!\pm\sqrt{\frac{\a^{\,\prime}\!\!}{2}}\,\right)^{\,3}}\nonumber\\
+ &&\!\!\!\!\!\!\!\!\!\!\!\!\!\!\left(\!\pm\,\sqrt{\frac{\a^{\,\prime}\!\!}{2}}\,\right)^{\,3}\ \frac{1}{2}\ \partial_{\xi_{\,2}}\cdot\partial_{\xi_{\,2}}\,(1\,+\,\partial_{\xi_{\,1}}\cdot\partial_{\xi_{\,2}})\nonumber\\
&&\quad\ \ \times\,\left[(\partial_{\xi_{\,3}}\cdot\partial_{\xi_{\,1}}\,+\,1)\,i\,\cD_{\,1}\,-\,\frac{1}{4}\,\partial_{\xi_{\,1}} \cdot\partial_{\xi_{\,1}}\,(p_{\,12}\cdot\partial_{\xi_{\,3}}\,-\,p_{\,3}\cdot\partial_{\xi_{\,3}})\right]\nonumber\\ \vphantom{\left(\!\pm\sqrt{\frac{\a^{\,\prime}\!\!}{2}}\,\right)^{\,3}}
&&\left.\quad\ \ \times\,\left[-\,\frac{1}{2}\ (p_{\,12}\cdot\partial_{\xi_{\,3}}\,+\,p_{\,3}\cdot\partial_{\xi_{\,3}})\,i\,\cD_{\,3}\,-\,\frac{1}{4}\,\partial_{\xi_{\,3}} \cdot\partial_{\xi_{\,3}}\,(p_{\,1}^{\,2}\,-\,p_{\,3}^{\,2}\,-\,p_{\,2}^2)\right]
\right\}\nonumber\\&&\times
\,\phi_{\,1}(\xi_{\,1})\,\phi_{\,2}(\xi_{\,2})\,\phi_{\,3}(\xi_{\,3})\ \Bigg|_{\xi_{\,i}\,=\,0}\ ,
\end{eqnarray}}
\!\!up to terms proportional to $ \cF(p\,,\,\xi)$, which can be eliminated by
suitable redefinitions. A similar but slightly more complicated result would
obtain in the unconstrained formalism of \cite{FranciaSagnotti,minimal},
starting from
\be
p^{\,2}\,\phi(p\,,\,\xi)\,=\,-\,\cA(p\,,\,\xi)\,+\,i\,(p\cdot \xi)\,\left(\cD(p\,,\,\xi)\,\phi(p\,,\,\xi)\,-\,\frac{1}{2}\,(p\cdot\xi)^2\,\a\right)\ ,
\ee
where $\cA$ is the extended Fronsdal operator introduced in \cite{minimal} and
$\a$ is the compensator of \cite{FranciaSagnotti}, and taking into account the
gauge variations
\be
\begin{split}
\delta \Big[\,\cD(p\,,\,\xi)\,\phi(p\,,\,\xi)\,-\,\frac{1}{2}\,(p\cdot \xi)^{\,2}\,\a(p\,,\,\xi)\Big]\, &= \,-\, p^{\,2}\,\L(p\,,\,\xi)\ , \\
\d\Big[\partial_{\xi}\cdot\partial_{\xi}\,\phi(p\,,\,\xi)\,+\,i\,(p\cdot\xi)\,\a(p\,,\,\xi)\Big]\, &=\,-\, 2\, i\, p\cdot\partial_{\xi}\,\L(p\,,\,\xi) \ .
\end{split}
\ee
One can thus prove that in the unconstrained formalism the full off--shell result can be
obtained effecting in eq.~\eqref{offcubic} the replacements
\be
\begin{split}
\cD(p\,,\,\xi)\,\phi(p\,,\,\xi)\,&\ra\,\cD(p\,,\,\xi)\,\phi(p\,,\,\xi)\,-\,\frac{1}{2}\,(p\cdot \xi)^{\,2}\,\a(p\,,\,\xi)\ ,\\
\partial_{\xi}\cdot\partial_{\xi}\,\phi(p\,,\,\xi)\,&\ra\,\partial_{\xi}\cdot\partial_{\xi}\, \phi(p\,,\,\xi)\,+\,i\,(p\cdot\xi)\,\a(p\,,\,\xi)\ .
\end{split}
\ee

A further addition to our results of Section \ref{sec:couplings} is an educated guess on
corresponding fermionic conserved currents. Up to divergences and $\g$--traces, the
starting point is provided by the natural generating function of Fermi--Fermi--Bose
couplings, that takes the relatively simple form
\begin{multline}
\cA_F^{\,[0]\, \pm}\,=\,\exp (\pm\,\cG )\ \bar{\psi}_{\,1}\left(p_{\,1}\,,\,\xi_{\,1}\,\pm\,\sqrt{\frac{\a^{\, \prime}\!\!}{2}}\ p_{\,23}\right)
[1\,+\,\slashed{\partial}_{\,\xi_{\,3}}]\,{\psi_{\,2}}\left(p_{\,2}\,,\,\xi_{\,2}\,\pm\,\sqrt{\frac{\a^{\, \prime}\!\!}{2}}\
p_{\,31}\right)\\ \times \ \phi_{\,3}\left(p_{\,3}\,,\,\xi_{\,3}\,\pm\,\sqrt{\frac{\a^{\, \prime}\!\!}{2}}\
p_{\,12}\right)\Bigg|_{\xi_{\,i}\,=\,0}\ ,\label{fermionic}
\end{multline}
since the introduction of additional $\g$--matrices is inessential on--shell,
where $\cG$ is again the operator of eq.~\eqref{G}. One can now extract from
eq.~\eqref{fermionic} the currents
\begin{multline} \label{fermi_current1}
J_{FF}^{\,\text{[0]} \, \pm}(x\,;\,\xi)\,=\, \exp\left(\mp\, i\ \sqrt{\frac{\a^{\,
\prime}\!\!}{2}}\ \xi_{\,\a}
\left[\partial_{\zeta_1}\cdot\partial_{\zeta_2}\,\partial^{\,\a}_{\,12}-2\,\partial^{\,\a}_{\zeta_1}\,
\partial_{\zeta_2}\cdot \partial_{\,1}
+\,2\,\partial^{\,\a}_{\zeta_2}\,\partial_{\zeta_1}\cdot \partial_{\,2}\right]\right)\\\times\,\bar{\Psi}_{\,1}\left(x\,\mp\,i\ \sqrt{\frac{\a^{\, \prime}\!\!}{2}}\  \xi\, ,\,\zeta_1\,\mp\,i\, \sqrt{2\, \a^{\,\prime}}\, \partial_{\,2}\right)\,\Big[1\,+\,\slashed{\xi}\Big]\,\Psi_{\,2}\left(x\,\pm\,i\ \sqrt{\frac{\a^{\, \prime}\!\!}{2}}\ \xi \,,\,\zeta_2\,\pm\,i\,\sqrt{2\, \a^{\,\prime}}\, \partial_{\,1}\right)\Bigg|_{\zeta_i\,=\,0}\ .
\end{multline}
and
\begin{multline} \label{fermi_current2}
J_{BF}^{\,\text{[0]} \, \pm}(x\,;\,\xi)\,=\, \exp\left(\mp\, i\ \sqrt{\frac{\a^{\,
\prime}\!\!}{2}}\ \xi_{\,\a}
\left[\partial_{\zeta_1}\cdot\partial_{\zeta_2}\,\partial^{\,\a}_{\,12}-2\,\partial^{\,\a}_{\zeta_1}\,
\partial_{\zeta_2}\cdot \partial_{\,1}
+\,2\,\partial^{\,\a}_{\zeta_2}\,\partial_{\zeta_1}\cdot \partial_{\,2}\right]\right)\\\times\,\Big[1\,+\,\slashed{\partial}_{\zeta_{\,2}}\Big]\,\Psi_{\,1}\left(x\,\mp\,i\ \sqrt{\frac{\a^{\, \prime}\!\!}{2}}\ \xi \,,\,\zeta_1\,\mp\,i\,\sqrt{2\, \a^{\,\prime}}\, \partial_{\,2}\right)\,\Phi_{\,2}\left(x\,\pm\,i\sqrt{\frac{\a^{\,\prime}\!\!}{2}}\ \xi \,,\,\zeta_2\,\pm\,i\,\sqrt{2\a^{\,\prime}}\,\partial_{\,1}\right)\Bigg|_{\zeta_i\,=\,0}\ .
\end{multline}
that are conserved if the bosonic Fierz--Pauli conditions
\be
\begin{split}
\Box\, \Phi(x\,,\,\xi)\,&=\,0\ ,\\
\partial\cdot\partial_{\xi}\,\Phi(x\,,\,\xi)\,&=\,0 \ , \\
\partial_{\xi}\cdot\partial_{\xi}\,\Psi(x\,,\,\xi)\,&=\,0
\end{split}
\ee
and their fermionic counterparts
\be
\begin{split}
\slashed{\partial}\,\Psi(x\,,\,\xi)\,&=\,0\ ,\\
\partial\cdot\partial_{\xi}\,\Psi(x\,,\,\xi)\,&=\,0\\
\slashed{\partial}_{\xi}\Psi(x\,,\,\xi)\,&=\,0\ .
\end{split}
\ee
are enforced.

Again, eqs.~\eqref{fermi_current1} and \eqref{fermi_current2} do not contain any divergences or $\g$--traces, but these contributions can be reinstated proceeding as in the preceding discussion of bosonic couplings. To this end, in the Fang--Fronsdal constrained formulation one should start from
\be
i\,\slashed{p}\,\psi(p\,,\,\xi)\,=\,-\,\cS(p\,,\,\xi)\,-\,i\,(p\cdot\xi)\,\slashed{\partial}_\xi\,\psi(p\,,\,\xi) \ ,
\ee
where $\cS(p\,,\,\xi)$ is the Fang--Fronsdal operator, to be combined with
\be
\d\psisl(p\,,\,\xi)\,=\,-\,i\,\slashed{p}\ \e(p\,,\,\xi)\ ,
\ee
that holds provided $\slashed{\e}(p\,,\,\xi)\,=\,0$. The final result rests on
the same iterative structure of eqs.~\eqref{commutators}, and reads
{\allowdisplaybreaks
\begin{eqnarray}\label{offcubic_fermi}
\cA^{\, [0]\,\pm}&=&\exp\Bigg\{\pm\ \sqrt{\frac{\a^{\, \prime}\!\!}{2}}\ \Big[(\partial_{\xi_{\,1}}\cdot\partial_{\xi_{\,2}}+1)
(\partial_{\xi_{\,3}}\cdot p_{\,12})\,+\,(\partial_{\xi_{\,2}}\cdot\partial_{\xi_{\,3}}+1)(\partial_{\xi_{\,1}}\cdot p_{\,23})\nonumber\\&&+\,(\partial_{\xi_{\,3}}\cdot\partial_{\xi_{\,1}}+1)(\partial_{\xi_{\,2}}\cdot
p_{\,31})\Big]\Bigg\}\,(1\,+\,\slashed{\partial}_{\xi_{\,3}})^{\,\a\b}\nonumber\\\vphantom{\left(\!\pm\sqrt{\frac{\a^{\,\prime}\!\!}{2}}\,\right)^{\,3}}
\times &&\!\!\!\!\!\!\!\!\!\!\!\!\!\!\left\{\vphantom{\left(\!\pm\sqrt{\frac{\a^{\,\prime}\!\!}{2}}\,\right)^{\,3}}
1\,+\,\frac{\a^{\,\prime}\!\!}{2}
\left[(\partial_{\xi_{\,1}}\cdot\partial_{\xi_{\,2}}\,+\,1)\,i\,\cD_{\,2}\,-\,\frac{1}{4}\,\partial_{\xi_{\,2}} \cdot\partial_{\xi_{\,2}}\,(p_{\,23}\cdot\partial_{\xi_{\,1}}\,-\,p_{\,1}\cdot\partial_{\xi_{\,1}})\right]\right.\nonumber\\ \vphantom{\left(\!\pm\sqrt{\frac{\a^{\,\prime}\!\!}{2}}\,\right)^{\,3}}&&\ \ \times \left[(\partial_{\xi_{\,3}}\cdot\partial_{\xi_{\,1}}\,+\,1)\,i\,\cD_{\,3}\,+\,\frac{1}{4}\,\partial_{\xi_{\,3}} \cdot\partial_{\xi_{\,3}}\,(p_{\,23}\cdot\partial_{\xi_{\,1}}\,+\,p_{\,1}\cdot\partial_{\xi_{\,1}})\right]\nonumber\\ \vphantom{\left(\!\pm\sqrt{\frac{\a^{\,\prime}\!\!}{2}}\,\right)^{\,3}}
&&+\qquad\text{cyclic}\nonumber\\
+ &&\!\!\!\!\!\!\!\!\!\!\!\!\!\!\left(\!\pm\,\sqrt{\frac{\a^{\,\prime}\!\!}{2}}\,\right)^{\,3}
\left[(\partial_{\xi_{\,1}}\cdot\partial_{\xi_{\,2}}\,+\,1)\,i\,\cD_{\,1}\,+\,\frac{1}{4}\,\partial_{\xi_{\,1}} \cdot\partial_{\xi_{\,1}}\,(p_{\,31}\cdot\partial_{\xi_{\,2}}\,+\,p_{\,2}\cdot\partial_{\xi_{\,2}})\right]\nonumber\\ \vphantom{\left(\!\pm\sqrt{\frac{\a^{\,\prime}\!\!}{2}}\,\right)^{\,3}}
&&\quad\ \ \times\,\left[(\partial_{\xi_{\,2}}\cdot\partial_{\xi_{\,3}}\,+\,1)\,i\,\cD_{\,2}\,+\,\frac{1}{4}\,\partial_{\xi_{\,2}} \cdot\partial_{\xi_{\,2}}\,(p_{\,12}\cdot\partial_{\xi_{\,3}}\,+\,p_{\,3}\cdot\partial_{\xi_{\,3}})\right]\nonumber\\ \vphantom{\left(\!\pm\sqrt{\frac{\a^{\,\prime}\!\!}{2}}\,\right)^{\,3}}
&&\quad\ \ \times\,\left[(\partial_{\xi_{\,3}}\cdot\partial_{\xi_{\,1}}\,+\,1)\,i\,\cD_{\,3}\,+\,\frac{1}{4}\,\partial_{\xi_{\,3}} \cdot\partial_{\xi_{\,3}}\,(p_{\,23}\cdot\partial_{\xi_{\,1}}\,+\,p_{\,1}\cdot\partial_{\xi_{\,1}})\right]\nonumber\\
\vphantom{\left(\!\pm\sqrt{\frac{\a^{\,\prime}\!\!}{2}}\,\right)^{\,3}}
- &&\!\!\!\!\!\!\!\!\!\!\!\!\!\!\left(\!\pm\,\sqrt{\frac{\a^{\,\prime}\!\!}{2}}\,\right)^{\,3}\ \frac{1}{2}\ \partial_{\xi_{\,1}}\cdot\partial_{\xi_{\,1}}\,(1\,+\,\partial_{\xi_{\,1}}\cdot\partial_{\xi_{\,2}})\nonumber\\
&&\quad\ \ \times\,\left[(\partial_{\xi_{\,2}}\cdot\partial_{\xi_{\,3}}\,+\,1)\,i\,\cD_{\,2}\,+\,\frac{1}{4}\,\partial_{\xi_{\,2}} \cdot\partial_{\xi_{\,2}}\,(p_{\,12}\cdot\partial_{\xi_{\,3}}\,+\,p_{\,3}\cdot\partial_{\xi_{\,3}})\right]\nonumber\\ \vphantom{\left(\!\pm\sqrt{\frac{\a^{\,\prime}\!\!}{2}}\,\right)^{\,3}}
&&\quad\ \ \times\,\left[\frac{1}{2}\ (p_{\,12}\cdot\partial_{\xi_{\,3}}\,-\,p_{\,3}\cdot\partial_{\xi_{\,3}})\,i\,\cD_{\,3}\,+\,\frac{1}{4}\,\partial_{\xi_{\,3}} \cdot\partial_{\xi_{\,3}}\,(p_{\,3}^{\,2}\,-\,p_{\,2}^{\,2}\,+\,p_{\,1}^2)\right]\nonumber\\
\vphantom{\left(\!\pm\sqrt{\frac{\a^{\,\prime}\!\!}{2}}\,\right)^{\,3}}
- &&\!\!\!\!\!\!\!\!\!\!\!\!\!\!\left(\!\pm\,\sqrt{\frac{\a^{\,\prime}\!\!}{2}}\,\right)^{\,3}
\left[(\partial_{\xi_{\,1}}\cdot\partial_{\xi_{\,2}}\,+\,1)\,i\,\cD_{\,2}\,-\,\frac{1}{4}\,\partial_{\xi_{\,2}} \cdot\partial_{\xi_{\,2}}\,(p_{\,23}\cdot\partial_{\xi_{\,1}}\,-\,p_{\,1}\cdot\partial_{\xi_{\,1}})\right]\nonumber\\ \vphantom{\left(\!\pm\sqrt{\frac{\a^{\,\prime}\!\!}{2}}\,\right)^{\,3}}
&&\quad\ \ \times\,\left[(\partial_{\xi_{\,3}}\cdot\partial_{\xi_{\,1}}\,+\,1)\,i\,\cD_{\,1}\,-\,\frac{1}{4}\,\partial_{\xi_{\,1}} \cdot\partial_{\xi_{\,1}}\,(p_{\,12}\cdot\partial_{\xi_{\,3}}\,-\,p_{\,3}\cdot\partial_{\xi_{\,3}})\right]\nonumber\\ \vphantom{\left(\!\pm\sqrt{\frac{\a^{\,\prime}\!\!}{2}}\,\right)^{\,3}}
&&\quad\ \ \times\,\left[(\partial_{\xi_{\,2}}\cdot\partial_{\xi_{\,3}}\,+\,1)\,i\,\cD_{\,3}\,-\,\frac{1}{4}\,\partial_{\xi_{\,3}} \cdot\partial_{\xi_{\,3}}\,(p_{\,31}\cdot\partial_{\xi_{\,2}}\,-\,p_{\,2}\cdot\partial_{\xi_{\,2}})\right]
\vphantom{\left(\!\pm\sqrt{\frac{\a^{\,\prime}\!\!}{2}}\,\right)^{\,3}}\nonumber\\
+ &&\!\!\!\!\!\!\!\!\!\!\!\!\!\!\left(\!\pm\,\sqrt{\frac{\a^{\,\prime}\!\!}{2}}\,\right)^{\,3}\ \frac{1}{2}\ \partial_{\xi_{\,2}}\cdot\partial_{\xi_{\,2}}\,(1\,+\,\partial_{\xi_{\,1}}\cdot\partial_{\xi_{\,2}})\nonumber\\
&&\quad\ \ \times\,\left[(\partial_{\xi_{\,3}}\cdot\partial_{\xi_{\,1}}\,+\,1)\,i\,\cD_{\,1}\,-\,\frac{1}{4}\,\partial_{\xi_{\,1}} \cdot\partial_{\xi_{\,1}}\,(p_{\,12}\cdot\partial_{\xi_{\,3}}\,-\,p_{\,3}\cdot\partial_{\xi_{\,3}})\right]\nonumber\\ \vphantom{\left(\!\pm\sqrt{\frac{\a^{\,\prime}\!\!}{2}}\,\right)^{\,3}}
&&\left.\quad\ \ \times\,\left[-\,\frac{1}{2}\ (p_{\,12}\cdot\partial_{\xi_{\,3}}\,+\,p_{\,3}\cdot\partial_{\xi_{\,3}})\,i\,\cD_{\,3}\,-\,\frac{1}{4}\,\partial_{\xi_{\,3}} \cdot\partial_{\xi_{\,3}}\,(p_{\,1}^{\,2}\,-\,p_{\,3}^{\,2}\,-\,p_{\,2}^2)\right]
\right\}\nonumber\\&&\times
\,\bar{\psi}_{\,1\,\a}(\xi_{\,1})\,\psi_{\,2\,\b}(\xi_{\,2})\,\phi_{\,3}(\xi_{\,3})\ \Bigg|_{\xi_{\,i}\,=\,0}\nonumber\\
&+&\exp\Bigg\{\pm\ \sqrt{\frac{\a^{\, \prime}\!\!}{2}}\ \Big[(\partial_{\xi_{\,1}}\cdot\partial_{\xi_{\,2}}+1)
(\partial_{\xi_{\,3}}\cdot p_{\,12})\,+\,(\partial_{\xi_{\,2}}\cdot\partial_{\xi_{\,3}}+1)(\partial_{\xi_{\,1}}\cdot p_{\,23})\nonumber\\&&+\,(\partial_{\xi_{\,3}}\cdot\partial_{\xi_{\,1}}+1)(\partial_{\xi_{\,2}}\cdot
p_{\,31})\Big]\Bigg\}\nonumber\\
&&\!\!\!\!\!\!\!\!\!\!\!\!\!\!\!\!\!\!\!\!\!\!\!
\times\left\{\left(\!\pm\,\sqrt{\frac{\a^{\,\prime}\!\!}{2}}\,\right)^{\phantom{\,2}}\,\slashed{\partial}_{\xi_{\,1}}
\left[(\partial_{\xi_{\,2}}\cdot\partial_{\xi_{\,3}}\,+\,1)\,i\,\cD_{\,3}\,-\,\frac{1}{4}\,\partial_{\xi_{\,3}} \cdot\partial_{\xi_{\,3}}\,(p_{\,31}\cdot\partial_{\xi_{\,2}}\,-\,p_{\,2}\cdot\partial_{\xi_{\,2}})\right] \nonumber\right.\\ \vphantom{\left(\!\pm\sqrt{\frac{\a^{\,\prime}\!\!}{2}}\,\right)^{\,3}}
&&\!\!\!\!\!\!\!\!\!\!\!\!\!\!\!\!\!\!\!-\,\left(\!\pm\,\sqrt{\frac{\a^{\,\prime}\!\!}{2}}\,\right)^{\phantom{\,2}}\, \slashed{\partial}_{\xi_{\,2}}
\left[(\partial_{\xi_{\,3}}\cdot\partial_{\xi_{\,1}}\,+\,1)\,i\,\cD_{\,3}\,+\,\frac{1}{4}\,\partial_{\xi_{\,3}} \cdot\partial_{\xi_{\,3}}\,(p_{\,23}\cdot\partial_{\xi_{\,1}}\,+\,p_{\,1}\cdot\partial_{\xi_{\,1}})\right] \nonumber\\ \vphantom{\left(\!\pm\sqrt{\frac{\a^{\,\prime}\!\!}{2}}\,\right)^{\,3}}
&&\!\!\!\!\!\!\!\!\!\!\!\!\!\!\!+\ \ \frac{\a^{\,\prime}\!\!}{2}\ \ \slashed{\partial}_{\xi_{\,1}}
\left[(\partial_{\xi_{\,2}}\cdot\partial_{\xi_{\,3}}\,+\,1)\,i\,\cD_{\,2}\,+\,\frac{1}{4}\,\partial_{\xi_{\,2}} \cdot\partial_{\xi_{\,2}}\,(p_{\,12}\cdot\partial_{\xi_{\,3}}\,+\,p_{\,3}\cdot\partial_{\xi_{\,3}})\right] \nonumber\\ \vphantom{\left(\!\pm\sqrt{\frac{\a^{\,\prime}\!\!}{2}}\,\right)^{\,3}}
&&\qquad\quad \ \ \ \times\left[(\partial_{\xi_{\,3}}\cdot\partial_{\xi_{\,1}}\,+\,1)\,i\,\cD_{\,3}\,+\,\frac{1}{4}\,\partial_{\xi_{\,3}} \cdot\partial_{\xi_{\,3}}\,(p_{\,23}\cdot\partial_{\xi_{\,1}}\,+\,p_{\,1}\cdot\partial_{\xi_{\,1}})\right] \nonumber\\ \vphantom{\left(\!\pm\sqrt{\frac{\a^{\,\prime}\!\!}{2}}\,\right)^{\,3}}
&&\!\!\!\!\!\!\!\!\!\!\!\!\!\!\!+\ \ \frac{\a^{\,\prime}\!\!}{2}\ \ \slashed{\partial}_{\xi_{\,2}}
\left[(\partial_{\xi_{\,3}}\cdot\partial_{\xi_{\,1}}\,+\,1)\,i\,\cD_{\,1}\,-\,\frac{1}{4}\,\partial_{\xi_{\,1}} \cdot\partial_{\xi_{\,1}}\,(p_{\,12}\cdot\partial_{\xi_{\,3}}\,-\,p_{\,3}\cdot\partial_{\xi_{\,3}})\right] \nonumber\\ \vphantom{\left(\!\pm\sqrt{\frac{\a^{\,\prime}\!\!}{2}}\,\right)^{\,3}}
&&\qquad\quad \ \ \ \times\left[(\partial_{\xi_{\,2}}\cdot\partial_{\xi_{\,3}}\,+\,1)\,i\,\cD_{\,3}\,-\,\frac{1}{4}\,\partial_{\xi_{\,3}} \cdot\partial_{\xi_{\,3}}\,(p_{\,31}\cdot\partial_{\xi_{\,2}}\,-\,p_{\,2}\cdot\partial_{\xi_{\,2}})\right] \nonumber\\ && \left.\vphantom{\left(\!\pm\sqrt{\frac{\a^{\,\prime}\!\!}{2}}\,\right)^{\,3}}\right\}^{\,\a\b}\
\,\bar{\psi}_{\,1\,\a}(\xi_{\,1})\,\psi_{\,2\,\b}(\xi_{\,2})\,\phi_{\,3}(\xi_{\,3})\ \Bigg|_{\xi_{\,i}\,=\,0}
\ ,
\end{eqnarray}}
\!\!that again can be extended to the unconstrained forms proceeding as for the
bosonic vertices.

The advantage of leaving out divergences and traces can be further appreciated considering the deformations of the gauge algebra induced by the cubic couplings that we have described and their implications for quartic couplings. Leaving aside, for brevity, all Chan--Paton factors and focussing on the terms proportional to $p^{\,2}$ that are present in the gauge variation of
\begin{multline}
\cA^{\,[0]}\,=\,\exp\, \left\{\sqrt{\frac{\a^{\, \prime}\!\!}{2}}\ \Big[(\partial_{\xi_{\,1}}\cdot\partial_{\xi_{\,2}}\,+\,1)\,\partial_{\xi_{\,3}}\cdot p_{\,12}\,+\,(\partial_{\xi_{\,2}}\cdot\partial_{\xi_{\,3}}\,+\,1)\,\partial_{\xi_{\,1}}\cdot p_{\,23}\,\right.\\\left.+\,(\partial_{\xi_{\,3}}\cdot\partial_{\xi_{\,1}}\,+\,1)\,\partial_{\xi_{\,2}}\cdot p_{\,31}\Big]\vphantom{\sqrt{\frac{\a^{\, \prime}\!\!}{2}}}\right\} \, \phi_{\,1}(p_{\,1}\,,\,\xi_{\,1})\,\phi_{\,2}(p_{\,2}\,,\,\xi_{\,2})\,\phi_{\,3}(p_{\,3}\,,\,\xi_{\,3})\ ,
\end{multline}
one can identify the first on--shell non--linear complement of the free gauge transformation,
\be
\delta\cA^{\,[0]}\,=\,\underbrace{\sqrt{\frac{\a^{\, \prime}\!\!}{2}}\ \exp(\cG)\ (\partial_{\xi_{\,2}}\cdot\partial_{\xi_{\,3}}\,+\,1)\,\L_{\,1}(p_{\,1}\,,\,\xi_{\,1})
\,\phi_{\,2}(p_{\,2}\,,\,\xi_{\,2})}_{-\,\delta^{(1)} \phi_{\,3}(-\,p_{\,3}\,,\,\xi_{\,3})}\,p_{\,3}^{\,2}\,\phi_{\,3}(p_{\,3}\,,\,\xi_{\,3})\ .
\ee

For brevity, we have displayed this result in a schematic form, leaving the contributions associated to different permutations of the fields and their Chan--Paton factors implicit and not distinguishing trivial deformations that can be absorbed by field redefinitions from non--trivial ones. Still, this result provides a starting point to extend the N\"oether procedure to quartic order, again up to divergences and traces. In fact, having identified the quadratic gauge transformations
\begin{multline}
\delta^{(1)} \phi\,(p\,,\,\xi)\,=\,\sqrt{\frac{\a^{\, \prime}\!\!}{2}}\ \exp\left\{\sqrt{\frac{\a^{\, \prime}\!\!}{2}}\ \Big[(\partial_{\xi_{\,3}}\cdot\partial_{\xi_{\,4}}\,+\,1)\,\xi\cdot p_{\,34}\,+\,2\,(\partial_{\xi_{\,4}}\cdot\xi\,+\,1)\,\partial_{\xi_{\,3}}\cdot p_{\,4}\right. \\ \left. -\,2\,(\xi\cdot \partial_{\xi_{\,3}}\,+\,1)\,\partial_{\xi_{\,4}}\cdot p_{\,3}\Big]\vphantom{\sqrt{\frac{\a^{\, \prime}\!\!}{2}}}\right\} \,(\xi\cdot\partial_{\xi_{\,4}}\,+\,1)\,\phi_{\,3}(p_{\,3}\,,\,\xi_{\,3})\,\L_{\,4}(p_{\,4}\,,\,\xi_{\,4})\ ,
\end{multline}
here displayed schematically and without Chan--Paton factors, the corresponding gauge variation of the cubic coupling reads
{\allowdisplaybreaks
\begin{eqnarray}
\delta^{(1)}\cA_3&=&\sqrt{\frac{\a^{\, \prime}\!\!}{2}}\ \exp\left\{\frac{\a^{\,\prime}\!\!}{2}\,\Big[(1\,+\,\partial_{\xi_{\,1}}\cdot\partial_{\xi_{\,2}})(1\,+\,\partial_{\xi_{\,3}}
\cdot\partial_{\xi_{\,4}}) \,p_{\,12}\cdot p_{\,34}\right.+\nonumber\\
&+&2\,(1\,+\,\partial_{\xi_{\,1}}\cdot\partial_{\xi_{\,2}})\,\Big[p_{\,12}\cdot \partial_{\xi_{\,4}}\,\partial_{\xi_{\,3}}\cdot p_{\,4}\,-\,p_{\,12}\cdot \partial_{\xi_{\,3}}\,\partial_{\xi_{\,4}}\cdot p_{\,3}\Big]\nonumber\\
&+&2\,(1\,+\,\partial_{\xi_{\,3}}\cdot\partial_{\xi_{\,4}})\,\Big[p_{\,34}\cdot \partial_{\xi_{\,2}}\,\partial_{\xi_{\,1}}\cdot p_{\,2}\,-\,p_{\,34}\cdot \partial_{\xi_{\,1}}\,\partial_{\xi_{\,2}}\cdot p_{\,1}\Big]\nonumber\\
&+&4\,\Big(\partial_{\xi_{\,1}}\cdot\partial_{\xi_{\,3}}\, \partial_{\xi_{\,2}}\cdot p_{\,1}\, \partial_{\xi_{\,4}}\cdot p_{\,3}\,+\,\partial_{\xi_{\,2}}\cdot\partial_{\xi_{\,4}} \, \partial_{\xi_{\,1}}\cdot p_{\,2} \, \partial_{\xi_{\,3}}\cdot p_{\,4}\nonumber\\
&&-\,\partial_{\xi_{\,1}}\cdot\partial_{\xi_{\,4}} \, \partial_{\xi_{\,2}}\cdot p_{\,1} \, \partial_{\xi_{\,3}}\cdot p_{\,4}\,-\,\partial_{\xi_{\,2}}\cdot\partial_{\xi_{\,3}} \, \partial_{\xi_{\,1}}\cdot p_{\,2} \, \partial_{\xi_{\,4}}\cdot p_{\,3}\Big)\Big]\nonumber\\
&+&\left. \sqrt{2 \a^{\,\prime}}\,\Big(\partial_{\xi_{\,1}}\cdot p_{\,2}\,-\,\partial_{\xi_{\,2}}\cdot p_{\,1}\,+\,\partial_{\xi_{\,3}}\cdot p_{\,4}\,-\,\partial_{\xi_{\,4}}\cdot p_{\,3}\Big)\vphantom{\frac{\a^\prime}{2}}\right\}\nonumber\\
&\times&\left\{1\,+\,\sqrt{\frac{\a^{\, \prime}\!\!}{2}}\ \Big[1\,+\,\partial_{\xi_{\,1}}\cdot\partial_{\xi_{\,2}})\, p_{\,12}\cdot\partial_{\xi_{\,3}}
\,+\,2\,\partial_{\xi_{\,2}}\cdot\partial_{\xi_{\,3}}\,\partial_{\xi_{\,1}}\cdot p_{\,2}\,-\,2\,\partial_{\xi_{\,1}}\cdot\partial_{\xi_{\,3}}\,\partial_{\xi_{\,2}}\cdot p_{\,1}\Big]\right\}\nonumber\\
&& \times \ \phi_{\,1}(p_{\,1}\,,\,\xi_{\,1})\,\phi_{\,2}(p_{\,2}\,,\,\xi_{\,2})\,\phi_{\,3}(p_{\,3}\,,\,\xi_{\,3})\,\L_{\,4}(p_{\,4}\,,\,\xi_{\,4})\ .\label{delta1}
\end{eqnarray}}
\!\!If one chooses not to add auxiliary fields, a relatively simple \emph{non--local} quartic coupling is determined by the condition that its linear gauge variation cancel exactly this contribution, and reads
{\allowdisplaybreaks
\begin{eqnarray}
\cA_4^{(s)}&=&\frac{1}{s}\ \sqrt{\frac{2}{\a^{\, \prime}\!\!}}\ \ \exp\left\{\frac{\a^{\,\prime}\!\!}{2}\,\Big[(1\,+\,\partial_{\xi_{\,1}}\cdot\partial_{\xi_{\,2}})(1\,+\,\partial_{\xi_{\,3}}
\cdot\partial_{\xi_{\,4}}) \,p_{\,12}\cdot p_{\,34}\right.+\nonumber\\
&+&2\,(1\,+\,\partial_{\xi_{\,1}}\cdot\partial_{\xi_{\,2}})\,\Big[p_{\,12}\cdot \partial_{\xi_{\,4}}\,\partial_{\xi_{\,3}}\cdot p_{\,4}\,-\,p_{\,12}\cdot \partial_{\xi_{\,3}}\,\partial_{\xi_{\,4}}\cdot p_{\,3}\Big]\nonumber\\
&+&2\,(1\,+\,\partial_{\xi_{\,3}}\cdot\partial_{\xi_{\,4}})\,\Big[p_{\,34}\cdot \partial_{\xi_{\,2}}\,\partial_{\xi_{\,1}}\cdot p_{\,2}\,-\,p_{\,34}\cdot \partial_{\xi_{\,1}}\,\partial_{\xi_{\,2}}\cdot p_{\,1}\Big]\nonumber\\
&+&4\,\Big(\partial_{\xi_{\,1}}\cdot\partial_{\xi_{\,3}}\, \partial_{\xi_{\,2}}\cdot p_{\,1}\, \partial_{\xi_{\,4}}\cdot p_{\,3}\,+\,\partial_{\xi_{\,2}}\cdot\partial_{\xi_{\,4}} \, \partial_{\xi_{\,1}}\cdot p_{\,2} \, \partial_{\xi_{\,3}}\cdot p_{\,4}\nonumber\\
&&-\,\partial_{\xi_{\,1}}\cdot\partial_{\xi_{\,4}} \, \partial_{\xi_{\,2}}\cdot p_{\,1} \, \partial_{\xi_{\,3}}\cdot p_{\,4}\,-\,\partial_{\xi_{\,2}}\cdot\partial_{\xi_{\,3}} \, \partial_{\xi_{\,1}}\cdot p_{\,2} \, \partial_{\xi_{\,4}}\cdot p_{\,3}\Big)\Big]\nonumber\\
&+&\left. \sqrt{2 \a^{\,\prime}}\,\Big(\partial_{\xi_{\,1}}\cdot p_{\,2}\,-\,\partial_{\xi_{\,2}}\cdot p_{\,1}\,+\,\partial_{\xi_{\,3}}\cdot p_{\,4}\,-\,\partial_{\xi_{\,4}}\cdot p_{\,3}\Big)\vphantom{\frac{\a^\prime}{2}}\right\} \nonumber\\
&&\times \ \phi_{\,1}(p_{\,1}\,,\,\xi_{\,1})\,\phi_{\,2}(p_{\,2}\,,\,\xi_{\,2})\,\phi_{\,3}(p_{\,3}\,,\,\xi_{\,3})\, \phi_{\,4}(p_{\,4}\,,\,\xi_{\,4})\ ,
\end{eqnarray}}
\!\!This should be combined with two additional contributions,
\be
\cA_4^{(t)}\,=\,\cA_4^{(s)}(2\ra 3,3\ra 2)\qquad \cA_4^{(u)}\,=\,\cA_4^{(s)}(2\ra 4,4\ra 2)\ ,
\ee
so that the full answer is
\be
\cA_{\,4}\,=\,\frac{1}{3}\,\left(\cA_{\,4}^{(s)}\,+\,\cA_{\,4}^{(t)}\,+\,\cA_{\,4}^{(u)}\right)\ ,
\ee
where the three contributions can be associated to the three Mandelstam variables $s$, $t$ and $u$.
Eventually, one should also reinstate Chan-Paton factors along the lines of the string result \eqref{fouramplitude}.

Remarkably, this quartic coupling affords a simple presentation in terms of the $\star$--product, so that for instance
\begin{multline}
\cA_{\,4}^{(s)}\,=\,\sqrt{\frac{2}{\a^{\, \prime}\!\!}}\ \ \frac{1}{(p_{\,1}\,+\,p_{\,2})^{\,2}}\ \, e^{\,\cG_{\,12}(\xi)}\,\star\ e^{\,\cG_{\,34}(\xi)}\\\times\,\phi_{\,1}(p_{\,1}\,,\,\xi_{\,1})\,\phi_{\,2}(p_{\,2}\,,\,\xi_{\,2})\,\phi_{\,3}(p_{\,3}\,,\,\xi_{\,3}) \,\phi_{\,4}(p_{\,4}\,,\,\xi_{\,4})\Big|_{\,\xi_{\,i}\,=\,0}\ ,
\end{multline}
where
\be
\cG_{\,ij}(\xi)\,=\,\sqrt{\frac{\a^{\, \prime}\!\!}{2}}\ \ \Big[ (\partial_{\xi_{\,i}}\cdot\partial_{\xi_{\,j}}\,+\,1)\,\xi\cdot p_{\,ij}\,+\,2\,(\partial_{\xi_{\,j}}\cdot\xi\,+\,1)\,\partial_{\xi_{\,i}}\cdot p_{\,j}\,-\,2\,(\xi\cdot\partial_{\xi_{\,i}}\,+\,1)\,\partial_{\xi_{\,j}}\cdot p_{\,i}\Big]\ ,
\ee
again with Chan--Paton factors left implicit. Different contributions with various orderings of Chan--Paton factors are actually encoded in this expression, provided $\phi(\xi)$ is taken to be matrix valued.

This way of presenting the result can shed some light on generalizations to higher--point functions and also on the off--shell extension of this coupling, since it rests recursively on the cubic coupling. With these considerations, exploiting the off--shell completion \eqref{offcubic}, the off--shell quartic coupling would be schematically of the form
\be
\cA_{\,4}^{\,\text{off--shell}}\,=\,\cA_{\,3}^{\,\text{off--shell}}\,\frac{\star}{\square}\ \cA_{\,3}^{\,\text{off--shell}}\ .
\ee
We leave a more detailed analysis of this important issue and of other related developments for the future.

\end{appendix}
\newpage

\end{document}